\documentclass[%
 reprint,
 amsmath,amssymb,
 aps,
 prd,
]{revtex4-2}

\usepackage{graphicx}
\usepackage{dcolumn}
\usepackage{bm}
\usepackage[dvipsnames, usenames]{xcolor}
\usepackage{acronym}
\usepackage{hyperref}

\newcommand{\chieff}{\chi_{\rm eff}}

\newcommand{\dataset}{\{d_i\}_{i=1}^{N_{\rm obs}}}
\newcommand{\beq}{\begin{equation}}
\newcommand{\eeq}{\end{equation}}
\newcommand{\nn}{\nonumber}
\newcommand{\parint}[1]{\int {\rm d}#1}
\newcommand{\thetaint}{\parint{\theta}}
\newcommand{\dataint}{\int_{\mathcal{D}_{\uparrow}} {\rm d}d}
\newcommand{\Msun}{\;M_\odot}
\newcommand{\pp}{{\sc Power-Law + Peak}}
\newcommand{\pr}{{\sc Power-Law Redshift}}

\newcommand{\LIGOlabMIT}{\affiliation{LIGO, Massachusetts Institute of Technology, 77 Massachusetts Avenue, Cambridge, MA 02139, USA}}
\newcommand{\MKI}{\affiliation{Kavli Institute for Astrophysics and Space Research and Department of Physics, Massachusetts Institute of Technology, 77 Massachusetts Avenue, Cambridge, MA 02139, USA}}
\newcommand{\CIERA}{\affiliation{Center for Interdisciplinary Exploration and Research in Astrophysics (CIERA), Northwestern University, 1800 Sherman Ave, Evanston, IL 60201, USA}}

\begin{document}

\title{Probing Correlations in the Binary Black Hole Population with Flexible Models}

\author{Jack Heinzel}
\email{heinzelj@mit.edu}\LIGOlabMIT\MKI
\author{Sylvia Biscoveanu}%
\email{NASA Einstein Fellow}\LIGOlabMIT\MKI\CIERA
\author{Salvatore Vitale}\LIGOlabMIT\MKI
\email{}
\date{\today}

\begin{abstract}
The astrophysical formation channels of binary black hole systems predict correlations between their mass, spin, and redshift distributions, which can be probed with gravitational-wave observations. Population-level analysis of the latest LIGO-Virgo-KAGRA catalog of binary black hole mergers has identified evidence for such correlations assuming linear evolution of the mean and width of the effective spin distribution as a function of the binary mass ratio and merger redshift. However, the complex astrophysical processes at play in compact binary formation do not necessarily predict linear relationships between the distributions of these parameters.
In this work, we relax the assumption of linearity and instead search for correlations using a more flexible cubic spline model. Our results suggest a nonlinear correlation between the width of the effective spin distribution and redshift. We also show that the LIGO-Virgo-Kagra collaborations may find convincing Bayesian evidence for nonlinear correlations by the end of the fourth observing run, O4. This highlights the valuable role of flexible models in population analyses of compact-object binaries in the era of growing catalogs. 

\end{abstract}

\maketitle

\acrodef{GWTC-3}[GWTC-3]{third gravitational-wave transient catalog}
\acrodef{BBH}[BBH]{binary black hole}
\acrodef{LVK}[LVK]{LIGO-Virgo-Kagra collaboration}
\acrodef{KL}[KL]{Kullback-Leibler}
\acrodef{JS}[JS]{Jensen-Shannon}
\acrodef{GW}[GW]{gravitational wave}


\section{Introduction}
The growing catalog of gravitational-wave observations of \acf{BBH} mergers has allowed for increasingly detailed probes of the population properties of these systems, with the ultimate goal of revealing how they form and evolve. Analysis of the \acf{GWTC-3}~\cite{LIGOScientific:2021djp} of the \acf{LVK}~\cite{TheLIGOScientific:2014jea, TheVirgo:2014hva, Aso:2013eba, Somiya:2011np, KAGRA:2020tym} found evidence for substructure in the \ac{BBH} primary mass distribution~\cite{KAGRA:2021duu, Tiwari:2020otp, Edelman:2021zkw} beyond a single power-law~\cite{Fishbach:2017zga} plus Gaussian peak~\cite{Talbot:2018cva} and for preferentially equal-mass mergers~\cite{Fishbach:2019bbm, Farah:2023swu}. The black hole spin distribution favors small~\cite{Wysocki:2018} but likely nonzero spins~\cite{Biscoveanu:2020are, Callister:2022qwb, Tong:2022iws, Mould:2022xeu, Galaudage:2021rkt, Roulet:2021hcu}, and the distribution of the spin tilts relative to the orbital angular momentum is consistent with isotropy~\cite{Roulet:2021hcu, KAGRA:2021duu, Vitale:2022dpa}. The merger rate evolves with redshift at a rate consistent with the cosmic star formation rate~\cite{Fishbach:2021yvy}. Taken altogether, these constraints imply that there are probably multiple formation channels~\cite{Zevin:2020gbd, Cheng:2023ddt, Wong:2020ise} shaping the \ac{BBH} population, although no definitive evidence for sub-populations with different properties has been identified~\cite{Baibhav:2022qxm, Godfrey:2023oxb, Wang:2022gnx}.

While most previous analyses of the \ac{BBH} population have relied on phenomenological population models with simple parametric functional forms, recent work has explored the use of ``non-parametric'' models like splines~\cite{Edelman:2022ydv, Golomb:2022bon}, Dirichlet processes~\cite{Rinaldi:2021bhm}, generic mixture models~\cite{Tiwari:2020otp,Tiwari:2020vym,Tiwari:2021yvr}, and autoregressive models~\cite{Callister:2023tgi}. Despite including many more free parameters than the phenomenological models, these non-parametric models offer increased flexibility to fit the data without imposing a specific functional form. This avoids the issue of model misspecification~\cite[e.g.,][]{Romero-Shaw:2022ctb} at the cost of a clear mapping between features observed in the data and those predicted by astrophysical theory.

As the observed population of \ac{BBH} mergers grows, population analyses have moved beyond modeling the mass, spin, and redshift distributions independently towards searching for correlations between these parameters.
Such correlations are expected both within individual formation channels and due to the superposition of sub-populations forming via distinct channels~\cite{Bavera:2022mef, Bavera:2020inc,McKernan:2012rf, Stone:2016wzz, McKernan:2019beu, McKernan:2021nwk, Tagawa:2019osr,Zevin:2022wrw}. For example, \acp{BBH} formed via isolated binary evolution may exhibit correlations due to the relationship between metallicity and the efficiency of angular momentum transport via stellar winds, which remove mass and spin down the progenitor, meaning that more massive systems would preferentially have high spins and be found at high redshift where stellar winds are less efficient because of lower metallicities~\cite{Belczynski:2017gds}.
A mass-spin correlation is also expected for systems formed dynamically via hierarchical mergers in dense environments, since remnants of previous mergers that go on to merge again will be both more massive and rapidly spinning~\cite[e.g.,][]{PortegiesZwart:2002iks, Rodriguez:2015oxa, Gerosa:2021mno}. 

Evidence for a correlation between \ac{BBH} mass and spin was obtained using a phenomenological model where the mean and width of the distribution of effective aligned spin ($\chieff$) vary linearly as a function of the mass ratio, such that systems with more unequal mass ratios have larger effective aligned spins~\citep{Callister:2021fpo, KAGRA:2021duu}. An approach using copula density functions that ensure fixed marginal distributions in the presence of a correlation identifies this correlation at similar significance~\cite{Adamcewicz:2022hce, Adamcewicz:2023mov}. Weaker evidence using the linear correlation model also hints at a potential correlation between the effective spin distribution and the total mass and primary mass~\cite{Safarzadeh:2020mlb, Franciolini:2022iaa, Biscoveanu:2022qac}, although a broadening of the spin distribution at the highest masses can be explained due to the relative dearth of observations in this regime, leading to a more uncertain measurement~\cite{Tiwari:2021yvr}. A likely correlation between the width of the effective spin distribution and redshift has also been identified~\cite{Biscoveanu:2022qac}. Recently, using a non-parametric method, Ref.~\cite{Rinaldi:2023bbd} reports a correlation in primary mass-redshift space, arising from two sub-populations. However, some results of this work are in tension with previous population analyses, including both parametric and nonparametric methods \cite[e.g.][]{vanSon:2021zpk, Tiwari:2021yvr, Edelman:2022ydv, Callister:2023tgi}. 

In this work, we build upon previous methods looking for correlations between individual pairs of parameters describing the \ac{BBH} mass, spin, and redshift distributions but adopt a more flexible model for the shape of the correlation. Specifically, we assume the \ac{BBH} effective spin distribution is described by a Gaussian with an unknown mean and width, both of which may correlate with either the mass or redshift such that the shape of the correlation is given by a cubic spline. Our model recovers evidence for the previously-identified correlations between effective spin and mass ratio and redshift but prefers a nonlinear shape for the spin-redshift correlation. This result highlights the important role that flexible population models will play in identifying model misspecification as the catalog of \ac{BBH} merger observations grows. The rest of this work is structured as follows. 

In \S \ref{sec:probe_correlations} we briefly describe our statistical assumptions, which are standard in \acf{GW} population inference, then describe our model for probing correlations in more detail. In \S \ref{sec:GWTC3_results} we present the constraints on the BBH data collected thus far \cite{LIGOScientific:2021djp}, which tend to be broadly consistent with previous studies apart from some evidence for nonlinearity in the $\chieff-z$ correlation. 
In \S \ref{sec:o4_projections} we make projections for the future: how well can nonlinearity be measured with future observations? In a pair of simulated Universes with nonlinear correlations in $\chieff-q$ and $\chieff-z$, nonlinearity does not reveal itself in the $\chieff-q$ distribution with 400 detections, but there is strong evidence for nonlinearity in the $\chieff-z$ distribution with 400 detections. While the detectability of nonlinearity ultimately is subject to how nonlinear the true correlated distribution is, we show it is possible to detect nonlinearity in the near future. 
Finally, we conclude in \S \ref{sec:conclusion}.

\section{Probing Correlations: Priors and Parameterization}
\label{sec:probe_correlations}

The goal of population modeling is to infer the distribution from which an ensemble of observations is drawn. This can be accomplished using hierarchical Bayesian inference, which takes a multi-stage approach by first characterizing individual observations and then combining them on a population level.
In a \ac{GW} context, these individual observations are noisy, so the statistical likelihood of observing the data given a population model $p(\theta | \Lambda)$ parameterized by $\Lambda$ must be marginalized over the possible \ac{GW} parameters \cite{Loredo_2004}
\beq
\mathcal{L}(d | \Lambda) = \thetaint \mathcal{L}(d|\theta) p(\theta | \Lambda).
\label{eq:marginalized_like}
\eeq
Here, $d$ represents the detected data, and $\theta$ represents the unknown source parameters, like the binary masses, spins, and redshift. 

Furthermore, BBHs suffer a Malmquist bias; they are not all equally detectable. To account for this bias, we must define the detection efficiency
\beq
\alpha(\Lambda) = \dataint \thetaint \mathcal{L}(d|\theta) p(\theta | \Lambda),
\label{eq:selection_eff}
\eeq
which is the fraction of events in the population $p(\theta | \Lambda)$ which probabilistically generate detectable data ($d \in \mathcal{D}_{\uparrow}$). Assuming the events are distributed in time by a Poisson process and marginalizing over the Poisson rate parameter with a uniform-in-$\log$ prior, we obtain the rate-marginalized hierarchical likelihood \citep{Loredo_2004, Farr:2013yna, Foreman-Mackey_2014, Mandel_2019, Vitale:2020aaz, Essick:2023upv}.
In particular, with a collection of $N_{\rm obs}$ events with data $\dataset$ passing a detection threshold 
\beq
\mathcal{L}(\dataset | \Lambda) \propto \prod_{i=1}^{N_{\rm obs}} \frac{\mathcal{L}(d_i|\Lambda)}{\alpha(\Lambda)}.
\label{eq:hierarchical_like}
\eeq
We use this likelihood to sample from the hyperparameters $\Lambda$. In practice, we estimate the marginal integrals in Eqs. \ref{eq:marginalized_like} and \ref{eq:selection_eff} with Monte Carlo estimators, see Refs. \cite{Farr_2019, Essick:2022ojx, Talbot:2023pex} for details.

A common approach tackles 
the hierarchical inference problem by phenomenologically parameterizing the unknown source distribution as e.g. power-laws, Gaussians, etc., and inferring these hyperparameters given the data observations. Previous studies have identified simple and successful parameterization schemes; we choose the astrophysically motivated parameterizations \pp~\cite{Talbot:2018cva}
and \pr~\cite{Fishbach:2018edt} for the mass $p(m_1, q|\Lambda)$ and redshift $p(z|\Lambda)$ distributions respectively. The primary mass distribution is parameterized as a smoothed power-law plus a gaussian component, and a smoothed power-law for the mass ratio (hyperparameters are the minimum and maximum BH mass--assumed to be the same for both the primary and secondary--power-law index, low mass smoothing parameter, mean and width of the Gaussian, fraction of BBHs in the Gaussian component, and the power-law index for the mass ratio). The redshift distribution is modelled as a power-law with a single power-law index hyperparameter. 

For the spins, we project the 6 dimensional spin distribution to a one dimensional parameter $\chieff$ describing the leading order spin effect on the inspiral evolution of the binary \cite{Damour:2001,Racine:2008qv,Ajith:2009bn}, which is often the most well measured spin parameter \cite{Vitale:2016avz, Purrer:2015nkh}. 
We then model the $\chieff$ population as a truncated Gaussian distribution with a variable mean and width~\cite{Miller:2020zox, Roulet:2018jbe}
\beq 
p(\chieff|\theta, \Lambda) = \frac{1}{\sigma(\theta, \Lambda)}\frac{\phi\left(\frac{\chieff - \mu(\theta, \Lambda)}{\sigma(\theta, \Lambda)}\right)}{\Phi\left(\frac{1-\mu(\theta, \Lambda)}{\sigma(\theta, \Lambda)}\right) - \Phi\left(\frac{-1-\mu(\theta, \Lambda)}{\sigma(\theta, \Lambda)}\right)}
\label{eq:general_chieff_correlation}
\eeq
where the truncation restricts the support of the distribution to be $[-1,1]$, which reflects the physical constraint that the magnitude of $\chieff$ cannot exceed 1 for BH spins bounded by the Kerr limit. $\phi$ is the standard Gaussian and $\Phi$ is the standard error function, 
\beq
\phi(x) = \frac{e^{-x^2/2}}{\sqrt{2\pi}} \qquad \Phi(x) = \int_{-\infty}^x \phi(s)ds.
\label{eq:standard_gaus}
\eeq
This is the approach used by Refs.~\cite{KAGRA:2021duu,Safarzadeh:2020mlb,Callister:2021fpo,Biscoveanu:2022qac} to explore models where $\mu(\theta)$ and $\sigma(\theta)$ are linear functions of primary mass ($\theta = m_1$) mass ratio ($\theta = q$) and redshift ($\theta = z$) respectively. Building on this previous work, we model $\mu(\theta)$ and $\sigma(\theta)$ with spline functions.

Spline functions are becoming increasingly popular in \ac{GW} population inference, primarily because they are innately flexible and fast to evaluate, but also because they are easily parameterized by their nodes. While splines do not assume much structure, they do fail to probe structure below the scale of the node separation length. For this reason, one should include the number of nodes as a model hyperparameter or repeat the analysis varying the choice of the number of nodes. 

In this work, we use a cubic spline model, where nodes are interpolated using cubic polynomials which preserve continuity in the function and first and second derivatives at the nodes ($\mathcal{C}^2$ functions). Given node locations, this provides all but two conditions to set the four coefficients for each cubic polynomial; the final two conditions are given at the endpoints, typically imposed by setting the second derivative to zero. Defined in this way, the node positions fully determine the spline curve. Splines also have the advantage of approximate locality, meaning they can fit a structure in one region of parameter space independently of the behavior of the spline far away (separated by many nodes)~\cite{DeBoor_1976}. 

For all the inferences we present below, we use priors shown in Table \ref{tab:prior_table}. We use the {\tt Overall} samples for GWTC1 events~\cite{gwtc1_pe}, {\tt PrecessingSpinIMRHM} for the events first identified in GWTC2~\cite{gwtc2_pe} and the {\tt C01:IMRPhenomXPHM} samples for the GWTC2.1 and GWTC3 events~\cite{ligo_scientific_collaboration_and_virgo_2022_6513631, ligo_scientific_collaboration_and_virgo_2023_8177023}, and the search sensitivity estimates provided in Ref.~\cite{ligo_scientific_collaboration_and_virgo_2023_7890398}. We use the {\tt GWPopulation} package for constructing the hierarchical likelihood~\cite{Talbot:2019}, and compute Bayesian evidences while sampling the hyperposterior using the {\tt Dynesty} implementation in {\tt Bilby}~\cite{Speagle_2020,Ashton:2018jfp}. To efficiently evaluate spline functions, we use the {\tt cached\_interpolate} package introduced in Ref.~\cite{Golomb:2022bon}. In addition, because we estimate the population likelihood (Eq. \ref{eq:hierarchical_like}) with Monte Carlo integrals, there is inherent uncertainty associated with each likelihood estimate. To avoid biased inference, we cut hyperposterior samples with $\log$-likelihood uncertainty greater than 1, following the recommendation of Ref.~\cite{Talbot:2023pex}.

\renewcommand{\arraystretch}{1.3}
\renewcommand{\tabcolsep}{0.3cm}
\begin{table*}
    \centering
    \begin{tabular}{|l||ll|}
       \hline Hyperparameter & Description & Prior \\
       \hline $\alpha$ & $m_1$ power-law index & U(-4, 12) \\
       $\beta$ & $q$ power-law index & U(-4, 7) \\
       $m_{\rm max}$ & maximum BH mass & U$(60\Msun, 100\Msun)$ \\
       $m_{\rm min}$ & minimum BH mass & U$(2\Msun, 10\Msun)$ \\
       $\delta_m$ & low-mass smoothing parameter & U$(0\Msun, 10\Msun)$ \\
       $\mu_{m}$ & $m_1$ Gaussian component mean & U$(20\Msun, 50\Msun)$ \\
       $\sigma_{m}$ & $m_1$ Gaussian component standard deviation  & U$(1\Msun, 10\Msun)$ \\
       $\lambda$ & fraction of BBHs in Gaussian component & U(0, 0.2) \\
       $\lambda_z$ & $z$ power-law index & U(-2, 10) \\
       \hline $\mu_{\chieff:n}$ & $n^{\rm th}$ spline node for the $\chieff$ mean & U$(-1,1)$ \\
       $\ln\sigma_{\chieff:n}$ & $n^{\rm th}$ spline node for the $\chieff$ standard deviation & U$(-5,0)$
       \\ \hline
       
    \end{tabular}
    \caption{Priors for each hyperparameter used in our model. The upper panel describes the standard priors, except where they are reduced to improve the sampling efficiency without cutting off the hyperposterior ($m_{\rm max}$ and $\lambda$). The lower panel describes the prior assumed for our spline parameters.}
    \label{tab:prior_table}
\end{table*}

\section{Correlations in \ac{GWTC-3}}
\label{sec:GWTC3_results}

Using the catalog of 69 BBHs described in Ref.~\cite{KAGRA:2021duu}  passing a detection-pipeline-computed false alarm rate threshold of 1 yr${}^{-1}$, we search for evidence of nonlinear correlations between the spin distribution and the mass ratio, redshift, and the source frame primary mass (the latter is described in appendix \S~\ref{app:chieff_m}). 
To do this, we model the $\chieff$ distribution with Eq. \ref{eq:general_chieff_correlation}, and present the results of the individual analyses below.


\subsection{Effective Spin Distribution and Mass Ratio}

The first correlation we explore is between $\chieff$ and mass ratio $q$, setting the mean and standard deviations to spline functions of mass ratio. We place the nodes uniformly between $q=0$ and $q=1$~\footnote{Though $q\to0$ is an unphysical region of parameter space with no observations, the node at $q=0$ should be thought of as simply a parameter necessary to ensure the model is defined over the entire space (any distribution must be defined over the full parameter space), and not an \textit{a priori} statement that BBHs exist here; indeed the $p(q\;|\;m_1, \Lambda)$ smoothed powerlaw is always zero at $q=0$.},
\begin{align}
\mu(\theta) &= S\left(q\;|\;(0,\mu_{\chieff:0}), ..., (1,\mu_{\chieff:N})\right) \nn \\
\ln \sigma(\theta) &= S\left(q\,|\,(0,\ln \sigma_{\chieff:0}), ..., (1,\ln \sigma_{\chieff:N})\right),
\label{eq:chieff_q_spline}
\end{align}
where $S(x\;|\;(X_1, Y_1), ..., (X_N, Y_N))$ represents the cubic spline function in the variable $x$ passing through the nodes with $x$ coordinates $X_1 < X_2 < ... < X_N$ and corresponding $y$ coordinates $Y_1, Y_2, ..., Y_N$. In all our models, we fix the $x$ coordinates to reduce the dimension of the inference.

\begin{figure}
    \includegraphics[width=0.9\linewidth]{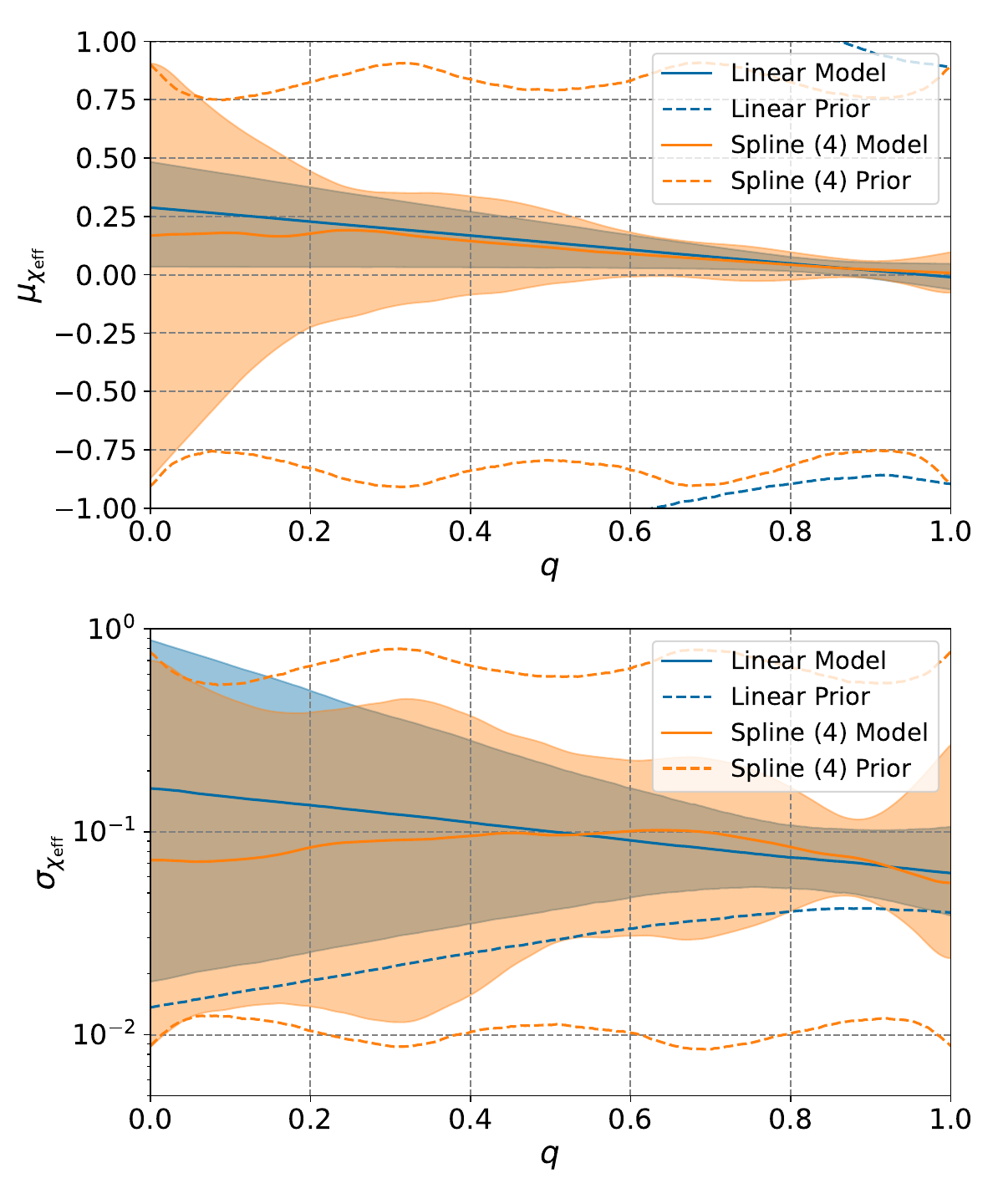}
    \caption{A comparison between the linear model and 4 node spline model for correlation between $\chieff$ and mass ratio, inferred using \ac{GWTC-3}. Solid lines represent the median, while the shaded region represents the central 90\% credible interval. Dashed lines show the upper and lower boundaries of the prior 90\% interval. The upper panel shows the mean of the $\chieff$ Gaussian as a function of mass ratio $q$, while the lower panel is the standard deviation of the $\chieff$ distribution.}
    \label{fig:chieff_q_s4_v_linear}
\end{figure}

In Fig. \ref{fig:chieff_q_s4_v_linear}, we show a comparison between our results with the linear model of Ref.~\cite{Callister:2021fpo} to the spline model with 4 nodes. While our spline model results are broadly consistent with the linear model, they generically feature broader credible intervals towards extreme mass ratios $q \to 0$. We argue that this is an advantage of the spline model, as the data should have less information about BBHs with unequal mass ratios as they are less common in the detected population~\cite[e.g.,][]{Fishbach:2019bbm, Farah:2023swu} and be completely uninformative about events with mass ratio $q \to 0$. All of the 69 BBHs in the catalog are inconsistent with vanishing mass ratios, and what's more, our population model with hyperparameter $m_{\rm min} \ge 2\Msun$ (the minimum BH mass must exceed $2\Msun$, see Ref. \cite{Talbot:2018cva}) requires zero support at $q\to0$. Being an approximately local model, the spline model can fit the structure at near-equal mass ratios, while simultaneously saying nothing about the behaviour of extreme mass ratio events. Hence, the posterior approaches the prior as $q\to 0$. We also repeat the analysis with 3-6 nodes and obtain results consistent with the 4-node analysis presented here, see appendix \S \ref{app:extra_analysis}.

\begin{figure}
    \centering
    \includegraphics[width=0.9\linewidth]{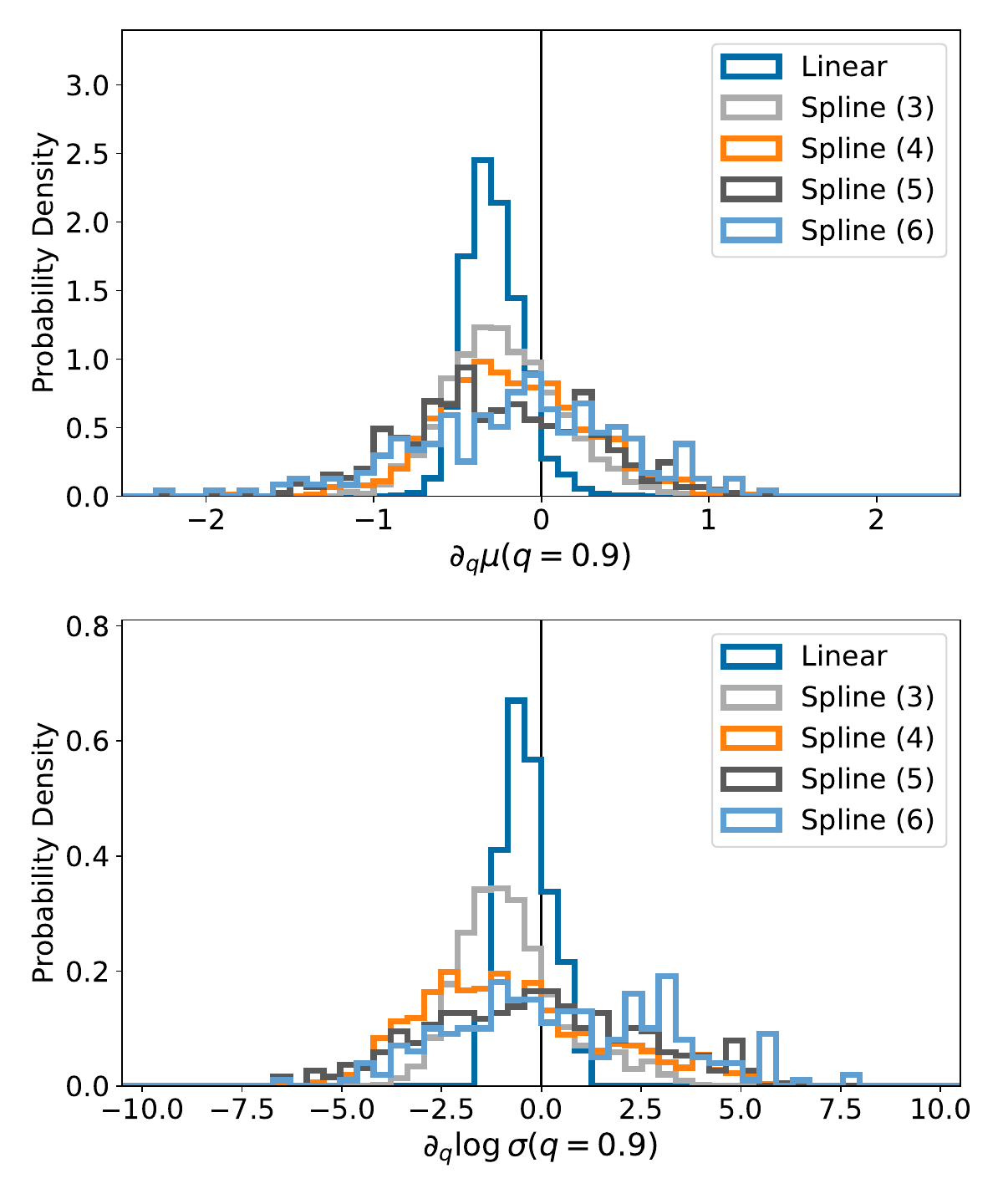}
    \caption{Posteriors on the derivative of the mean and width of the $\chieff$ distribution as a function of mass ratio, calculated at $q^* = 0.9$, and inferred using GWTC 3. The linear model concludes that the mean of the $\chieff$ distribution decreases as mass ratio approaches $1$, with a significance of $\sim 98\%$. The spline models are more agnostic, finding significances of $\sim 60-75\%$.}
    \label{fig:chieff_q_slope}
\end{figure}

To compute the significance of any evolution with mass ratio, we compute the derivative of the inferred evolution with respect to mass ratio. Spline functions are easily differentiable, and so we can compute the slope of the spline function at an arbitrary point $q^*$. We choose the fiducial value of $q^* = 0.9$, as this appears to be a  well constrained region and a good proxy for understanding the evolution of the spin population in the region of near-equal mass ratios. This gives us posteriors on the slope at this point $q^*$ for each model, which we show in Fig. \ref{fig:chieff_q_slope}. We then compute the significance of an evolution in the mean of the Gaussian as a function of mass ratio from the fraction of the posterior support with positive/negative slope. The linear model has a negative slope with significance $\sim 98\%$, as initially reported in Ref. \cite{Callister:2021fpo}. The spline models has negative slope with more inconclusive significance: $\sim 70-80\%$. We show all calculated slope significances at their fiducial values in Table \ref{tab:slope_significance} in appendix \S \ref{app:extra_analysis}.

\subsection{Effective Spin Distribution and Redshift}

Next, we turn our attention to the correlation between $\chieff$ and redshift $z$. Ref.~\cite{Biscoveanu:2022qac} 
examined a correlation between redshift and the effective spin distribution, and discovered evidence for a broadening in the effective spin distribution, a positive correlation between the width $\sigma(z)$ of the $\chieff$ Gaussian and the redshift, and no evidence for any trend in the mean of the Gaussian. In their analysis, Ref.~\cite{Biscoveanu:2022qac} parameterized $\mu(z) = \mu_0 + \delta\mu_z (z-0.5)$ and $\log_{10}\sigma(z) = \log\sigma_0 + \delta\log\sigma_z (z-0.5)$ as linear models and quantified the significance of the measured broadening using the posterior on the slope $\delta\log\sigma_z$.

In a similar approach, we model the $\chieff$ correlation with redshift using Eq. \ref{eq:general_chieff_correlation}, only we use 
\begin{align}
\mu(\theta) &= S\left(z\;|\;(0,\mu_{\chieff:0}), ..., (2.3,\mu_{\chieff:N})\right) \nn \\
\ln \sigma(\theta) &= S\left(z\;|\; (0,\ln \sigma_{\chieff:0}), ..., (2.3,\ln \sigma_{\chieff:N})\right).
\label{eq:chieff_z_spline}
\end{align}
In our model, the first and last nodes are placed at redshifts $z=0$ and $z=2.3$, the maximum redshift we assume in the \pr~model~\footnote{This is somewhat at odds with the maximum redshift considered in the sensitivity injections of~\cite{ligo_scientific_collaboration_and_virgo_2023_7890398}, with $z_{\rm max} = 1.9$. We verified our results are unchanged after considering this adjustment.}. The hyperparameters associated to the splines are the $y$ coordinates of the nodes. We explored models that fix the nodes uniformly between the first and last nodes; however, these resulted in nodes too coarsely spread at small redshift to optimally fit the structure. Additionally, there is limited information in the data thus far to constrain the population much beyond redshift $z \gtrsim 1$, so a better node spacing should place nodes tightly at small redshift and more loosely at high redshift. Heuristically, we found that a linear spacing in $z^{1/2}$ places nodes satisfactorily. 

Using these models to infer the BBH population hyperparameters given the \ac{GWTC-3} dataset, we infer the evolution of the mean and width of the $\chieff$ Gaussian as a function of redshift. We show a comparison between a linear model and the spline model with 4 nodes in Fig. \ref{fig:chieff_z_s4_v_linear}. We show the results for a collection of analyses with 3-6 spline nodes in appendix \S \ref{app:extra_analysis} in Fig. \ref{fig:chieff_correlations} and the associated model evidences in Fig. \ref{fig:chieff_evidences}.

\begin{figure}
    \centering
    \includegraphics[width=0.9\linewidth]{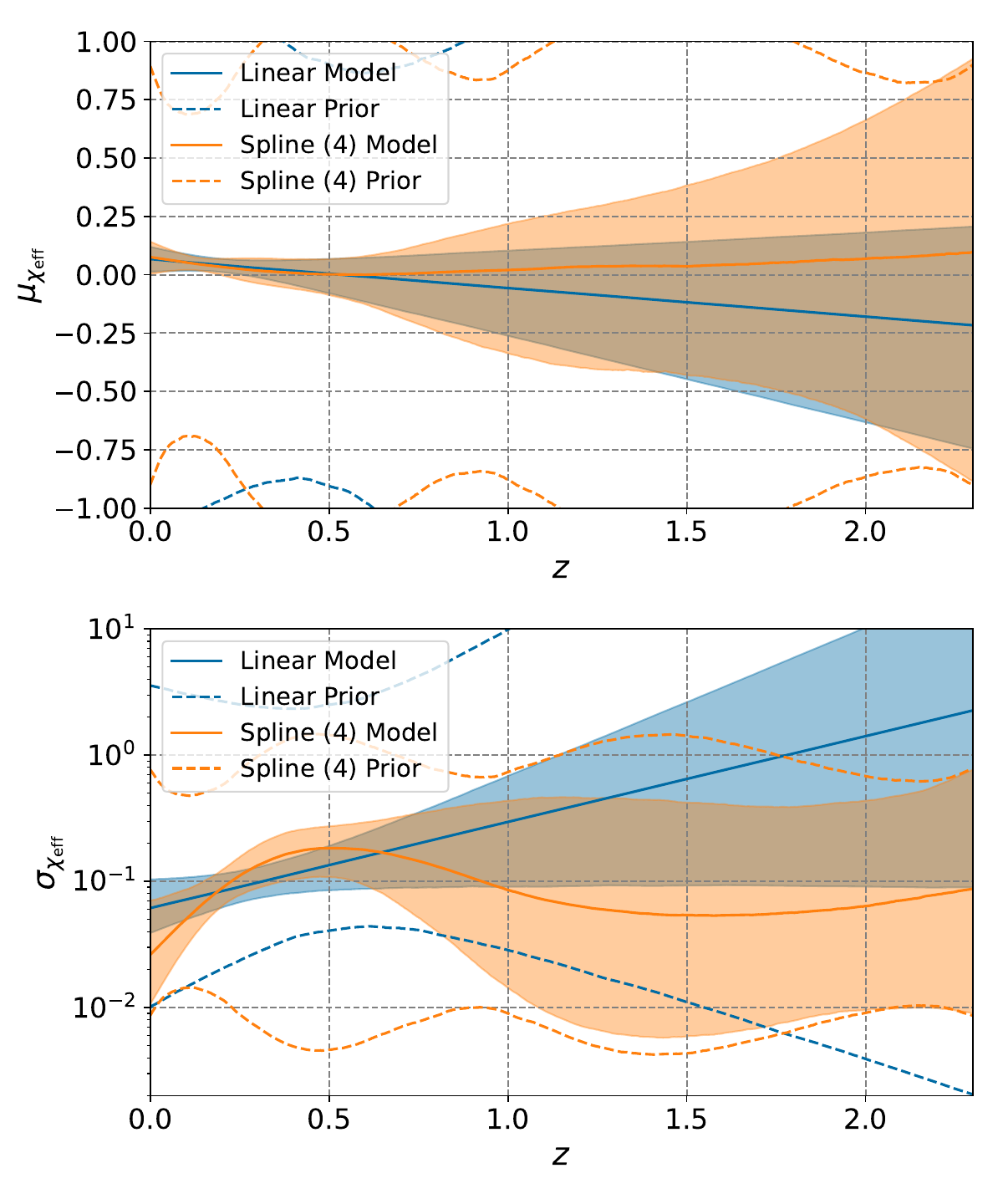}
    \caption{A comparison between the linear model and 4 node spline model for correlation between $\chieff$ and redshift, inferred using \ac{GWTC-3}. Solid lines represent the median, while the shaded region represents the central 90\% credible interval, and the dashed lines show the boundary of the prior 90\% interval. The upper panel shows the mean of the $\chieff$ Gaussian as a function of redshift $z$, while the lower panel represents the standard deviation of the $\chieff$ distribution.}
    \label{fig:chieff_z_s4_v_linear}
\end{figure}

First, we quantify the confidence of a broadening slope for each model. To this end, we compute the slope of the width of the Gaussian at a fiducial value $z^*$, $\partial \log \sigma(z){z=z^*}$ for all models. 
We choose the fiducial value $z^* = 0.2$ (see appendix \S \ref{app:nonlinear_z} for a discussion on this choice), and show histograms of the slopes for each model in appendix \S \ref{app:nonlinear_z} in Fig. \ref{fig:chieff_z_slope}. We can then quantify the significance of the increase by the proportion of the posterior with slope greater than zero. 

There are varying degrees of evidence that the $\chieff$ distribution is broadening as a function of redshift at $z^*=0.2$, depending on the model used. The mean is decreasing at $\sim 80-95\%$ confidence (the posterior support with slope less than 0), depending on the model assumed. The width is increasing at $90-98.6\%$ confidence, consistent with the finding of Ref.~\cite{Biscoveanu:2022qac} that the width of the $\chieff$ distribution broadens with increasing redshift. We collect all significances calculated at fiducial values in Table \ref{tab:slope_significance} in appendix \S \ref{app:extra_analysis}.

\begin{figure*}
    \centering
    \includegraphics[width=0.9\linewidth]{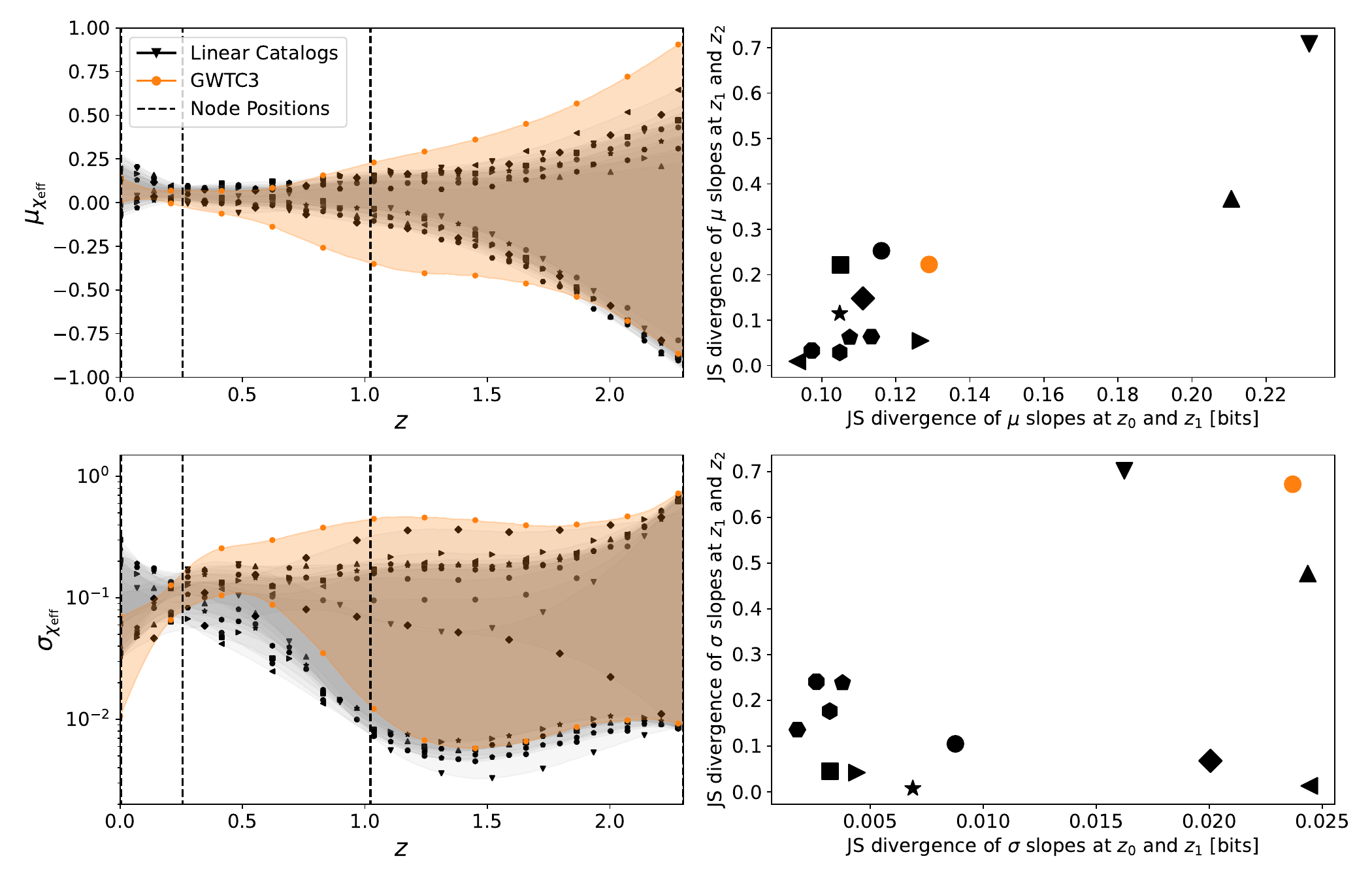}
    \caption{Comparison between the spline inference on a linearly correlated Universe and the inference on \ac{GWTC-3}. The first row shows the mean and the second row the standard deviation. The left column shows the 90\% credible intervals inferred using \ac{GWTC-3} (orange) and the 12 synthetic catalogs of 69 events drawn from a linearly correlated Universe (gray). The right column shows the \acf{JS} divergences (in bits) between slope posteriors taken at the first two nodes on the x-axis, and between the second and third nodes on the y-axis. For each linear catalog there is a black point on the scatter plot, and the orange point corresponds to the \ac{GWTC-3} divergences. This represents the posterior difference between the two slopes, quantifying the nonlinearity. The further the point is from the origin, the more confident we are in the inferred nonlinearity. Notice there is no inference on a linear Universe which is more nonlinear than \ac{GWTC-3} in both dimensions.}
    \label{fig:nonlinearity_quantification}
\end{figure*}

It also appears that there is some nonlinearity in the evolution of the width as a function of redshift. To understand if this degree of inferred nonlinearity is expected in a Universe with a linear correlation, we perform the spline model inference with 4 nodes on 12 catalogs of 69 events drawn from a linearly correlated Universe, (see appendix \S \ref{app:nonlinear_z} for details on the selection procedure and the generation of the synthetic catalogs). We then compute the derivative at the first three nodes $z_0$, $z_1$ and $z_2$ (ignoring the last node since the posterior is essentially the prior there) and calculate the \acf{JS} divergence~\cite{kullback1951, Lin:1991zzm} between the slope posteriors. A linearly correlated Universe would theoretically have slope posteriors all consistent with the true value, however there will be some random scatter between the posteriors, measured by the \ac{JS} divergence between them. Looking at a scatter plot of the divergence between the slope at $z_0$ and $z_1$ on the x-axis and between $z_1$ and $z_2$ on the y-axis in Fig. \ref{fig:nonlinearity_quantification}, notice that there are no divergences from a linearly correlated Universe which is more extreme than the \ac{GWTC-3} divergences in both axes. This points towards a nonlinear trend in the width of the $\chieff$ distribution as a function of redshift, though the Bayesian evidence ($\mathcal{Z} = \int \pi(\Lambda) \mathcal{L}(\{d_i\}|\Lambda) d\Lambda$) is not yet conclusive (see Fig. \ref{fig:chieff_evidences} in appendix \S \ref{app:extra_analysis})




\section{Future Prospects: Can We Detect Nonlinearity in O4?}
\label{sec:o4_projections}
Flexible spline models stand in contrast to linear models, where the correlation around each spline node is independently inferred, rather than enforcing a consistent slope across the whole parameter space. 
At the conclusion of O3, these flexible models highlight our lack of knowledge about poorly constrained regions. While we have some hints of a nonlinear correlation in $\chieff-z$, there is not yet a definitive preference in the Bayesian evidence for a nonlinear correlation. This may change by the end of the fourth observing run, O4. 

To predict how well we may actually detect nonlinear correlations in the future, we produced two synthetic catalogs of 200 events and 400 events for two different possible Universes (so four catalogs total). The events are detected by a network of LIGO interferometers (located at Hanford and Livingston), assuming the fiducial O4 noise spectra with an average BNS inspiral range of 160 Mpc~\cite{O3_psds}. We draw \ac{GW} events from a few different populations with true hyperparameters given in Table \ref{tab:true_hyper}. These populations are consistent with the data collected by the \ac{LVK} thus far, analyzed with the models we presented above. We use the waveform model \texttt{IMRPhenomXP} with \texttt{PrecVersion=104}~\cite{Pratten:2020ceb}, and use the heterodyning/relative binning scheme of Refs.~\cite{Cornish:2010kf, Cornish:2021lje, Zackay:2018qdy} to efficiently sample the \ac{GW} event parameter posteriors.

\renewcommand{\arraystretch}{1.25}
\begin{table*}
    \centering
    \begin{tabular}{|l||lll|}
       \hline Hyperparameter & Description & $q-\chieff$ Correlation & $z-\chieff$ Correlation \\
       \hline $\alpha$ & $m_1$ power-law index & $3$ & $3$ \\
       $\beta$ & $q$ power-law index & 1 & 1 \\
       $m_{\rm max}$ & maximum BH mass & $85\Msun$ & $85 \Msun$ \\
       $m_{\rm min}$ & minimum BH mass & $5\Msun$ & $5 \Msun$ \\
       $\delta_m$ & low-mass smoothing parameter & $3\Msun$ & $3\Msun$ \\
       $\mu_{m}$ & $m_1$ Gaussian component mean & $35\Msun$ & $35 \Msun$ \\
       $\sigma_{m}$ & $m_1$ Gaussian component standard deviation  & $5\Msun$ & $5\Msun$ \\
       $\lambda$ & fraction of BBHs in Gaussian component & $0.03$ & $0.03$ \\
       $\lambda_z$ & $z$ power-law index & 2 & 2 \\ \hline
       $(x_0, \mu_{\chieff:0})$ & first mean spline node coordinates & $(0, 0.4)$ & $(0, 0)$ \\
       $(x_1, \mu_{\chieff:1})$ & second mean spline node coordinates & $(0.4, 0.3)$ & $(0.3, 0)$ \\
       $(x_2, \mu_{\chieff:2})$ & third mean spline node coordinates & $(0.8, 0.05)$ & $(0.65, 0)$ \\
       $(x_3, \mu_{\chieff:3})$ & fourth mean spline node coordinates & $(1, 0.02)$ & $(2.3, 0)$ \\
       $(x_0, \ln\sigma_{\chieff:0})$ & first standard deviation spline node coordinates & $(0, -2.5)$ & $(0, -3.5)$ \\
       $(x_1, \ln\sigma_{\chieff:1})$ & second standard deviation spline node coordinates & $(0.4, -2.5)$ & $(0.3, -2)$ \\
       $(x_2, \ln\sigma_{\chieff:2})$ & third standard deviation spline node coordinates & $(0.8, -2.5)$ & $(0.65, -1.5)$ \\
       $(x_3, \ln\sigma_{\chieff:3})$ & fourth standard deviation spline node coordinates & $(1, -2.5)$ & $(2.3, -1.25)$ \\ \hline
    \end{tabular}
    \caption{Hyperparameters for the simulated Universes with nonlinear correlation. In the left column are the hyperparameters for the Universe with a correlation between $\chieff$ and mass ratio, and in the right column is the correlation with redshift. The correlation functions are themselves splines, with node placements given by e.g. the $(z_N, \mu_{\chieff:N})$ coordinate pairs.}
    \label{tab:true_hyper}
\end{table*}

We then select detected events on a network matched-filter SNR $\rho_{\rm mf, HL} > 9$, and generate a set of Monte Carlo injections for estimating the selection efficiency consistent with this detection criterion \cite{Essick:2023toz}. 

We recover each simulated catalog assuming both a linear model and the spline model with 4 nodes. We restrict to one spline analysis using 4 nodes to limit the computational expense. 

We decided to explore two node placement options. First, we recover using our original node placement scheme, placed in the same manner as described above (linear in $q$ and $z^{1/2}$). This node placement does not match the true node placement (Table \ref{tab:true_hyper}), and the inference finds more ``bumps'' in the correlation than are truly there. To this end, we also recover using the true node placement scheme. See appendix \S \ref{app:node_placement_bias} for a discussion on node placement and bias in the inference.

\begin{figure}
    \centering
    \includegraphics[width=0.9\linewidth]{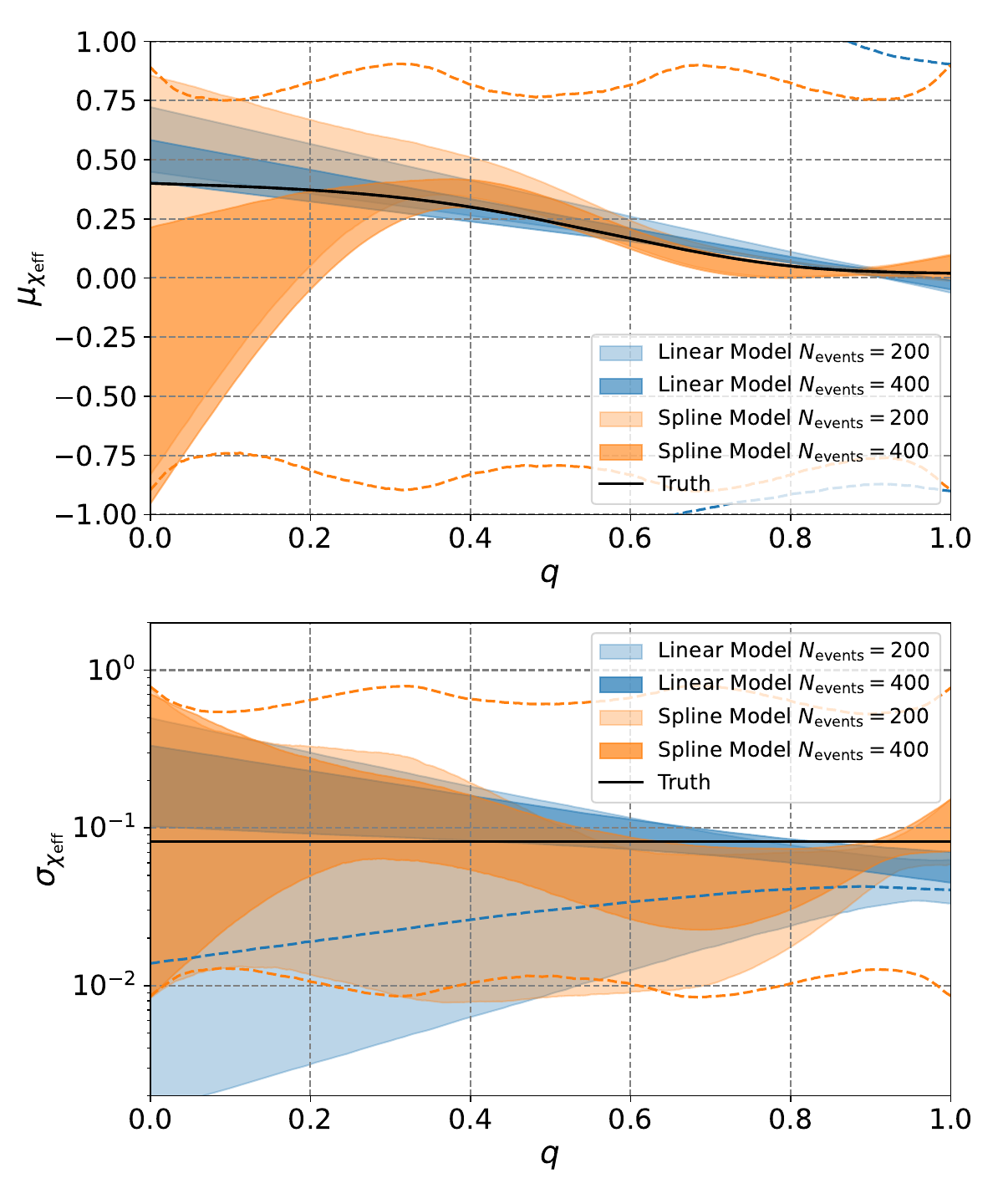}
    \caption{Inferred mean and width of the $\chieff$ distribution as a function of mass ratio $q$ on two mock catalogs. In blue are the 90\% credible intervals when assuming a linear model for the correlation, while in orange are the 90\% intervals inferred under the spline model with 4 nodes. The lighter shades represent the inference on a catalog with 200 events, while the darker shade is with 400 events. Dashed lines represent the prior 90\% region, and in solid black is the true correlation.}
    \label{fig:chieff_q_mock_catalog_s4_v_linear}
\end{figure}

We show our results comparing the linear model and the spline model with our original node placement (linear in $q$ and $z^{1/2}$) in Fig. \ref{fig:chieff_q_mock_catalog_s4_v_linear}.

In a catalog of 200 detections, the model evidences indicate no significant preference for a linear model or the spline model. Indeed, the Bayes factor of the spline model over the linear model with 200 events is $\log_{\rm 10}\mathcal{B}_{s|l}(200) = 0.00 \pm 0.13$. However, when we increase the number of detections to 400, the data begins to indicate a slight preference for the spline model, although nothing yet conclusive; $\log_{\rm 10}\mathcal{B}_{s|l}(400) = 0.28 \pm 0.14$. We emphasize that this holds for a particular choice of a nonlinear correlation. In truth, the correlation may be more or less linear than the one we simulated, which would make evidence of nonlinearity correspondingly more or less definitive at these numbers of events.

\begin{figure}
    \centering
    \includegraphics[width=0.9\linewidth]{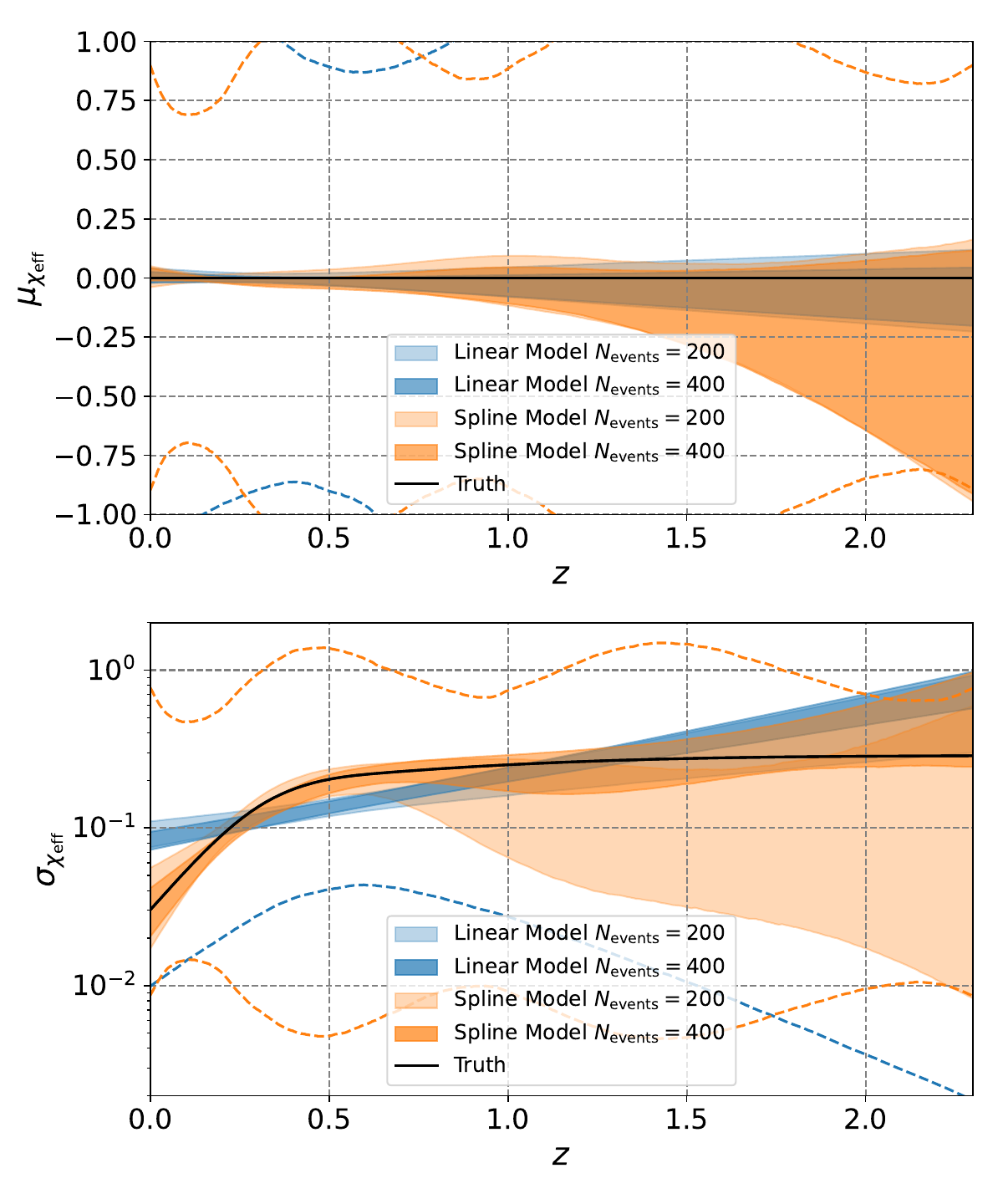}
    \caption{Inferred mean and width of the $\chieff$ distribution as a function of redshift $z$ on two mock catalogs. In blue are the 90\% credible intervals when assuming a linear model for the correlation, while in orange are the 90\% intervals inferred under the spline model with 4 nodes. The lighter shades are the inference using a catalog of 200 events, while the darker shade is with 400 events. Dashed lines represent the prior 90\% region, and in solid black is the true correlation.}
    \label{fig:chieff_z_mock_catalog_s4_v_linear}
\end{figure}

To understand how we may constrain the correlation between the spin population and the redshift in the future, we also simulated a Universe with a nonlinear correlation between the width of the $\chieff$ distribution and redshift (see Table \ref{tab:true_hyper}), where the model hyperparameters chosen are consistent with the catalog through the end of the third observing run. 

We once again produced mock catalogs with 200 and 400 detections, and we show the results of the inferences in Fig. \ref{fig:chieff_z_mock_catalog_s4_v_linear}. This time, even a false node placement scheme does a good job at fitting the correlation. Furthermore, model evidences become conclusively in favor of the nonlinear model with a catalog of 400 detections. In particular, for the redshift correlation $\log_{\rm 10}\mathcal{B}_{s|l}(200) = 1.06 \pm 0.14$ and $\log_{\rm 10}\mathcal{B}_{s|l}(400) = 4.09 \pm 0.16$.
We also checked that a spline model with nodes placed in the correct locations produce consistent results, as expected.

We reiterate that the results of this projection study depend on the assumed true correlation. However, since the correlation we chose is consistent with the data collected thus far, this demonstrates that we \textit{can} detect nonlinearity in the spin correlation with a catalog of $\sim400$ detections. 

\section{Conclusions}
\label{sec:conclusion}
In this paper we presented a flexible model for understanding the correlation between the spin population in $\chieff$ and primary mass $m_1$, mass ratio $q$ or redshift $z$. On the \ac{LVK} data published thus far, we obtain results broadly consistent with previous analyses \cite{Safarzadeh:2020mlb,Callister:2021fpo,KAGRA:2021duu,Biscoveanu:2022qac,Franciolini:2022iaa}. Furthermore, because of their flexibility, these spline models highlight the regions of parameter space that drive a measured correlation, and the regions of parameter space which are more uncertain. 

In particular, we find that the mean of the $\chieff$ distribution likely increases with mass ratio, the width likely broadens with redshift, and may also broaden for more massive binaries. Importantly, it is also possible that not all of these claims are true at the same time. Perhaps a correlation with one parameter may masquerade as a correlation with another parameter when analyzed under the false hypothesis. Ref. \cite{Biscoveanu:2022qac} studied this possibility using linear models which simultaneously fit the correlation with $m$, $q$ and $z$, and found that the data is not yet informative enough to answer this question, though it appears possible that all correlations are real. While we do not consider simultaneous models in this work, we do observe the Bayes factors all appear consistent, suggesting the data is not yet informative enough to pick out any mismodeled correlations.
 

Each of these potential correlations may prove to be important probes of the astrophysical environments which produce BBHs. The observed anticorrelation between the mean of the $\chieff$ distribution with mass ratio may indicate that the BBHs observed come from binaries which experience mass ratio reversal using an optimistic common envelope (CE) prescription \cite{Broekgaarden:2022nst}, or isolated binaries which either undergo a CE phase with large CE efficiency or stable mass transfer with super-Eddington accretion \cite{Bavera:2020uch, Zevin:2022wrw}. It is also possible that hierarchical mergers in a dense environment can produce the observed correlation; we would expect hierarchical mergers involving just one second-generation black hole from a previous merger to have more extreme mass ratios and higher spins~\cite[e.g.,][]{Gerosa:2021mno}. However, in an environment with isotropic spin symmetry, the $\chieff$ distribution must be centered on zero, and so one would expect a \textit{broadening} of the distribution at small mass ratios, not an increase in the mean. In order to produce the observed correlation, then, the environment must break isotropy symmetry somehow \cite{Santini:2023ukl}, e.g., in an AGN disk \cite{McKernan:2021nwk, Santini:2023ukl}. Finally, if the observed BBHs do not originate from one channnel, but a superposition of multiple \cite[e.g.,][]{Zevin:2020gbd,Cheng:2023ddt}, a (linear) correlation may arise from a Simpson's-type paradox, where multiple populations naturally separated in $\chieff-q$ space are 
interpreted as a correlation \cite{Baibhav:2022qxm}. Especially in this last scenario, a flexible nonlinear model will help shed light on the origin of the $\chieff-q$ correlation as we move into O4. 

The correlation with redshift probes evidence for other kinds of pathways toward merging stellar mass BBHs. This observation is commonly explained by connecting the spin-up of a BH or its progenitor with the delay time to merger. If the spin-up mechanism of BH progenitors is stronger for closer separations, the remnant \ac{BBH} system will radiate energy to GWs more rapidly, and hence merge earlier. Because \acp{BBH} are born at a higher rate at higher redshifts, this results in a positive correlation between spin magnitude and redshift \cite{Zaldarriaga:2017qkw,Bavera:2020inc,Bavera:2021evk,Fuller:2022ysb}. One potential mechanism is tidal torques: the spin-up due to tidal torques is amplified if the binary is at a smaller initial separation. If the stellar progenitors retain some angular momentum upon collapse, the BH spins should be correlated with the observed redshift at merger \cite{Qin:2018vaa, Fuller:2019sxi, Bavera:2022mef}. This is complicated by the expectation that formation in the field produces nearly aligned spins. In this scenario, then there should be a positive correlation between redshift and the mean of the Gaussian; increasing the width requires larger spins with a more isotropic tilt distribution. If there are strong supernova kicks, however, this can even out the spin tilts and give a $\chieff$ distribution more centered on zero \cite{Rodriguez:2016vmx, Gerosa:2018wbw, Callister:2020vyz, Stevenson:2022hmi}.

Finally, a correlation between $\chieff$ and the primary mass is most naturally understood as a signature of hierarchical mergers. In this paper, we observe a broadening of the $\chieff$ distribution as the primary mass increases, with varying levels of confidence depending on the model (see appendix \S \ref{app:chieff_m}). This is precisely the expectation of a hierarchical merger picture in an environment endowed with isotropy symmetry \cite{Gerosa:2021mno, Gerosa:2017kvu, Fishbach:2017dwv, Tagawa:2021ofj}. That said, the most recent studies which directly searched for signatures of hierarchical mergers in the \ac{LVK} catalog strongly disfavor scenarios where all the \ac{LVK} BBHs are formed hierarchically \cite{Fishbach:2022lzq, Doctor:2019ruh, Kimball:2020opk}, though they cannot be ruled out as a subpopulation.

To understand how we might probe nonlinear correlations in the future, we also simulated nonlinearly correlated Universes consistent with the data collected thus far. In a Universe with a nonlinear correlation between $\chieff$ and mass ratio, we found that a linear model may still be appropriate with $\sim 400$ detections, although this is of course heavily dependent on how nonlinear the true correlation is. For a nonlinear correlation in the $z-\chieff$ plane, we find that a linear model may be significantly disfavored as we approach $\sim 400$ detections. Because the nonlinear correlations we assumed were consistent with \ac{LVK} data collected thus far, it is possible that linear models for correlation will begin to fail by the end of O4. 

However, we also encountered a few drawbacks with spline functions. For one, flexible models are intrinsically high dimensional, and thus sampling from the hyperposterior can become significantly more expensive. Second, spline models are perhaps too good at finding nonlinearities, to the point of confidently finding nonlinear features even when they are not present (see appendix \S \ref{app:node_placement_bias}). To be confident in a spline-feature, one should recover the feature varying the number of nodes or the locations of the nodes, or even directly estimate the false alarm probability as in Ref.~\cite{Farah:2023vsc}.

In this paper, we argue that flexible models for correlation offer a complementary but equally important perspective compared to strongly parameterized models. Foremost, spline models can effectively fit a much wider range of potential correlations. We do not \textit{a-priori} expect nature to provide us with linear correlations, or even correlations that can be well-approximated by a line~\cite[e.g.,][]{Belczynski:2017gds, PortegiesZwart:2002iks, Rodriguez:2015oxa, Gerosa:2021mno}.  The correlations in nature may be strongly nonlinear, or perhaps form as a result of a superposition of subpopulations, and these can only be observed when analyzed with a sufficiently flexible model. Of course, it is possible the correlations will indeed turn out to be linear, however we can only observe such a phenomenon by allowing for the alternative.



\section{Acknowledgements}

We thank Tom Callister, Matthew Mould, and Colm Talbot for helpful suggestions and relevant expertise. We also thank Christian Adamcewicz for reviewing this work and for providing helpful comments.
This material is based upon work supported by NSF's LIGO Laboratory which is a major facility fully funded by the National Science Foundation. LIGO was constructed by the California Institute of Technology and Massachusetts Institute of Technology with funding from the National Science Foundation and operates under cooperative agreement PHY-0757058. J.H. and S.B. are supported by the NSF Graduate Research Fellowship under Grant No. DGE-1122374. S.B. is also supported by NASA through the NASA Hubble Fellowship grant HST-HF2-51524.001-A awarded by the Space Telescope Science Institute, which is operated by the Association of Universities for Research in Astronomy, Inc., for NASA, under contract NAS5-26555.
S.V. is also supported by NSF PHY-2045740. The authors are grateful for computational resources provided by the Caltech LIGO Laboratory and supported by NSF PHY-0757058 and PHY-0823459.

\bibliography{pnp_review}

\providecommand{\noopsort}[1]{}\providecommand{\singleletter}[1]{#1}%
\begin{thebibliography}{111}%
\makeatletter
\providecommand \@ifxundefined [1]{%
 \@ifx{#1\undefined}
}%
\providecommand \@ifnum [1]{%
 \ifnum #1\expandafter \@firstoftwo
 \else \expandafter \@secondoftwo
 \fi
}%
\providecommand \@ifx [1]{%
 \ifx #1\expandafter \@firstoftwo
 \else \expandafter \@secondoftwo
 \fi
}%
\providecommand \natexlab [1]{#1}%
\providecommand \enquote  [1]{``#1''}%
\providecommand \bibnamefont  [1]{#1}%
\providecommand \bibfnamefont [1]{#1}%
\providecommand \citenamefont [1]{#1}%
\providecommand \href@noop [0]{\@secondoftwo}%
\providecommand \href [0]{\begingroup \@sanitize@url \@href}%
\providecommand \@href[1]{\@@startlink{#1}\@@href}%
\providecommand \@@href[1]{\endgroup#1\@@endlink}%
\providecommand \@sanitize@url [0]{\catcode `\\12\catcode `\$12\catcode
  `\&12\catcode `\#12\catcode `\^12\catcode `\_12\catcode `\%12\relax}%
\providecommand \@@startlink[1]{}%
\providecommand \@@endlink[0]{}%
\providecommand \url  [0]{\begingroup\@sanitize@url \@url }%
\providecommand \@url [1]{\endgroup\@href {#1}{\urlprefix }}%
\providecommand \urlprefix  [0]{URL }%
\providecommand \Eprint [0]{\href }%
\providecommand \doibase [0]{https://doi.org/}%
\providecommand \selectlanguage [0]{\@gobble}%
\providecommand \bibinfo  [0]{\@secondoftwo}%
\providecommand \bibfield  [0]{\@secondoftwo}%
\providecommand \translation [1]{[#1]}%
\providecommand \BibitemOpen [0]{}%
\providecommand \bibitemStop [0]{}%
\providecommand \bibitemNoStop [0]{.\EOS\space}%
\providecommand \EOS [0]{\spacefactor3000\relax}%
\providecommand \BibitemShut  [1]{\csname bibitem#1\endcsname}%
\let\auto@bib@innerbib\@empty
\bibitem [{\citenamefont {Abbott}\ \emph {et~al.}(2021)\citenamefont {Abbott}
  \emph {et~al.}}]{LIGOScientific:2021djp}%
  \BibitemOpen
  \bibfield  {author} {\bibinfo {author} {\bibfnamefont {R.}~\bibnamefont
  {Abbott}} \emph {et~al.} (\bibinfo {collaboration} {LIGO Scientific, VIRGO,
  KAGRA}),\ }\href@noop {} {\bibinfo {title} {{GWTC-3: Compact Binary
  Coalescences Observed by LIGO and Virgo During the Second Part of the Third
  Observing Run}}} (\bibinfo {year} {2021}),\ \Eprint
  {https://arxiv.org/abs/2111.03606} {arXiv:2111.03606 [gr-qc]} \BibitemShut
  {NoStop}%
\bibitem [{\citenamefont {Aasi}\ \emph {et~al.}(2015)\citenamefont {Aasi} \emph
  {et~al.}}]{TheLIGOScientific:2014jea}%
  \BibitemOpen
  \bibfield  {author} {\bibinfo {author} {\bibfnamefont {J.}~\bibnamefont
  {Aasi}} \emph {et~al.} (\bibinfo {collaboration} {LIGO Scientific}),\
  }\bibfield  {title} {\bibinfo {title} {{Advanced LIGO}},\ }\href
  {https://doi.org/10.1088/0264-9381/32/7/074001} {\bibfield  {journal}
  {\bibinfo  {journal} {Class. Quant. Grav.}\ }\textbf {\bibinfo {volume}
  {32}},\ \bibinfo {pages} {074001} (\bibinfo {year} {2015})},\ \Eprint
  {https://arxiv.org/abs/1411.4547} {arXiv:1411.4547 [gr-qc]} \BibitemShut
  {NoStop}%
\bibitem [{\citenamefont {Acernese}\ \emph {et~al.}(2015)\citenamefont
  {Acernese} \emph {et~al.}}]{TheVirgo:2014hva}%
  \BibitemOpen
  \bibfield  {author} {\bibinfo {author} {\bibfnamefont {F.}~\bibnamefont
  {Acernese}} \emph {et~al.} (\bibinfo {collaboration} {VIRGO}),\ }\bibfield
  {title} {\bibinfo {title} {{Advanced Virgo: a second-generation
  interferometric gravitational wave detector}},\ }\href
  {https://doi.org/10.1088/0264-9381/32/2/024001} {\bibfield  {journal}
  {\bibinfo  {journal} {Class. Quant. Grav.}\ }\textbf {\bibinfo {volume}
  {32}},\ \bibinfo {pages} {024001} (\bibinfo {year} {2015})},\ \Eprint
  {https://arxiv.org/abs/1408.3978} {arXiv:1408.3978 [gr-qc]} \BibitemShut
  {NoStop}%
\bibitem [{\citenamefont {Aso}\ \emph {et~al.}(2013)\citenamefont {Aso},
  \citenamefont {Michimura}, \citenamefont {Somiya}, \citenamefont {Ando},
  \citenamefont {Miyakawa}, \citenamefont {Sekiguchi}, \citenamefont
  {Tatsumi},\ and\ \citenamefont {Yamamoto}}]{Aso:2013eba}%
  \BibitemOpen
  \bibfield  {author} {\bibinfo {author} {\bibfnamefont {Y.}~\bibnamefont
  {Aso}}, \bibinfo {author} {\bibfnamefont {Y.}~\bibnamefont {Michimura}},
  \bibinfo {author} {\bibfnamefont {K.}~\bibnamefont {Somiya}}, \bibinfo
  {author} {\bibfnamefont {M.}~\bibnamefont {Ando}}, \bibinfo {author}
  {\bibfnamefont {O.}~\bibnamefont {Miyakawa}}, \bibinfo {author}
  {\bibfnamefont {T.}~\bibnamefont {Sekiguchi}}, \bibinfo {author}
  {\bibfnamefont {D.}~\bibnamefont {Tatsumi}},\ and\ \bibinfo {author}
  {\bibfnamefont {H.}~\bibnamefont {Yamamoto}} (\bibinfo {collaboration}
  {KAGRA}),\ }\bibfield  {title} {\bibinfo {title} {{Interferometer design of
  the KAGRA gravitational wave detector}},\ }\href
  {https://doi.org/10.1103/PhysRevD.88.043007} {\bibfield  {journal} {\bibinfo
  {journal} {Phys. Rev. D}\ }\textbf {\bibinfo {volume} {88}},\ \bibinfo
  {pages} {043007} (\bibinfo {year} {2013})},\ \Eprint
  {https://arxiv.org/abs/1306.6747} {arXiv:1306.6747 [gr-qc]} \BibitemShut
  {NoStop}%
\bibitem [{\citenamefont {Somiya}(2012)}]{Somiya:2011np}%
  \BibitemOpen
  \bibfield  {author} {\bibinfo {author} {\bibfnamefont {K.}~\bibnamefont
  {Somiya}} (\bibinfo {collaboration} {KAGRA}),\ }\bibfield  {title} {\bibinfo
  {title} {{Detector configuration of KAGRA: The Japanese cryogenic
  gravitational-wave detector}},\ }\href
  {https://doi.org/10.1088/0264-9381/29/12/124007} {\bibfield  {journal}
  {\bibinfo  {journal} {Class. Quant. Grav.}\ }\textbf {\bibinfo {volume}
  {29}},\ \bibinfo {pages} {124007} (\bibinfo {year} {2012})},\ \Eprint
  {https://arxiv.org/abs/1111.7185} {arXiv:1111.7185 [gr-qc]} \BibitemShut
  {NoStop}%
\bibitem [{\citenamefont {Akutsu}\ \emph {et~al.}(2021)\citenamefont {Akutsu}
  \emph {et~al.}}]{KAGRA:2020tym}%
  \BibitemOpen
  \bibfield  {author} {\bibinfo {author} {\bibfnamefont {T.}~\bibnamefont
  {Akutsu}} \emph {et~al.} (\bibinfo {collaboration} {KAGRA}),\ }\bibfield
  {title} {\bibinfo {title} {{Overview of KAGRA: Detector design and
  construction history}},\ }\href {https://doi.org/10.1093/ptep/ptaa125}
  {\bibfield  {journal} {\bibinfo  {journal} {PTEP}\ }\textbf {\bibinfo
  {volume} {2021}},\ \bibinfo {pages} {05A101} (\bibinfo {year} {2021})},\
  \Eprint {https://arxiv.org/abs/2005.05574} {arXiv:2005.05574
  [physics.ins-det]} \BibitemShut {NoStop}%
\bibitem [{\citenamefont {Abbott}\ \emph {et~al.}(2023)\citenamefont {Abbott}
  \emph {et~al.}}]{KAGRA:2021duu}%
  \BibitemOpen
  \bibfield  {author} {\bibinfo {author} {\bibfnamefont {R.}~\bibnamefont
  {Abbott}} \emph {et~al.} (\bibinfo {collaboration} {KAGRA, VIRGO, LIGO
  Scientific}),\ }\bibfield  {title} {\bibinfo {title} {{Population of Merging
  Compact Binaries Inferred Using Gravitational Waves through GWTC-3}},\ }\href
  {https://doi.org/10.1103/PhysRevX.13.011048} {\bibfield  {journal} {\bibinfo
  {journal} {Phys. Rev. X}\ }\textbf {\bibinfo {volume} {13}},\ \bibinfo
  {pages} {011048} (\bibinfo {year} {2023})},\ \Eprint
  {https://arxiv.org/abs/2111.03634} {arXiv:2111.03634 [astro-ph.HE]}
  \BibitemShut {NoStop}%
\bibitem [{\citenamefont {Tiwari}\ and\ \citenamefont
  {Fairhurst}(2021)}]{Tiwari:2020otp}%
  \BibitemOpen
  \bibfield  {author} {\bibinfo {author} {\bibfnamefont {V.}~\bibnamefont
  {Tiwari}}\ and\ \bibinfo {author} {\bibfnamefont {S.}~\bibnamefont
  {Fairhurst}},\ }\bibfield  {title} {\bibinfo {title} {{The Emergence of
  Structure in the Binary Black Hole Mass Distribution}},\ }\href
  {https://doi.org/10.3847/2041-8213/abfbe7} {\bibfield  {journal} {\bibinfo
  {journal} {Astrophys. J. Lett.}\ }\textbf {\bibinfo {volume} {913}},\
  \bibinfo {pages} {L19} (\bibinfo {year} {2021})},\ \Eprint
  {https://arxiv.org/abs/2011.04502} {arXiv:2011.04502 [astro-ph.HE]}
  \BibitemShut {NoStop}%
\bibitem [{\citenamefont {Edelman}\ \emph {et~al.}(2022)\citenamefont
  {Edelman}, \citenamefont {Doctor}, \citenamefont {Godfrey},\ and\
  \citenamefont {Farr}}]{Edelman:2021zkw}%
  \BibitemOpen
  \bibfield  {author} {\bibinfo {author} {\bibfnamefont {B.}~\bibnamefont
  {Edelman}}, \bibinfo {author} {\bibfnamefont {Z.}~\bibnamefont {Doctor}},
  \bibinfo {author} {\bibfnamefont {J.}~\bibnamefont {Godfrey}},\ and\ \bibinfo
  {author} {\bibfnamefont {B.}~\bibnamefont {Farr}},\ }\bibfield  {title}
  {\bibinfo {title} {{Ain\textquoteright{}t No Mountain High Enough:
  Semiparametric Modeling of LIGO\textendash{}Virgo\textquoteright{}s Binary
  Black Hole Mass Distribution}},\ }\href
  {https://doi.org/10.3847/1538-4357/ac3667} {\bibfield  {journal} {\bibinfo
  {journal} {Astrophys. J.}\ }\textbf {\bibinfo {volume} {924}},\ \bibinfo
  {pages} {101} (\bibinfo {year} {2022})},\ \Eprint
  {https://arxiv.org/abs/2109.06137} {arXiv:2109.06137 [astro-ph.HE]}
  \BibitemShut {NoStop}%
\bibitem [{\citenamefont {Fishbach}\ and\ \citenamefont
  {Holz}(2017)}]{Fishbach:2017zga}%
  \BibitemOpen
  \bibfield  {author} {\bibinfo {author} {\bibfnamefont {M.}~\bibnamefont
  {Fishbach}}\ and\ \bibinfo {author} {\bibfnamefont {D.~E.}\ \bibnamefont
  {Holz}},\ }\bibfield  {title} {\bibinfo {title} {{Where Are LIGO's Big Black
  Holes?}},\ }\href {https://doi.org/10.3847/2041-8213/aa9bf6} {\bibfield
  {journal} {\bibinfo  {journal} {Astrophys. J. Lett.}\ }\textbf {\bibinfo
  {volume} {851}},\ \bibinfo {pages} {L25} (\bibinfo {year} {2017})},\ \Eprint
  {https://arxiv.org/abs/1709.08584} {arXiv:1709.08584 [astro-ph.HE]}
  \BibitemShut {NoStop}%
\bibitem [{\citenamefont {Talbot}\ and\ \citenamefont
  {Thrane}(2018)}]{Talbot:2018cva}%
  \BibitemOpen
  \bibfield  {author} {\bibinfo {author} {\bibfnamefont {C.}~\bibnamefont
  {Talbot}}\ and\ \bibinfo {author} {\bibfnamefont {E.}~\bibnamefont
  {Thrane}},\ }\bibfield  {title} {\bibinfo {title} {{Measuring the binary
  black hole mass spectrum with an astrophysically motivated
  parameterization}},\ }\href {https://doi.org/10.3847/1538-4357/aab34c}
  {\bibfield  {journal} {\bibinfo  {journal} {Astrophys. J.}\ }\textbf
  {\bibinfo {volume} {856}},\ \bibinfo {pages} {173} (\bibinfo {year}
  {2018})},\ \Eprint {https://arxiv.org/abs/1801.02699} {arXiv:1801.02699
  [astro-ph.HE]} \BibitemShut {NoStop}%
\bibitem [{\citenamefont {Fishbach}\ and\ \citenamefont
  {Holz}(2020)}]{Fishbach:2019bbm}%
  \BibitemOpen
  \bibfield  {author} {\bibinfo {author} {\bibfnamefont {M.}~\bibnamefont
  {Fishbach}}\ and\ \bibinfo {author} {\bibfnamefont {D.~E.}\ \bibnamefont
  {Holz}},\ }\bibfield  {title} {\bibinfo {title} {{Picky Partners: The Pairing
  of Component Masses in Binary Black Hole Mergers}},\ }\href
  {https://doi.org/10.3847/2041-8213/ab7247} {\bibfield  {journal} {\bibinfo
  {journal} {Astrophys. J. Lett.}\ }\textbf {\bibinfo {volume} {891}},\
  \bibinfo {pages} {L27} (\bibinfo {year} {2020})},\ \Eprint
  {https://arxiv.org/abs/1905.12669} {arXiv:1905.12669 [astro-ph.HE]}
  \BibitemShut {NoStop}%
\bibitem [{\citenamefont {Farah}\ \emph
  {et~al.}(2023{\natexlab{a}})\citenamefont {Farah}, \citenamefont {Fishbach},\
  and\ \citenamefont {Holz}}]{Farah:2023swu}%
  \BibitemOpen
  \bibfield  {author} {\bibinfo {author} {\bibfnamefont {A.~M.}\ \bibnamefont
  {Farah}}, \bibinfo {author} {\bibfnamefont {M.}~\bibnamefont {Fishbach}},\
  and\ \bibinfo {author} {\bibfnamefont {D.~E.}\ \bibnamefont {Holz}},\
  }\href@noop {} {\bibinfo {title} {{Two of a Kind: Comparing big and small
  black holes in binaries with gravitational waves}}} (\bibinfo {year}
  {2023}{\natexlab{a}}),\ \Eprint {https://arxiv.org/abs/2308.05102}
  {arXiv:2308.05102 [astro-ph.HE]} \BibitemShut {NoStop}%
\bibitem [{\citenamefont {Wysocki}\ \emph {et~al.}(2019)\citenamefont
  {Wysocki}, \citenamefont {Lange},\ and\ \citenamefont
  {O'Shaughnessy}}]{Wysocki:2018}%
  \BibitemOpen
  \bibfield  {author} {\bibinfo {author} {\bibfnamefont {D.}~\bibnamefont
  {Wysocki}}, \bibinfo {author} {\bibfnamefont {J.}~\bibnamefont {Lange}},\
  and\ \bibinfo {author} {\bibfnamefont {R.}~\bibnamefont {O'Shaughnessy}},\
  }\bibfield  {title} {\bibinfo {title} {Reconstructing phenomenological
  distributions of compact binaries via gravitational wave observations},\
  }\href {https://doi.org/10.1103/PhysRevD.100.043012} {\bibfield  {journal}
  {\bibinfo  {journal} {Phys. Rev. D}\ }\textbf {\bibinfo {volume} {100}},\
  \bibinfo {pages} {043012} (\bibinfo {year} {2019})}\BibitemShut {NoStop}%
\bibitem [{\citenamefont {Biscoveanu}\ \emph {et~al.}(2021)\citenamefont
  {Biscoveanu}, \citenamefont {Isi}, \citenamefont {Vitale},\ and\
  \citenamefont {Varma}}]{Biscoveanu:2020are}%
  \BibitemOpen
  \bibfield  {author} {\bibinfo {author} {\bibfnamefont {S.}~\bibnamefont
  {Biscoveanu}}, \bibinfo {author} {\bibfnamefont {M.}~\bibnamefont {Isi}},
  \bibinfo {author} {\bibfnamefont {S.}~\bibnamefont {Vitale}},\ and\ \bibinfo
  {author} {\bibfnamefont {V.}~\bibnamefont {Varma}},\ }\bibfield  {title}
  {\bibinfo {title} {{New Spin on LIGO-Virgo Binary Black Holes}},\ }\href
  {https://doi.org/10.1103/PhysRevLett.126.171103} {\bibfield  {journal}
  {\bibinfo  {journal} {Phys. Rev. Lett.}\ }\textbf {\bibinfo {volume} {126}},\
  \bibinfo {pages} {171103} (\bibinfo {year} {2021})},\ \Eprint
  {https://arxiv.org/abs/2007.09156} {arXiv:2007.09156 [astro-ph.HE]}
  \BibitemShut {NoStop}%
\bibitem [{\citenamefont {Callister}\ \emph {et~al.}(2022)\citenamefont
  {Callister}, \citenamefont {Miller}, \citenamefont {Chatziioannou},\ and\
  \citenamefont {Farr}}]{Callister:2022qwb}%
  \BibitemOpen
  \bibfield  {author} {\bibinfo {author} {\bibfnamefont {T.~A.}\ \bibnamefont
  {Callister}}, \bibinfo {author} {\bibfnamefont {S.~J.}\ \bibnamefont
  {Miller}}, \bibinfo {author} {\bibfnamefont {K.}~\bibnamefont
  {Chatziioannou}},\ and\ \bibinfo {author} {\bibfnamefont {W.~M.}\
  \bibnamefont {Farr}},\ }\bibfield  {title} {\bibinfo {title} {{No Evidence
  that the Majority of Black Holes in Binaries Have Zero Spin}},\ }\href
  {https://doi.org/10.3847/2041-8213/ac847e} {\bibfield  {journal} {\bibinfo
  {journal} {Astrophys. J. Lett.}\ }\textbf {\bibinfo {volume} {937}},\
  \bibinfo {pages} {L13} (\bibinfo {year} {2022})},\ \Eprint
  {https://arxiv.org/abs/2205.08574} {arXiv:2205.08574 [astro-ph.HE]}
  \BibitemShut {NoStop}%
\bibitem [{\citenamefont {Tong}\ \emph {et~al.}(2022)\citenamefont {Tong},
  \citenamefont {Galaudage},\ and\ \citenamefont {Thrane}}]{Tong:2022iws}%
  \BibitemOpen
  \bibfield  {author} {\bibinfo {author} {\bibfnamefont {H.}~\bibnamefont
  {Tong}}, \bibinfo {author} {\bibfnamefont {S.}~\bibnamefont {Galaudage}},\
  and\ \bibinfo {author} {\bibfnamefont {E.}~\bibnamefont {Thrane}},\
  }\bibfield  {title} {\bibinfo {title} {{Population properties of spinning
  black holes using the gravitational-wave transient catalog 3}},\ }\href
  {https://doi.org/10.1103/PhysRevD.106.103019} {\bibfield  {journal} {\bibinfo
   {journal} {Phys. Rev. D}\ }\textbf {\bibinfo {volume} {106}},\ \bibinfo
  {pages} {103019} (\bibinfo {year} {2022})},\ \Eprint
  {https://arxiv.org/abs/2209.02206} {arXiv:2209.02206 [astro-ph.HE]}
  \BibitemShut {NoStop}%
\bibitem [{\citenamefont {Mould}\ \emph {et~al.}(2022)\citenamefont {Mould},
  \citenamefont {Gerosa}, \citenamefont {Broekgaarden},\ and\ \citenamefont
  {Steinle}}]{Mould:2022xeu}%
  \BibitemOpen
  \bibfield  {author} {\bibinfo {author} {\bibfnamefont {M.}~\bibnamefont
  {Mould}}, \bibinfo {author} {\bibfnamefont {D.}~\bibnamefont {Gerosa}},
  \bibinfo {author} {\bibfnamefont {F.~S.}\ \bibnamefont {Broekgaarden}},\ and\
  \bibinfo {author} {\bibfnamefont {N.}~\bibnamefont {Steinle}},\ }\bibfield
  {title} {\bibinfo {title} {{Which black hole formed first? Mass-ratio
  reversal in massive binary stars from gravitational-wave data}},\ }\href
  {https://doi.org/10.1093/mnras/stac2859} {\bibfield  {journal} {\bibinfo
  {journal} {Mon. Not. Roy. Astron. Soc.}\ }\textbf {\bibinfo {volume} {517}},\
  \bibinfo {pages} {2738} (\bibinfo {year} {2022})},\ \Eprint
  {https://arxiv.org/abs/2205.12329} {arXiv:2205.12329 [astro-ph.HE]}
  \BibitemShut {NoStop}%
\bibitem [{\citenamefont {Galaudage}\ \emph {et~al.}(2021)\citenamefont
  {Galaudage} \emph {et~al.}}]{Galaudage:2021rkt}%
  \BibitemOpen
  \bibfield  {author} {\bibinfo {author} {\bibfnamefont {S.}~\bibnamefont
  {Galaudage}} \emph {et~al.},\ }\bibfield  {title} {\bibinfo {title}
  {{Building Better Spin Models for Merging Binary Black Holes: Evidence for
  Nonspinning and Rapidly Spinning Nearly Aligned Subpopulations}},\ }\href
  {https://doi.org/10.3847/2041-8213/ac2f3c} {\bibfield  {journal} {\bibinfo
  {journal} {Astrophys. J. Lett.}\ }\textbf {\bibinfo {volume} {921}},\
  \bibinfo {pages} {L15} (\bibinfo {year} {2021})},\ \bibinfo {note} {[Erratum:
  Astrophys.J.Lett. 936, L18 (2022), Erratum: Astrophys.J. 936, L18 (2022)]},\
  \Eprint {https://arxiv.org/abs/2109.02424} {arXiv:2109.02424 [gr-qc]}
  \BibitemShut {NoStop}%
\bibitem [{\citenamefont {Roulet}\ \emph {et~al.}(2021)\citenamefont {Roulet},
  \citenamefont {Chia}, \citenamefont {Olsen}, \citenamefont {Dai},
  \citenamefont {Venumadhav}, \citenamefont {Zackay},\ and\ \citenamefont
  {Zaldarriaga}}]{Roulet:2021hcu}%
  \BibitemOpen
  \bibfield  {author} {\bibinfo {author} {\bibfnamefont {J.}~\bibnamefont
  {Roulet}}, \bibinfo {author} {\bibfnamefont {H.~S.}\ \bibnamefont {Chia}},
  \bibinfo {author} {\bibfnamefont {S.}~\bibnamefont {Olsen}}, \bibinfo
  {author} {\bibfnamefont {L.}~\bibnamefont {Dai}}, \bibinfo {author}
  {\bibfnamefont {T.}~\bibnamefont {Venumadhav}}, \bibinfo {author}
  {\bibfnamefont {B.}~\bibnamefont {Zackay}},\ and\ \bibinfo {author}
  {\bibfnamefont {M.}~\bibnamefont {Zaldarriaga}},\ }\bibfield  {title}
  {\bibinfo {title} {{Distribution of effective spins and masses of binary
  black holes from the LIGO and Virgo O1\textendash{}O3a observing runs}},\
  }\href {https://doi.org/10.1103/PhysRevD.104.083010} {\bibfield  {journal}
  {\bibinfo  {journal} {Phys. Rev. D}\ }\textbf {\bibinfo {volume} {104}},\
  \bibinfo {pages} {083010} (\bibinfo {year} {2021})},\ \Eprint
  {https://arxiv.org/abs/2105.10580} {arXiv:2105.10580 [astro-ph.HE]}
  \BibitemShut {NoStop}%
\bibitem [{\citenamefont {Vitale}\ \emph {et~al.}(2022)\citenamefont {Vitale},
  \citenamefont {Biscoveanu},\ and\ \citenamefont {Talbot}}]{Vitale:2022dpa}%
  \BibitemOpen
  \bibfield  {author} {\bibinfo {author} {\bibfnamefont {S.}~\bibnamefont
  {Vitale}}, \bibinfo {author} {\bibfnamefont {S.}~\bibnamefont {Biscoveanu}},\
  and\ \bibinfo {author} {\bibfnamefont {C.}~\bibnamefont {Talbot}},\
  }\bibfield  {title} {\bibinfo {title} {{Spin it as you like: The (lack of a)
  measurement of the spin tilt distribution with LIGO-Virgo-KAGRA binary black
  holes}},\ }\href {https://doi.org/10.1051/0004-6361/202245084} {\bibfield
  {journal} {\bibinfo  {journal} {Astron. Astrophys.}\ }\textbf {\bibinfo
  {volume} {668}},\ \bibinfo {pages} {L2} (\bibinfo {year} {2022})},\ \Eprint
  {https://arxiv.org/abs/2209.06978} {arXiv:2209.06978 [astro-ph.HE]}
  \BibitemShut {NoStop}%
\bibitem [{\citenamefont {Fishbach}\ \emph {et~al.}(2021)\citenamefont
  {Fishbach}, \citenamefont {Doctor}, \citenamefont {Callister}, \citenamefont
  {Edelman}, \citenamefont {Ye}, \citenamefont {Essick}, \citenamefont {Farr},
  \citenamefont {Farr},\ and\ \citenamefont {Holz}}]{Fishbach:2021yvy}%
  \BibitemOpen
  \bibfield  {author} {\bibinfo {author} {\bibfnamefont {M.}~\bibnamefont
  {Fishbach}}, \bibinfo {author} {\bibfnamefont {Z.}~\bibnamefont {Doctor}},
  \bibinfo {author} {\bibfnamefont {T.}~\bibnamefont {Callister}}, \bibinfo
  {author} {\bibfnamefont {B.}~\bibnamefont {Edelman}}, \bibinfo {author}
  {\bibfnamefont {J.}~\bibnamefont {Ye}}, \bibinfo {author} {\bibfnamefont
  {R.}~\bibnamefont {Essick}}, \bibinfo {author} {\bibfnamefont {W.~M.}\
  \bibnamefont {Farr}}, \bibinfo {author} {\bibfnamefont {B.}~\bibnamefont
  {Farr}},\ and\ \bibinfo {author} {\bibfnamefont {D.~E.}\ \bibnamefont
  {Holz}},\ }\bibfield  {title} {\bibinfo {title} {{When Are
  LIGO/Virgo\textquoteright{}s Big Black Hole Mergers?}},\ }\href
  {https://doi.org/10.3847/1538-4357/abee11} {\bibfield  {journal} {\bibinfo
  {journal} {Astrophys. J.}\ }\textbf {\bibinfo {volume} {912}},\ \bibinfo
  {pages} {98} (\bibinfo {year} {2021})},\ \Eprint
  {https://arxiv.org/abs/2101.07699} {arXiv:2101.07699 [astro-ph.HE]}
  \BibitemShut {NoStop}%
\bibitem [{\citenamefont {Zevin}\ \emph {et~al.}(2021)\citenamefont {Zevin},
  \citenamefont {Bavera}, \citenamefont {Berry}, \citenamefont {Kalogera},
  \citenamefont {Fragos}, \citenamefont {Marchant}, \citenamefont {Rodriguez},
  \citenamefont {Antonini}, \citenamefont {Holz},\ and\ \citenamefont
  {Pankow}}]{Zevin:2020gbd}%
  \BibitemOpen
  \bibfield  {author} {\bibinfo {author} {\bibfnamefont {M.}~\bibnamefont
  {Zevin}}, \bibinfo {author} {\bibfnamefont {S.~S.}\ \bibnamefont {Bavera}},
  \bibinfo {author} {\bibfnamefont {C.~P.~L.}\ \bibnamefont {Berry}}, \bibinfo
  {author} {\bibfnamefont {V.}~\bibnamefont {Kalogera}}, \bibinfo {author}
  {\bibfnamefont {T.}~\bibnamefont {Fragos}}, \bibinfo {author} {\bibfnamefont
  {P.}~\bibnamefont {Marchant}}, \bibinfo {author} {\bibfnamefont {C.~L.}\
  \bibnamefont {Rodriguez}}, \bibinfo {author} {\bibfnamefont {F.}~\bibnamefont
  {Antonini}}, \bibinfo {author} {\bibfnamefont {D.~E.}\ \bibnamefont {Holz}},\
  and\ \bibinfo {author} {\bibfnamefont {C.}~\bibnamefont {Pankow}},\
  }\bibfield  {title} {\bibinfo {title} {{One Channel to Rule Them All?
  Constraining the Origins of Binary Black Holes Using Multiple Formation
  Pathways}},\ }\href {https://doi.org/10.3847/1538-4357/abe40e} {\bibfield
  {journal} {\bibinfo  {journal} {Astrophys. J.}\ }\textbf {\bibinfo {volume}
  {910}},\ \bibinfo {pages} {152} (\bibinfo {year} {2021})},\ \Eprint
  {https://arxiv.org/abs/2011.10057} {arXiv:2011.10057 [astro-ph.HE]}
  \BibitemShut {NoStop}%
\bibitem [{\citenamefont {Cheng}\ \emph {et~al.}(2023)\citenamefont {Cheng},
  \citenamefont {Zevin},\ and\ \citenamefont {Vitale}}]{Cheng:2023ddt}%
  \BibitemOpen
  \bibfield  {author} {\bibinfo {author} {\bibfnamefont {A.~Q.}\ \bibnamefont
  {Cheng}}, \bibinfo {author} {\bibfnamefont {M.}~\bibnamefont {Zevin}},\ and\
  \bibinfo {author} {\bibfnamefont {S.}~\bibnamefont {Vitale}},\ }\bibfield
  {title} {\bibinfo {title} {{What You Don\textquoteright{}t Know Can Hurt You:
  Use and Abuse of Astrophysical Models in Gravitational-wave Population
  Analyses}},\ }\href {https://doi.org/10.3847/1538-4357/aced98} {\bibfield
  {journal} {\bibinfo  {journal} {Astrophys. J.}\ }\textbf {\bibinfo {volume}
  {955}},\ \bibinfo {pages} {127} (\bibinfo {year} {2023})},\ \Eprint
  {https://arxiv.org/abs/2307.03129} {arXiv:2307.03129 [astro-ph.HE]}
  \BibitemShut {NoStop}%
\bibitem [{\citenamefont {Wong}\ \emph {et~al.}(2021)\citenamefont {Wong},
  \citenamefont {Breivik}, \citenamefont {Kremer},\ and\ \citenamefont
  {Callister}}]{Wong:2020ise}%
  \BibitemOpen
  \bibfield  {author} {\bibinfo {author} {\bibfnamefont {K.~W.~K.}\
  \bibnamefont {Wong}}, \bibinfo {author} {\bibfnamefont {K.}~\bibnamefont
  {Breivik}}, \bibinfo {author} {\bibfnamefont {K.}~\bibnamefont {Kremer}},\
  and\ \bibinfo {author} {\bibfnamefont {T.}~\bibnamefont {Callister}},\
  }\bibfield  {title} {\bibinfo {title} {{Joint constraints on the
  field-cluster mixing fraction, common envelope efficiency, and globular
  cluster radii from a population of binary hole mergers via deep learning}},\
  }\href {https://doi.org/10.1103/PhysRevD.103.083021} {\bibfield  {journal}
  {\bibinfo  {journal} {Phys. Rev. D}\ }\textbf {\bibinfo {volume} {103}},\
  \bibinfo {pages} {083021} (\bibinfo {year} {2021})},\ \Eprint
  {https://arxiv.org/abs/2011.03564} {arXiv:2011.03564 [astro-ph.HE]}
  \BibitemShut {NoStop}%
\bibitem [{\citenamefont {Baibhav}\ \emph {et~al.}(2023)\citenamefont
  {Baibhav}, \citenamefont {Doctor},\ and\ \citenamefont
  {Kalogera}}]{Baibhav:2022qxm}%
  \BibitemOpen
  \bibfield  {author} {\bibinfo {author} {\bibfnamefont {V.}~\bibnamefont
  {Baibhav}}, \bibinfo {author} {\bibfnamefont {Z.}~\bibnamefont {Doctor}},\
  and\ \bibinfo {author} {\bibfnamefont {V.}~\bibnamefont {Kalogera}},\
  }\bibfield  {title} {\bibinfo {title} {{Dropping Anchor: Understanding the
  Populations of Binary Black Holes with Random and Aligned-spin
  Orientations}},\ }\href {https://doi.org/10.3847/1538-4357/acbf4c} {\bibfield
   {journal} {\bibinfo  {journal} {Astrophys. J.}\ }\textbf {\bibinfo {volume}
  {946}},\ \bibinfo {pages} {50} (\bibinfo {year} {2023})},\ \Eprint
  {https://arxiv.org/abs/2212.12113} {arXiv:2212.12113 [astro-ph.HE]}
  \BibitemShut {NoStop}%
\bibitem [{\citenamefont {Godfrey}\ \emph {et~al.}(2023)\citenamefont
  {Godfrey}, \citenamefont {Edelman},\ and\ \citenamefont
  {Farr}}]{Godfrey:2023oxb}%
  \BibitemOpen
  \bibfield  {author} {\bibinfo {author} {\bibfnamefont {J.}~\bibnamefont
  {Godfrey}}, \bibinfo {author} {\bibfnamefont {B.}~\bibnamefont {Edelman}},\
  and\ \bibinfo {author} {\bibfnamefont {B.}~\bibnamefont {Farr}},\ }\href@noop
  {} {\bibinfo {title} {{Cosmic Cousins: Identification of a Subpopulation of
  Binary Black Holes Consistent with Isolated Binary Evolution}}} (\bibinfo
  {year} {2023}),\ \Eprint {https://arxiv.org/abs/2304.01288} {arXiv:2304.01288
  [astro-ph.HE]} \BibitemShut {NoStop}%
\bibitem [{\citenamefont {Wang}\ \emph {et~al.}(2022)\citenamefont {Wang},
  \citenamefont {Li}, \citenamefont {Vink}, \citenamefont {Fan}, \citenamefont
  {Tang}, \citenamefont {Qin},\ and\ \citenamefont {Wei}}]{Wang:2022gnx}%
  \BibitemOpen
  \bibfield  {author} {\bibinfo {author} {\bibfnamefont {Y.-Z.}\ \bibnamefont
  {Wang}}, \bibinfo {author} {\bibfnamefont {Y.-J.}\ \bibnamefont {Li}},
  \bibinfo {author} {\bibfnamefont {J.~S.}\ \bibnamefont {Vink}}, \bibinfo
  {author} {\bibfnamefont {Y.-Z.}\ \bibnamefont {Fan}}, \bibinfo {author}
  {\bibfnamefont {S.-P.}\ \bibnamefont {Tang}}, \bibinfo {author}
  {\bibfnamefont {Y.}~\bibnamefont {Qin}},\ and\ \bibinfo {author}
  {\bibfnamefont {D.-M.}\ \bibnamefont {Wei}},\ }\bibfield  {title} {\bibinfo
  {title} {{Potential Subpopulations and Assembling Tendency of the Merging
  Black Holes}},\ }\href {https://doi.org/10.3847/2041-8213/aca89f} {\bibfield
  {journal} {\bibinfo  {journal} {Astrophys. J. Lett.}\ }\textbf {\bibinfo
  {volume} {941}},\ \bibinfo {pages} {L39} (\bibinfo {year} {2022})},\ \Eprint
  {https://arxiv.org/abs/2208.11871} {arXiv:2208.11871 [astro-ph.HE]}
  \BibitemShut {NoStop}%
\bibitem [{\citenamefont {Edelman}\ \emph {et~al.}(2023)\citenamefont
  {Edelman}, \citenamefont {Farr},\ and\ \citenamefont
  {Doctor}}]{Edelman:2022ydv}%
  \BibitemOpen
  \bibfield  {author} {\bibinfo {author} {\bibfnamefont {B.}~\bibnamefont
  {Edelman}}, \bibinfo {author} {\bibfnamefont {B.}~\bibnamefont {Farr}},\ and\
  \bibinfo {author} {\bibfnamefont {Z.}~\bibnamefont {Doctor}},\ }\bibfield
  {title} {\bibinfo {title} {{Cover Your Basis: Comprehensive Data-driven
  Characterization of the Binary Black Hole Population}},\ }\href
  {https://doi.org/10.3847/1538-4357/acb5ed} {\bibfield  {journal} {\bibinfo
  {journal} {Astrophys. J.}\ }\textbf {\bibinfo {volume} {946}},\ \bibinfo
  {pages} {16} (\bibinfo {year} {2023})},\ \Eprint
  {https://arxiv.org/abs/2210.12834} {arXiv:2210.12834 [astro-ph.HE]}
  \BibitemShut {NoStop}%
\bibitem [{\citenamefont {Golomb}\ and\ \citenamefont
  {Talbot}(2023)}]{Golomb:2022bon}%
  \BibitemOpen
  \bibfield  {author} {\bibinfo {author} {\bibfnamefont {J.}~\bibnamefont
  {Golomb}}\ and\ \bibinfo {author} {\bibfnamefont {C.}~\bibnamefont
  {Talbot}},\ }\bibfield  {title} {\bibinfo {title} {{Searching for structure
  in the binary black hole spin distribution}},\ }\href
  {https://doi.org/10.1103/PhysRevD.108.103009} {\bibfield  {journal} {\bibinfo
   {journal} {Phys. Rev. D}\ }\textbf {\bibinfo {volume} {108}},\ \bibinfo
  {pages} {103009} (\bibinfo {year} {2023})},\ \Eprint
  {https://arxiv.org/abs/2210.12287} {arXiv:2210.12287 [astro-ph.HE]}
  \BibitemShut {NoStop}%
\bibitem [{\citenamefont {Rinaldi}\ and\ \citenamefont
  {Del~Pozzo}(2021)}]{Rinaldi:2021bhm}%
  \BibitemOpen
  \bibfield  {author} {\bibinfo {author} {\bibfnamefont {S.}~\bibnamefont
  {Rinaldi}}\ and\ \bibinfo {author} {\bibfnamefont {W.}~\bibnamefont
  {Del~Pozzo}},\ }\bibfield  {title} {\bibinfo {title} {{(H)DPGMM: a hierarchy
  of Dirichlet process Gaussian mixture models for the inference of the black
  hole mass function}},\ }\href {https://doi.org/10.1093/mnras/stab3224}
  {\bibfield  {journal} {\bibinfo  {journal} {Mon. Not. Roy. Astron. Soc.}\
  }\textbf {\bibinfo {volume} {509}},\ \bibinfo {pages} {5454} (\bibinfo {year}
  {2021})},\ \Eprint {https://arxiv.org/abs/2109.05960} {arXiv:2109.05960
  [astro-ph.IM]} \BibitemShut {NoStop}%
\bibitem [{\citenamefont {Tiwari}(2021)}]{Tiwari:2020vym}%
  \BibitemOpen
  \bibfield  {author} {\bibinfo {author} {\bibfnamefont {V.}~\bibnamefont
  {Tiwari}},\ }\bibfield  {title} {\bibinfo {title} {{VAMANA: modeling binary
  black hole population with minimal assumptions}},\ }\href
  {https://doi.org/10.1088/1361-6382/ac0b54} {\bibfield  {journal} {\bibinfo
  {journal} {Class. Quant. Grav.}\ }\textbf {\bibinfo {volume} {38}},\ \bibinfo
  {pages} {155007} (\bibinfo {year} {2021})},\ \Eprint
  {https://arxiv.org/abs/2006.15047} {arXiv:2006.15047 [astro-ph.HE]}
  \BibitemShut {NoStop}%
\bibitem [{\citenamefont {Tiwari}(2022)}]{Tiwari:2021yvr}%
  \BibitemOpen
  \bibfield  {author} {\bibinfo {author} {\bibfnamefont {V.}~\bibnamefont
  {Tiwari}},\ }\bibfield  {title} {\bibinfo {title} {{Exploring Features in the
  Binary Black Hole Population}},\ }\href
  {https://doi.org/10.3847/1538-4357/ac589a} {\bibfield  {journal} {\bibinfo
  {journal} {Astrophys. J.}\ }\textbf {\bibinfo {volume} {928}},\ \bibinfo
  {pages} {155} (\bibinfo {year} {2022})},\ \Eprint
  {https://arxiv.org/abs/2111.13991} {arXiv:2111.13991 [astro-ph.HE]}
  \BibitemShut {NoStop}%
\bibitem [{\citenamefont {Callister}\ and\ \citenamefont
  {Farr}(2023)}]{Callister:2023tgi}%
  \BibitemOpen
  \bibfield  {author} {\bibinfo {author} {\bibfnamefont {T.~A.}\ \bibnamefont
  {Callister}}\ and\ \bibinfo {author} {\bibfnamefont {W.~M.}\ \bibnamefont
  {Farr}},\ }\href@noop {} {\bibinfo {title} {{A Parameter-Free Tour of the
  Binary Black Hole Population}}} (\bibinfo {year} {2023}),\ \Eprint
  {https://arxiv.org/abs/2302.07289} {arXiv:2302.07289 [astro-ph.HE]}
  \BibitemShut {NoStop}%
\bibitem [{\citenamefont {Romero-Shaw}\ \emph {et~al.}(2022)\citenamefont
  {Romero-Shaw}, \citenamefont {Thrane},\ and\ \citenamefont
  {Lasky}}]{Romero-Shaw:2022ctb}%
  \BibitemOpen
  \bibfield  {author} {\bibinfo {author} {\bibfnamefont {I.~M.}\ \bibnamefont
  {Romero-Shaw}}, \bibinfo {author} {\bibfnamefont {E.}~\bibnamefont
  {Thrane}},\ and\ \bibinfo {author} {\bibfnamefont {P.~D.}\ \bibnamefont
  {Lasky}},\ }\bibfield  {title} {\bibinfo {title} {{When models fail: An
  introduction to posterior predictive checks and model misspecification in
  gravitational-wave astronomy}},\ }\href
  {https://doi.org/10.1017/pasa.2022.24} {\bibfield  {journal} {\bibinfo
  {journal} {Publ. Astron. Soc. Austral.}\ }\textbf {\bibinfo {volume} {39}},\
  \bibinfo {pages} {e025} (\bibinfo {year} {2022})},\ \Eprint
  {https://arxiv.org/abs/2202.05479} {arXiv:2202.05479 [astro-ph.IM]}
  \BibitemShut {NoStop}%
\bibitem [{\citenamefont {Bavera}\ \emph {et~al.}(2022)\citenamefont {Bavera},
  \citenamefont {Fishbach}, \citenamefont {Zevin}, \citenamefont {Zapartas},\
  and\ \citenamefont {Fragos}}]{Bavera:2022mef}%
  \BibitemOpen
  \bibfield  {author} {\bibinfo {author} {\bibfnamefont {S.~S.}\ \bibnamefont
  {Bavera}}, \bibinfo {author} {\bibfnamefont {M.}~\bibnamefont {Fishbach}},
  \bibinfo {author} {\bibfnamefont {M.}~\bibnamefont {Zevin}}, \bibinfo
  {author} {\bibfnamefont {E.}~\bibnamefont {Zapartas}},\ and\ \bibinfo
  {author} {\bibfnamefont {T.}~\bibnamefont {Fragos}},\ }\bibfield  {title}
  {\bibinfo {title} {{The \ensuremath{\chi}eff \ensuremath{-} z correlation of
  field binary black hole mergers and how 3G gravitational-wave detectors can
  constrain it}},\ }\href {https://doi.org/10.1051/0004-6361/202243724}
  {\bibfield  {journal} {\bibinfo  {journal} {Astron. Astrophys.}\ }\textbf
  {\bibinfo {volume} {665}},\ \bibinfo {pages} {A59} (\bibinfo {year}
  {2022})},\ \Eprint {https://arxiv.org/abs/2204.02619} {arXiv:2204.02619
  [astro-ph.HE]} \BibitemShut {NoStop}%
\bibitem [{\citenamefont {Bavera}\ \emph {et~al.}(2020)\citenamefont {Bavera},
  \citenamefont {Fragos}, \citenamefont {Qin}, \citenamefont {Zapartas},
  \citenamefont {Neijssel}, \citenamefont {Mandel}, \citenamefont {Batta},
  \citenamefont {Gaebel}, \citenamefont {Kimball},\ and\ \citenamefont
  {Stevenson}}]{Bavera:2020inc}%
  \BibitemOpen
  \bibfield  {author} {\bibinfo {author} {\bibfnamefont {S.~S.}\ \bibnamefont
  {Bavera}}, \bibinfo {author} {\bibfnamefont {T.}~\bibnamefont {Fragos}},
  \bibinfo {author} {\bibfnamefont {Y.}~\bibnamefont {Qin}}, \bibinfo {author}
  {\bibfnamefont {E.}~\bibnamefont {Zapartas}}, \bibinfo {author}
  {\bibfnamefont {C.~J.}\ \bibnamefont {Neijssel}}, \bibinfo {author}
  {\bibfnamefont {I.}~\bibnamefont {Mandel}}, \bibinfo {author} {\bibfnamefont
  {A.}~\bibnamefont {Batta}}, \bibinfo {author} {\bibfnamefont {S.~M.}\
  \bibnamefont {Gaebel}}, \bibinfo {author} {\bibfnamefont {C.}~\bibnamefont
  {Kimball}},\ and\ \bibinfo {author} {\bibfnamefont {S.}~\bibnamefont
  {Stevenson}},\ }\bibfield  {title} {\bibinfo {title} {{The origin of spin in
  binary black holes: Predicting the distributions of the main observables of
  Advanced LIGO}},\ }\href {https://doi.org/10.1051/0004-6361/201936204}
  {\bibfield  {journal} {\bibinfo  {journal} {Astron. Astrophys.}\ }\textbf
  {\bibinfo {volume} {635}},\ \bibinfo {pages} {A97} (\bibinfo {year}
  {2020})},\ \Eprint {https://arxiv.org/abs/1906.12257} {arXiv:1906.12257
  [astro-ph.HE]} \BibitemShut {NoStop}%
\bibitem [{\citenamefont {McKernan}\ \emph {et~al.}(2012)\citenamefont
  {McKernan}, \citenamefont {Ford}, \citenamefont {Lyra},\ and\ \citenamefont
  {Perets}}]{McKernan:2012rf}%
  \BibitemOpen
  \bibfield  {author} {\bibinfo {author} {\bibfnamefont {B.}~\bibnamefont
  {McKernan}}, \bibinfo {author} {\bibfnamefont {K.~E.~S.}\ \bibnamefont
  {Ford}}, \bibinfo {author} {\bibfnamefont {W.}~\bibnamefont {Lyra}},\ and\
  \bibinfo {author} {\bibfnamefont {H.~B.}\ \bibnamefont {Perets}},\ }\bibfield
   {title} {\bibinfo {title} {{Intermediate mass black holes in AGN disks: I.
  Production \& Growth}},\ }\href
  {https://doi.org/10.1111/j.1365-2966.2012.21486.x} {\bibfield  {journal}
  {\bibinfo  {journal} {Mon. Not. Roy. Astron. Soc.}\ }\textbf {\bibinfo
  {volume} {425}},\ \bibinfo {pages} {460} (\bibinfo {year} {2012})},\ \Eprint
  {https://arxiv.org/abs/1206.2309} {arXiv:1206.2309 [astro-ph.GA]}
  \BibitemShut {NoStop}%
\bibitem [{\citenamefont {Stone}\ \emph {et~al.}(2017)\citenamefont {Stone},
  \citenamefont {Metzger},\ and\ \citenamefont {Haiman}}]{Stone:2016wzz}%
  \BibitemOpen
  \bibfield  {author} {\bibinfo {author} {\bibfnamefont {N.~C.}\ \bibnamefont
  {Stone}}, \bibinfo {author} {\bibfnamefont {B.~D.}\ \bibnamefont {Metzger}},\
  and\ \bibinfo {author} {\bibfnamefont {Z.}~\bibnamefont {Haiman}},\
  }\bibfield  {title} {\bibinfo {title} {{Assisted inspirals of stellar mass
  black holes embedded in AGN discs: solving the \textquoteleft{}final au
  problem\textquoteright{}}},\ }\href {https://doi.org/10.1093/mnras/stw2260}
  {\bibfield  {journal} {\bibinfo  {journal} {Mon. Not. Roy. Astron. Soc.}\
  }\textbf {\bibinfo {volume} {464}},\ \bibinfo {pages} {946} (\bibinfo {year}
  {2017})},\ \Eprint {https://arxiv.org/abs/1602.04226} {arXiv:1602.04226
  [astro-ph.GA]} \BibitemShut {NoStop}%
\bibitem [{\citenamefont {McKernan}\ \emph {et~al.}(2020)\citenamefont
  {McKernan}, \citenamefont {Ford}, \citenamefont {O'Shaughnessy},\ and\
  \citenamefont {Wysocki}}]{McKernan:2019beu}%
  \BibitemOpen
  \bibfield  {author} {\bibinfo {author} {\bibfnamefont {B.}~\bibnamefont
  {McKernan}}, \bibinfo {author} {\bibfnamefont {K.~E.~S.}\ \bibnamefont
  {Ford}}, \bibinfo {author} {\bibfnamefont {R.}~\bibnamefont
  {O'Shaughnessy}},\ and\ \bibinfo {author} {\bibfnamefont {D.}~\bibnamefont
  {Wysocki}},\ }\bibfield  {title} {\bibinfo {title} {{Monte Carlo simulations
  of black hole mergers in AGN discs: Low $\chi_{\rm eff}$ mergers and
  predictions for LIGO}},\ }\href {https://doi.org/10.1093/mnras/staa740}
  {\bibfield  {journal} {\bibinfo  {journal} {Mon. Not. Roy. Astron. Soc.}\
  }\textbf {\bibinfo {volume} {494}},\ \bibinfo {pages} {1203} (\bibinfo {year}
  {2020})},\ \Eprint {https://arxiv.org/abs/1907.04356} {arXiv:1907.04356
  [astro-ph.HE]} \BibitemShut {NoStop}%
\bibitem [{\citenamefont {McKernan}\ \emph {et~al.}(2022)\citenamefont
  {McKernan}, \citenamefont {Ford}, \citenamefont {Callister}, \citenamefont
  {Farr}, \citenamefont {O'Shaughnessy}, \citenamefont {Smith}, \citenamefont
  {Thrane},\ and\ \citenamefont {Vajpeyi}}]{McKernan:2021nwk}%
  \BibitemOpen
  \bibfield  {author} {\bibinfo {author} {\bibfnamefont {B.}~\bibnamefont
  {McKernan}}, \bibinfo {author} {\bibfnamefont {K.~E.~S.}\ \bibnamefont
  {Ford}}, \bibinfo {author} {\bibfnamefont {T.}~\bibnamefont {Callister}},
  \bibinfo {author} {\bibfnamefont {W.~M.}\ \bibnamefont {Farr}}, \bibinfo
  {author} {\bibfnamefont {R.}~\bibnamefont {O'Shaughnessy}}, \bibinfo {author}
  {\bibfnamefont {R.}~\bibnamefont {Smith}}, \bibinfo {author} {\bibfnamefont
  {E.}~\bibnamefont {Thrane}},\ and\ \bibinfo {author} {\bibfnamefont
  {A.}~\bibnamefont {Vajpeyi}},\ }\bibfield  {title} {\bibinfo {title}
  {{LIGO\textendash{}Virgo correlations between mass ratio and effective
  inspiral spin: testing the active galactic nuclei channel}},\ }\href
  {https://doi.org/10.1093/mnras/stac1570} {\bibfield  {journal} {\bibinfo
  {journal} {Mon. Not. Roy. Astron. Soc.}\ }\textbf {\bibinfo {volume} {514}},\
  \bibinfo {pages} {3886} (\bibinfo {year} {2022})},\ \Eprint
  {https://arxiv.org/abs/2107.07551} {arXiv:2107.07551 [astro-ph.HE]}
  \BibitemShut {NoStop}%
\bibitem [{\citenamefont {Tagawa}\ \emph {et~al.}(2020)\citenamefont {Tagawa},
  \citenamefont {Haiman},\ and\ \citenamefont {Kocsis}}]{Tagawa:2019osr}%
  \BibitemOpen
  \bibfield  {author} {\bibinfo {author} {\bibfnamefont {H.}~\bibnamefont
  {Tagawa}}, \bibinfo {author} {\bibfnamefont {Z.}~\bibnamefont {Haiman}},\
  and\ \bibinfo {author} {\bibfnamefont {B.}~\bibnamefont {Kocsis}},\
  }\bibfield  {title} {\bibinfo {title} {{Formation and Evolution of Compact
  Object Binaries in AGN Disks}},\ }\href
  {https://doi.org/10.3847/1538-4357/ab9b8c} {\bibfield  {journal} {\bibinfo
  {journal} {Astrophys. J.}\ }\textbf {\bibinfo {volume} {898}},\ \bibinfo
  {pages} {25} (\bibinfo {year} {2020})},\ \Eprint
  {https://arxiv.org/abs/1912.08218} {arXiv:1912.08218 [astro-ph.GA]}
  \BibitemShut {NoStop}%
\bibitem [{\citenamefont {Zevin}\ and\ \citenamefont
  {Bavera}(2022)}]{Zevin:2022wrw}%
  \BibitemOpen
  \bibfield  {author} {\bibinfo {author} {\bibfnamefont {M.}~\bibnamefont
  {Zevin}}\ and\ \bibinfo {author} {\bibfnamefont {S.~S.}\ \bibnamefont
  {Bavera}},\ }\bibfield  {title} {\bibinfo {title} {{Suspicious Siblings: The
  Distribution of Mass and Spin across Component Black Holes in Isolated Binary
  Evolution}},\ }\href {https://doi.org/10.3847/1538-4357/ac6f5d} {\bibfield
  {journal} {\bibinfo  {journal} {Astrophys. J.}\ }\textbf {\bibinfo {volume}
  {933}},\ \bibinfo {pages} {86} (\bibinfo {year} {2022})},\ \Eprint
  {https://arxiv.org/abs/2203.02515} {arXiv:2203.02515 [astro-ph.HE]}
  \BibitemShut {NoStop}%
\bibitem [{\citenamefont {Belczynski}\ \emph {et~al.}(2020)\citenamefont
  {Belczynski} \emph {et~al.}}]{Belczynski:2017gds}%
  \BibitemOpen
  \bibfield  {author} {\bibinfo {author} {\bibfnamefont {K.}~\bibnamefont
  {Belczynski}} \emph {et~al.},\ }\bibfield  {title} {\bibinfo {title}
  {{Evolutionary roads leading to low effective spins, high black hole masses,
  and O1/O2 rates for LIGO/Virgo binary black holes}},\ }\href
  {https://doi.org/10.1051/0004-6361/201936528} {\bibfield  {journal} {\bibinfo
   {journal} {Astron. Astrophys.}\ }\textbf {\bibinfo {volume} {636}},\
  \bibinfo {pages} {A104} (\bibinfo {year} {2020})},\ \Eprint
  {https://arxiv.org/abs/1706.07053} {arXiv:1706.07053 [astro-ph.HE]}
  \BibitemShut {NoStop}%
\bibitem [{\citenamefont {Portegies~Zwart}\ and\ \citenamefont
  {McMillan}(2002)}]{PortegiesZwart:2002iks}%
  \BibitemOpen
  \bibfield  {author} {\bibinfo {author} {\bibfnamefont {S.~F.}\ \bibnamefont
  {Portegies~Zwart}}\ and\ \bibinfo {author} {\bibfnamefont {S.~L.~W.}\
  \bibnamefont {McMillan}},\ }\bibfield  {title} {\bibinfo {title} {{The
  Runaway growth of intermediate-mass black holes in dense star clusters}},\
  }\href {https://doi.org/10.1086/341798} {\bibfield  {journal} {\bibinfo
  {journal} {Astrophys. J.}\ }\textbf {\bibinfo {volume} {576}},\ \bibinfo
  {pages} {899} (\bibinfo {year} {2002})},\ \Eprint
  {https://arxiv.org/abs/astro-ph/0201055} {arXiv:astro-ph/0201055}
  \BibitemShut {NoStop}%
\bibitem [{\citenamefont {Rodriguez}\ \emph {et~al.}(2015)\citenamefont
  {Rodriguez}, \citenamefont {Morscher}, \citenamefont {Pattabiraman},
  \citenamefont {Chatterjee}, \citenamefont {Haster},\ and\ \citenamefont
  {Rasio}}]{Rodriguez:2015oxa}%
  \BibitemOpen
  \bibfield  {author} {\bibinfo {author} {\bibfnamefont {C.~L.}\ \bibnamefont
  {Rodriguez}}, \bibinfo {author} {\bibfnamefont {M.}~\bibnamefont {Morscher}},
  \bibinfo {author} {\bibfnamefont {B.}~\bibnamefont {Pattabiraman}}, \bibinfo
  {author} {\bibfnamefont {S.}~\bibnamefont {Chatterjee}}, \bibinfo {author}
  {\bibfnamefont {C.-J.}\ \bibnamefont {Haster}},\ and\ \bibinfo {author}
  {\bibfnamefont {F.~A.}\ \bibnamefont {Rasio}},\ }\bibfield  {title} {\bibinfo
  {title} {{Binary Black Hole Mergers from Globular Clusters: Implications for
  Advanced LIGO}},\ }\href {https://doi.org/10.1103/PhysRevLett.115.051101}
  {\bibfield  {journal} {\bibinfo  {journal} {Phys. Rev. Lett.}\ }\textbf
  {\bibinfo {volume} {115}},\ \bibinfo {pages} {051101} (\bibinfo {year}
  {2015})},\ \bibinfo {note} {[Erratum: Phys.Rev.Lett. 116, 029901 (2016)]},\
  \Eprint {https://arxiv.org/abs/1505.00792} {arXiv:1505.00792 [astro-ph.HE]}
  \BibitemShut {NoStop}%
\bibitem [{\citenamefont {Gerosa}\ and\ \citenamefont
  {Fishbach}(2021)}]{Gerosa:2021mno}%
  \BibitemOpen
  \bibfield  {author} {\bibinfo {author} {\bibfnamefont {D.}~\bibnamefont
  {Gerosa}}\ and\ \bibinfo {author} {\bibfnamefont {M.}~\bibnamefont
  {Fishbach}},\ }\bibfield  {title} {\bibinfo {title} {{Hierarchical mergers of
  stellar-mass black holes and their gravitational-wave signatures}},\ }\href
  {https://doi.org/10.1038/s41550-021-01398-w} {\bibfield  {journal} {\bibinfo
  {journal} {Nature Astron.}\ }\textbf {\bibinfo {volume} {5}},\ \bibinfo
  {pages} {8} (\bibinfo {year} {2021})},\ \Eprint
  {https://arxiv.org/abs/2105.03439} {arXiv:2105.03439 [astro-ph.HE]}
  \BibitemShut {NoStop}%
\bibitem [{\citenamefont {Callister}\ \emph
  {et~al.}(2021{\natexlab{a}})\citenamefont {Callister}, \citenamefont
  {Haster}, \citenamefont {Ng}, \citenamefont {Vitale},\ and\ \citenamefont
  {Farr}}]{Callister:2021fpo}%
  \BibitemOpen
  \bibfield  {author} {\bibinfo {author} {\bibfnamefont {T.~A.}\ \bibnamefont
  {Callister}}, \bibinfo {author} {\bibfnamefont {C.-J.}\ \bibnamefont
  {Haster}}, \bibinfo {author} {\bibfnamefont {K.~K.~Y.}\ \bibnamefont {Ng}},
  \bibinfo {author} {\bibfnamefont {S.}~\bibnamefont {Vitale}},\ and\ \bibinfo
  {author} {\bibfnamefont {W.~M.}\ \bibnamefont {Farr}},\ }\bibfield  {title}
  {\bibinfo {title} {{Who Ordered That? Unequal-mass Binary Black Hole Mergers
  Have Larger Effective Spins}},\ }\href
  {https://doi.org/10.3847/2041-8213/ac2ccc} {\bibfield  {journal} {\bibinfo
  {journal} {Astrophys. J. Lett.}\ }\textbf {\bibinfo {volume} {922}},\
  \bibinfo {pages} {L5} (\bibinfo {year} {2021}{\natexlab{a}})},\ \Eprint
  {https://arxiv.org/abs/2106.00521} {arXiv:2106.00521 [astro-ph.HE]}
  \BibitemShut {NoStop}%
\bibitem [{\citenamefont {Adamcewicz}\ and\ \citenamefont
  {Thrane}(2022)}]{Adamcewicz:2022hce}%
  \BibitemOpen
  \bibfield  {author} {\bibinfo {author} {\bibfnamefont {C.}~\bibnamefont
  {Adamcewicz}}\ and\ \bibinfo {author} {\bibfnamefont {E.}~\bibnamefont
  {Thrane}},\ }\bibfield  {title} {\bibinfo {title} {{Do unequal-mass binary
  black hole systems have larger \ensuremath{\chi}eff? Probing correlations
  with copulas in gravitational-wave astronomy}},\ }\href
  {https://doi.org/10.1093/mnras/stac2961} {\bibfield  {journal} {\bibinfo
  {journal} {Mon. Not. Roy. Astron. Soc.}\ }\textbf {\bibinfo {volume} {517}},\
  \bibinfo {pages} {3928} (\bibinfo {year} {2022})},\ \Eprint
  {https://arxiv.org/abs/2208.03405} {arXiv:2208.03405 [astro-ph.HE]}
  \BibitemShut {NoStop}%
\bibitem [{\citenamefont {Adamcewicz}\ \emph {et~al.}(2023)\citenamefont
  {Adamcewicz}, \citenamefont {Lasky},\ and\ \citenamefont
  {Thrane}}]{Adamcewicz:2023mov}%
  \BibitemOpen
  \bibfield  {author} {\bibinfo {author} {\bibfnamefont {C.}~\bibnamefont
  {Adamcewicz}}, \bibinfo {author} {\bibfnamefont {P.~D.}\ \bibnamefont
  {Lasky}},\ and\ \bibinfo {author} {\bibfnamefont {E.}~\bibnamefont
  {Thrane}},\ }\bibfield  {title} {\bibinfo {title} {{Evidence for a
  Correlation between Binary Black Hole Mass Ratio and Black Hole Spins}},\
  }\href {https://doi.org/10.3847/1538-4357/acf763} {\bibfield  {journal}
  {\bibinfo  {journal} {Astrophys. J.}\ }\textbf {\bibinfo {volume} {958}},\
  \bibinfo {pages} {13} (\bibinfo {year} {2023})},\ \Eprint
  {https://arxiv.org/abs/2307.15278} {arXiv:2307.15278 [astro-ph.HE]}
  \BibitemShut {NoStop}%
\bibitem [{\citenamefont {Safarzadeh}\ \emph {et~al.}(2020)\citenamefont
  {Safarzadeh}, \citenamefont {Farr},\ and\ \citenamefont
  {Ramirez-Ruiz}}]{Safarzadeh:2020mlb}%
  \BibitemOpen
  \bibfield  {author} {\bibinfo {author} {\bibfnamefont {M.}~\bibnamefont
  {Safarzadeh}}, \bibinfo {author} {\bibfnamefont {W.~M.}\ \bibnamefont
  {Farr}},\ and\ \bibinfo {author} {\bibfnamefont {E.}~\bibnamefont
  {Ramirez-Ruiz}},\ }\bibfield  {title} {\bibinfo {title} {{A trend in the
  effective spin distribution of LIGO binary black holes with mass}},\ }\href
  {https://doi.org/10.3847/1538-4357/ab80be} {\bibfield  {journal} {\bibinfo
  {journal} {Astrophys. J.}\ }\textbf {\bibinfo {volume} {894}},\ \bibinfo
  {pages} {129} (\bibinfo {year} {2020})},\ \Eprint
  {https://arxiv.org/abs/2001.06490} {arXiv:2001.06490 [gr-qc]} \BibitemShut
  {NoStop}%
\bibitem [{\citenamefont {Franciolini}\ and\ \citenamefont
  {Pani}(2022)}]{Franciolini:2022iaa}%
  \BibitemOpen
  \bibfield  {author} {\bibinfo {author} {\bibfnamefont {G.}~\bibnamefont
  {Franciolini}}\ and\ \bibinfo {author} {\bibfnamefont {P.}~\bibnamefont
  {Pani}},\ }\bibfield  {title} {\bibinfo {title} {{Searching for mass-spin
  correlations in the population of gravitational-wave events: The GWTC-3 case
  study}},\ }\href {https://doi.org/10.1103/PhysRevD.105.123024} {\bibfield
  {journal} {\bibinfo  {journal} {Phys. Rev. D}\ }\textbf {\bibinfo {volume}
  {105}},\ \bibinfo {pages} {123024} (\bibinfo {year} {2022})},\ \Eprint
  {https://arxiv.org/abs/2201.13098} {arXiv:2201.13098 [astro-ph.HE]}
  \BibitemShut {NoStop}%
\bibitem [{\citenamefont {Biscoveanu}\ \emph {et~al.}(2022)\citenamefont
  {Biscoveanu}, \citenamefont {Callister}, \citenamefont {Haster},
  \citenamefont {Ng}, \citenamefont {Vitale},\ and\ \citenamefont
  {Farr}}]{Biscoveanu:2022qac}%
  \BibitemOpen
  \bibfield  {author} {\bibinfo {author} {\bibfnamefont {S.}~\bibnamefont
  {Biscoveanu}}, \bibinfo {author} {\bibfnamefont {T.~A.}\ \bibnamefont
  {Callister}}, \bibinfo {author} {\bibfnamefont {C.-J.}\ \bibnamefont
  {Haster}}, \bibinfo {author} {\bibfnamefont {K.~K.~Y.}\ \bibnamefont {Ng}},
  \bibinfo {author} {\bibfnamefont {S.}~\bibnamefont {Vitale}},\ and\ \bibinfo
  {author} {\bibfnamefont {W.~M.}\ \bibnamefont {Farr}},\ }\bibfield  {title}
  {\bibinfo {title} {{The Binary Black Hole Spin Distribution Likely Broadens
  with Redshift}},\ }\href {https://doi.org/10.3847/2041-8213/ac71a8}
  {\bibfield  {journal} {\bibinfo  {journal} {Astrophys. J. Lett.}\ }\textbf
  {\bibinfo {volume} {932}},\ \bibinfo {pages} {L19} (\bibinfo {year}
  {2022})},\ \Eprint {https://arxiv.org/abs/2204.01578} {arXiv:2204.01578
  [astro-ph.HE]} \BibitemShut {NoStop}%
\bibitem [{\citenamefont {Rinaldi}\ \emph {et~al.}(2023)\citenamefont
  {Rinaldi}, \citenamefont {Del~Pozzo}, \citenamefont {Mapelli}, \citenamefont
  {Medina},\ and\ \citenamefont {Dent}}]{Rinaldi:2023bbd}%
  \BibitemOpen
  \bibfield  {author} {\bibinfo {author} {\bibfnamefont {S.}~\bibnamefont
  {Rinaldi}}, \bibinfo {author} {\bibfnamefont {W.}~\bibnamefont {Del~Pozzo}},
  \bibinfo {author} {\bibfnamefont {M.}~\bibnamefont {Mapelli}}, \bibinfo
  {author} {\bibfnamefont {A.~L.}\ \bibnamefont {Medina}},\ and\ \bibinfo
  {author} {\bibfnamefont {T.}~\bibnamefont {Dent}},\ }\href@noop {} {\bibinfo
  {title} {{Evidence for the evolution of black hole mass function with
  redshift}}} (\bibinfo {year} {2023}),\ \Eprint
  {https://arxiv.org/abs/2310.03074} {arXiv:2310.03074 [astro-ph.HE]}
  \BibitemShut {NoStop}%
\bibitem [{\citenamefont {van Son}\ \emph {et~al.}(2022)\citenamefont {van
  Son}, \citenamefont {de~Mink}, \citenamefont {Callister}, \citenamefont
  {Justham}, \citenamefont {Renzo}, \citenamefont {Wagg}, \citenamefont
  {Broekgaarden}, \citenamefont {Kummer}, \citenamefont {Pakmor},\ and\
  \citenamefont {Mandel}}]{vanSon:2021zpk}%
  \BibitemOpen
  \bibfield  {author} {\bibinfo {author} {\bibfnamefont {L.~A.~C.}\
  \bibnamefont {van Son}}, \bibinfo {author} {\bibfnamefont {S.~E.}\
  \bibnamefont {de~Mink}}, \bibinfo {author} {\bibfnamefont {T.}~\bibnamefont
  {Callister}}, \bibinfo {author} {\bibfnamefont {S.}~\bibnamefont {Justham}},
  \bibinfo {author} {\bibfnamefont {M.}~\bibnamefont {Renzo}}, \bibinfo
  {author} {\bibfnamefont {T.}~\bibnamefont {Wagg}}, \bibinfo {author}
  {\bibfnamefont {F.~S.}\ \bibnamefont {Broekgaarden}}, \bibinfo {author}
  {\bibfnamefont {F.}~\bibnamefont {Kummer}}, \bibinfo {author} {\bibfnamefont
  {R.}~\bibnamefont {Pakmor}},\ and\ \bibinfo {author} {\bibfnamefont
  {I.}~\bibnamefont {Mandel}},\ }\bibfield  {title} {\bibinfo {title} {{The
  Redshift Evolution of the Binary Black Hole Merger Rate: A Weighty Matter}},\
  }\href {https://doi.org/10.3847/1538-4357/ac64a3} {\bibfield  {journal}
  {\bibinfo  {journal} {Astrophys. J.}\ }\textbf {\bibinfo {volume} {931}},\
  \bibinfo {pages} {17} (\bibinfo {year} {2022})},\ \Eprint
  {https://arxiv.org/abs/2110.01634} {arXiv:2110.01634 [astro-ph.HE]}
  \BibitemShut {NoStop}%
\bibitem [{\citenamefont {Loredo}(2004)}]{Loredo_2004}%
  \BibitemOpen
  \bibfield  {author} {\bibinfo {author} {\bibfnamefont {T.~J.}\ \bibnamefont
  {Loredo}},\ }\bibfield  {title} {\bibinfo {title} {Accounting for source
  uncertainties in analyses of astronomical survey data},\ }in\ \href
  {https://doi.org/10.1063/1.1835214} {\emph {\bibinfo {booktitle} {Bayesian
  Inference and Maximum Entropy Methods in Science and Engineering ed R.
  Fischer, R. Preuss and U. V. Toussaint}}},\ \bibinfo {series} {AIP Conference
  Series}, Vol.\ \bibinfo {volume} {735}\ (\bibinfo  {publisher} {AIP},\
  \bibinfo {address} {Melville, NY},\ \bibinfo {year} {2004})\ p.\ \bibinfo
  {pages} {195}\BibitemShut {NoStop}%
\bibitem [{\citenamefont {Farr}\ \emph {et~al.}(2015)\citenamefont {Farr},
  \citenamefont {Gair}, \citenamefont {Mandel},\ and\ \citenamefont
  {Cutler}}]{Farr:2013yna}%
  \BibitemOpen
  \bibfield  {author} {\bibinfo {author} {\bibfnamefont {W.~M.}\ \bibnamefont
  {Farr}}, \bibinfo {author} {\bibfnamefont {J.~R.}\ \bibnamefont {Gair}},
  \bibinfo {author} {\bibfnamefont {I.}~\bibnamefont {Mandel}},\ and\ \bibinfo
  {author} {\bibfnamefont {C.}~\bibnamefont {Cutler}},\ }\bibfield  {title}
  {\bibinfo {title} {{Counting And Confusion: Bayesian Rate Estimation With
  Multiple Populations}},\ }\href {https://doi.org/10.1103/PhysRevD.91.023005}
  {\bibfield  {journal} {\bibinfo  {journal} {Phys. Rev. D}\ }\textbf {\bibinfo
  {volume} {91}},\ \bibinfo {pages} {023005} (\bibinfo {year} {2015})},\
  \Eprint {https://arxiv.org/abs/1302.5341} {arXiv:1302.5341 [astro-ph.IM]}
  \BibitemShut {NoStop}%
\bibitem [{\citenamefont {Foreman-Mackey}\ \emph {et~al.}(2014)\citenamefont
  {Foreman-Mackey}, \citenamefont {Hogg},\ and\ \citenamefont
  {Morton}}]{Foreman-Mackey_2014}%
  \BibitemOpen
  \bibfield  {author} {\bibinfo {author} {\bibfnamefont {D.}~\bibnamefont
  {Foreman-Mackey}}, \bibinfo {author} {\bibfnamefont {D.~W.}\ \bibnamefont
  {Hogg}},\ and\ \bibinfo {author} {\bibfnamefont {T.~D.}\ \bibnamefont
  {Morton}},\ }\bibfield  {title} {\bibinfo {title} {Exoplanet population
  inference and the abundance of earth analogs from noise, incomplete
  catalogs},\ }\href {https://doi.org/10.1088/0004-637X/795/1/64} {\bibfield
  {journal} {\bibinfo  {journal} {The Astrophysical Journal}\ }\textbf
  {\bibinfo {volume} {795}},\ \bibinfo {pages} {64} (\bibinfo {year}
  {2014})}\BibitemShut {NoStop}%
\bibitem [{\citenamefont {Mandel}\ \emph {et~al.}(2019)\citenamefont {Mandel},
  \citenamefont {Farr},\ and\ \citenamefont {Gair}}]{Mandel_2019}%
  \BibitemOpen
  \bibfield  {author} {\bibinfo {author} {\bibfnamefont {I.}~\bibnamefont
  {Mandel}}, \bibinfo {author} {\bibfnamefont {W.~M.}\ \bibnamefont {Farr}},\
  and\ \bibinfo {author} {\bibfnamefont {J.~R.}\ \bibnamefont {Gair}},\
  }\bibfield  {title} {\bibinfo {title} {{Extracting distribution parameters
  from multiple uncertain observations with selection biases}},\ }\href
  {https://doi.org/10.1093/mnras/stz896} {\bibfield  {journal} {\bibinfo
  {journal} {MNRAS}\ }\textbf {\bibinfo {volume} {486}},\ \bibinfo {pages}
  {1086} (\bibinfo {year} {2019})},\ \Eprint
  {https://arxiv.org/abs/http://oup.prod.sis.lan/mnras/article-pdf/486/1/1086/28390969/stz896.pdf}
  {http://oup.prod.sis.lan/mnras/article-pdf/486/1/1086/28390969/stz896.pdf}
  \BibitemShut {NoStop}%
\bibitem [{\citenamefont {Vitale}(2020)}]{Vitale:2020aaz}%
  \BibitemOpen
  \bibfield  {author} {\bibinfo {author} {\bibfnamefont {S.}~\bibnamefont
  {Vitale}},\ }\href@noop {} {\bibinfo {title} {{One, No One, and One Hundred
  Thousand -- Inferring the properties of a population in presence of selection
  effects}}} (\bibinfo {year} {2020}),\ \Eprint
  {https://arxiv.org/abs/2007.05579} {arXiv:2007.05579 [astro-ph.IM]}
  \BibitemShut {NoStop}%
\bibitem [{\citenamefont {Essick}\ and\ \citenamefont
  {Fishbach}(2023)}]{Essick:2023upv}%
  \BibitemOpen
  \bibfield  {author} {\bibinfo {author} {\bibfnamefont {R.}~\bibnamefont
  {Essick}}\ and\ \bibinfo {author} {\bibfnamefont {M.}~\bibnamefont
  {Fishbach}},\ }\href@noop {} {\bibinfo {title} {{DAGnabbit! Ensuring
  Consistency between Noise and Detection in Hierarchical Bayesian Inference}}}
  (\bibinfo {year} {2023}),\ \Eprint {https://arxiv.org/abs/2310.02017}
  {arXiv:2310.02017 [gr-qc]} \BibitemShut {NoStop}%
\bibitem [{\citenamefont {Farr}(2019)}]{Farr_2019}%
  \BibitemOpen
  \bibfield  {author} {\bibinfo {author} {\bibfnamefont {W.~M.}\ \bibnamefont
  {Farr}},\ }\bibfield  {title} {\bibinfo {title} {Accuracy requirements for
  empirically measured selection functions},\ }\href
  {https://doi.org/10.3847/2515-5172/ab1d5f} {\bibfield  {journal} {\bibinfo
  {journal} {Research Notes of the {AAS}}\ }\textbf {\bibinfo {volume} {3}},\
  \bibinfo {pages} {66} (\bibinfo {year} {2019})}\BibitemShut {NoStop}%
\bibitem [{\citenamefont {Essick}\ and\ \citenamefont
  {Farr}(2022)}]{Essick:2022ojx}%
  \BibitemOpen
  \bibfield  {author} {\bibinfo {author} {\bibfnamefont {R.}~\bibnamefont
  {Essick}}\ and\ \bibinfo {author} {\bibfnamefont {W.}~\bibnamefont {Farr}},\
  }\href@noop {} {\bibinfo {title} {{Precision Requirements for Monte Carlo
  Sums within Hierarchical Bayesian Inference}}} (\bibinfo {year} {2022}),\
  \Eprint {https://arxiv.org/abs/2204.00461} {arXiv:2204.00461 [astro-ph.IM]}
  \BibitemShut {NoStop}%
\bibitem [{\citenamefont {Talbot}\ and\ \citenamefont
  {Golomb}(2023)}]{Talbot:2023pex}%
  \BibitemOpen
  \bibfield  {author} {\bibinfo {author} {\bibfnamefont {C.}~\bibnamefont
  {Talbot}}\ and\ \bibinfo {author} {\bibfnamefont {J.}~\bibnamefont
  {Golomb}},\ }\bibfield  {title} {\bibinfo {title} {{Growing pains:
  understanding the impact of likelihood uncertainty on hierarchical Bayesian
  inference for gravitational-wave astronomy}},\ }\href
  {https://doi.org/10.1093/mnras/stad2968} {\bibfield  {journal} {\bibinfo
  {journal} {Mon. Not. Roy. Astron. Soc.}\ }\textbf {\bibinfo {volume} {526}},\
  \bibinfo {pages} {3495} (\bibinfo {year} {2023})},\ \Eprint
  {https://arxiv.org/abs/2304.06138} {arXiv:2304.06138 [astro-ph.IM]}
  \BibitemShut {NoStop}%
\bibitem [{\citenamefont {Fishbach}\ \emph {et~al.}(2018)\citenamefont
  {Fishbach}, \citenamefont {Holz},\ and\ \citenamefont
  {Farr}}]{Fishbach:2018edt}%
  \BibitemOpen
  \bibfield  {author} {\bibinfo {author} {\bibfnamefont {M.}~\bibnamefont
  {Fishbach}}, \bibinfo {author} {\bibfnamefont {D.~E.}\ \bibnamefont {Holz}},\
  and\ \bibinfo {author} {\bibfnamefont {W.~M.}\ \bibnamefont {Farr}},\
  }\bibfield  {title} {\bibinfo {title} {{Does the Black Hole Merger Rate
  Evolve with Redshift?}},\ }\href {https://doi.org/10.3847/2041-8213/aad800}
  {\bibfield  {journal} {\bibinfo  {journal} {Astrophys. J. Lett.}\ }\textbf
  {\bibinfo {volume} {863}},\ \bibinfo {pages} {L41} (\bibinfo {year}
  {2018})},\ \Eprint {https://arxiv.org/abs/1805.10270} {arXiv:1805.10270
  [astro-ph.HE]} \BibitemShut {NoStop}%
\bibitem [{\citenamefont {Damour}(2001)}]{Damour:2001}%
  \BibitemOpen
  \bibfield  {author} {\bibinfo {author} {\bibfnamefont {T.}~\bibnamefont
  {Damour}},\ }\bibfield  {title} {\bibinfo {title} {Coalescence of two
  spinning black holes: An effective one-body approach},\ }\href
  {https://doi.org/10.1103/PhysRevD.64.124013} {\bibfield  {journal} {\bibinfo
  {journal} {Phys. Rev. D}\ }\textbf {\bibinfo {volume} {64}},\ \bibinfo
  {pages} {124013} (\bibinfo {year} {2001})}\BibitemShut {NoStop}%
\bibitem [{\citenamefont {Racine}(2008)}]{Racine:2008qv}%
  \BibitemOpen
  \bibfield  {author} {\bibinfo {author} {\bibfnamefont {E.}~\bibnamefont
  {Racine}},\ }\bibfield  {title} {\bibinfo {title} {{Analysis of spin
  precession in binary black hole systems including quadrupole-monopole
  interaction}},\ }\href {https://doi.org/10.1103/PhysRevD.78.044021}
  {\bibfield  {journal} {\bibinfo  {journal} {Phys. Rev. D}\ }\textbf {\bibinfo
  {volume} {78}},\ \bibinfo {pages} {044021} (\bibinfo {year} {2008})},\
  \Eprint {https://arxiv.org/abs/0803.1820} {arXiv:0803.1820 [gr-qc]}
  \BibitemShut {NoStop}%
\bibitem [{\citenamefont {Ajith}\ \emph {et~al.}(2011)\citenamefont {Ajith}
  \emph {et~al.}}]{Ajith:2009bn}%
  \BibitemOpen
  \bibfield  {author} {\bibinfo {author} {\bibfnamefont {P.}~\bibnamefont
  {Ajith}} \emph {et~al.},\ }\bibfield  {title} {\bibinfo {title}
  {{Inspiral-merger-ringdown waveforms for black-hole binaries with
  non-precessing spins}},\ }\href
  {https://doi.org/10.1103/PhysRevLett.106.241101} {\bibfield  {journal}
  {\bibinfo  {journal} {Phys. Rev. Lett.}\ }\textbf {\bibinfo {volume} {106}},\
  \bibinfo {pages} {241101} (\bibinfo {year} {2011})},\ \Eprint
  {https://arxiv.org/abs/0909.2867} {arXiv:0909.2867 [gr-qc]} \BibitemShut
  {NoStop}%
\bibitem [{\citenamefont {Vitale}\ \emph {et~al.}(2017)\citenamefont {Vitale},
  \citenamefont {Lynch}, \citenamefont {Raymond}, \citenamefont {Sturani},
  \citenamefont {Veitch},\ and\ \citenamefont {Graff}}]{Vitale:2016avz}%
  \BibitemOpen
  \bibfield  {author} {\bibinfo {author} {\bibfnamefont {S.}~\bibnamefont
  {Vitale}}, \bibinfo {author} {\bibfnamefont {R.}~\bibnamefont {Lynch}},
  \bibinfo {author} {\bibfnamefont {V.}~\bibnamefont {Raymond}}, \bibinfo
  {author} {\bibfnamefont {R.}~\bibnamefont {Sturani}}, \bibinfo {author}
  {\bibfnamefont {J.}~\bibnamefont {Veitch}},\ and\ \bibinfo {author}
  {\bibfnamefont {P.}~\bibnamefont {Graff}},\ }\bibfield  {title} {\bibinfo
  {title} {{Parameter estimation for heavy binary-black holes with networks of
  second-generation gravitational-wave detectors}},\ }\href
  {https://doi.org/10.1103/PhysRevD.95.064053} {\bibfield  {journal} {\bibinfo
  {journal} {Phys. Rev. D}\ }\textbf {\bibinfo {volume} {95}},\ \bibinfo
  {pages} {064053} (\bibinfo {year} {2017})},\ \Eprint
  {https://arxiv.org/abs/1611.01122} {arXiv:1611.01122 [gr-qc]} \BibitemShut
  {NoStop}%
\bibitem [{\citenamefont {P\"urrer}\ \emph {et~al.}(2016)\citenamefont
  {P\"urrer}, \citenamefont {Hannam},\ and\ \citenamefont
  {Ohme}}]{Purrer:2015nkh}%
  \BibitemOpen
  \bibfield  {author} {\bibinfo {author} {\bibfnamefont {M.}~\bibnamefont
  {P\"urrer}}, \bibinfo {author} {\bibfnamefont {M.}~\bibnamefont {Hannam}},\
  and\ \bibinfo {author} {\bibfnamefont {F.}~\bibnamefont {Ohme}},\ }\bibfield
  {title} {\bibinfo {title} {{Can we measure individual black-hole spins from
  gravitational-wave observations?}},\ }\href
  {https://doi.org/10.1103/PhysRevD.93.084042} {\bibfield  {journal} {\bibinfo
  {journal} {Phys. Rev. D}\ }\textbf {\bibinfo {volume} {93}},\ \bibinfo
  {pages} {084042} (\bibinfo {year} {2016})},\ \Eprint
  {https://arxiv.org/abs/1512.04955} {arXiv:1512.04955 [gr-qc]} \BibitemShut
  {NoStop}%
\bibitem [{\citenamefont {Miller}\ \emph {et~al.}(2020)\citenamefont {Miller},
  \citenamefont {Callister},\ and\ \citenamefont {Farr}}]{Miller:2020zox}%
  \BibitemOpen
  \bibfield  {author} {\bibinfo {author} {\bibfnamefont {S.}~\bibnamefont
  {Miller}}, \bibinfo {author} {\bibfnamefont {T.~A.}\ \bibnamefont
  {Callister}},\ and\ \bibinfo {author} {\bibfnamefont {W.}~\bibnamefont
  {Farr}},\ }\bibfield  {title} {\bibinfo {title} {{The Low Effective Spin of
  Binary Black Holes and Implications for Individual Gravitational-Wave
  Events}},\ }\href {https://doi.org/10.3847/1538-4357/ab80c0} {\bibfield
  {journal} {\bibinfo  {journal} {Astrophys. J.}\ }\textbf {\bibinfo {volume}
  {895}},\ \bibinfo {pages} {128} (\bibinfo {year} {2020})},\ \Eprint
  {https://arxiv.org/abs/2001.06051} {arXiv:2001.06051 [astro-ph.HE]}
  \BibitemShut {NoStop}%
\bibitem [{\citenamefont {Roulet}\ and\ \citenamefont
  {Zaldarriaga}(2019)}]{Roulet:2018jbe}%
  \BibitemOpen
  \bibfield  {author} {\bibinfo {author} {\bibfnamefont {J.}~\bibnamefont
  {Roulet}}\ and\ \bibinfo {author} {\bibfnamefont {M.}~\bibnamefont
  {Zaldarriaga}},\ }\bibfield  {title} {\bibinfo {title} {{Constraints on
  binary black hole populations from LIGO--Virgo detections}},\ }\href
  {https://doi.org/10.1093/mnras/stz226} {\bibfield  {journal} {\bibinfo
  {journal} {Mon. Not. Roy. Astron. Soc.}\ }\textbf {\bibinfo {volume} {484}},\
  \bibinfo {pages} {4216} (\bibinfo {year} {2019})},\ \Eprint
  {https://arxiv.org/abs/1806.10610} {arXiv:1806.10610 [astro-ph.HE]}
  \BibitemShut {NoStop}%
\bibitem [{\citenamefont {{de Boor}}(1976)}]{DeBoor_1976}%
  \BibitemOpen
  \bibfield  {author} {\bibinfo {author} {\bibfnamefont {C.}~\bibnamefont {{de
  Boor}}},\ }\bibfield  {title} {\bibinfo {title} {On cubic spline functions
  that vanish at all knots},\ }\href
  {https://doi.org/https://doi.org/10.1016/0001-8708(76)90166-3} {\bibfield
  {journal} {\bibinfo  {journal} {Advances in Mathematics}\ }\textbf {\bibinfo
  {volume} {20}},\ \bibinfo {pages} {1} (\bibinfo {year} {1976})}\BibitemShut
  {NoStop}%
\bibitem [{\citenamefont {{LIGO Scientific Collaboration}}\ \emph
  {et~al.}(2020)\citenamefont {{LIGO Scientific Collaboration}}, \citenamefont
  {{Virgo Collaboration}},\ and\ \citenamefont {{KAGRA
  Collaboration}}}]{gwtc1_pe}%
  \BibitemOpen
  \bibfield  {author} {\bibinfo {author} {\bibnamefont {{LIGO Scientific
  Collaboration}}}, \bibinfo {author} {\bibnamefont {{Virgo Collaboration}}},\
  and\ \bibinfo {author} {\bibnamefont {{KAGRA Collaboration}}},\ }\href
  {https://dcc.ligo.org/LIGO-P1800370/public} {\emph {\bibinfo {title}
  {Parameter estimation sample release for GWTC-1}}},\ \bibinfo {type} {Tech.
  Rep.}\ \bibinfo {number} {P1800370}\ (\bibinfo  {institution} {LVK},\
  \bibinfo {year} {2020})\BibitemShut {NoStop}%
\bibitem [{\citenamefont {{LIGO Scientific Collaboration}}\ \emph
  {et~al.}(2021)\citenamefont {{LIGO Scientific Collaboration}}, \citenamefont
  {{Virgo Collaboration}},\ and\ \citenamefont {{KAGRA
  Collaboration}}}]{gwtc2_pe}%
  \BibitemOpen
  \bibfield  {author} {\bibinfo {author} {\bibnamefont {{LIGO Scientific
  Collaboration}}}, \bibinfo {author} {\bibnamefont {{Virgo Collaboration}}},\
  and\ \bibinfo {author} {\bibnamefont {{KAGRA Collaboration}}},\ }\href
  {https://dcc.ligo.org/P2000223/public} {\emph {\bibinfo {title} {GWTC-2 Data
  Release: Parameter Estimation Samples and Skymaps}}},\ \bibinfo {type} {Tech.
  Rep.}\ \bibinfo {number} {P2000223}\ (\bibinfo  {institution} {LVK},\
  \bibinfo {year} {2021})\BibitemShut {NoStop}%
\bibitem [{\citenamefont {{LIGO Scientific Collaboration}}\ and\ \citenamefont
  {{Virgo
  Collaboration}}(2022)}]{ligo_scientific_collaboration_and_virgo_2022_6513631}%
  \BibitemOpen
  \bibfield  {author} {\bibinfo {author} {\bibnamefont {{LIGO Scientific
  Collaboration}}}\ and\ \bibinfo {author} {\bibnamefont {{Virgo
  Collaboration}}},\ }\bibfield  {title} {\bibinfo {title} {{GWTC-2.1: Deep
  Extended Catalog of Compact Binary Coalescences Observed by LIGO and Virgo
  During the First Half of the Third Observing Run - Parameter Estimation Data
  Release}},\ }\href {https://doi.org/10.5281/zenodo.6513631}
  {10.5281/zenodo.6513631} (\bibinfo {year} {2022})\BibitemShut {NoStop}%
\bibitem [{\citenamefont {{LIGO Scientific Collaboration}}\ \emph
  {et~al.}(2023{\natexlab{a}})\citenamefont {{LIGO Scientific Collaboration}},
  \citenamefont {{Virgo Collaboration}},\ and\ \citenamefont {{KAGRA
  Collaboration}}}]{ligo_scientific_collaboration_and_virgo_2023_8177023}%
  \BibitemOpen
  \bibfield  {author} {\bibinfo {author} {\bibnamefont {{LIGO Scientific
  Collaboration}}}, \bibinfo {author} {\bibnamefont {{Virgo Collaboration}}},\
  and\ \bibinfo {author} {\bibnamefont {{KAGRA Collaboration}}},\ }\bibfield
  {title} {\bibinfo {title} {{GWTC-3: Compact Binary Coalescences Observed by
  LIGO and Virgo During the Second Part of the Third Observing Run —
  Parameter estimation data release}},\ }\href
  {https://doi.org/10.5281/zenodo.8177023} {10.5281/zenodo.8177023} (\bibinfo
  {year} {2023}{\natexlab{a}})\BibitemShut {NoStop}%
\bibitem [{\citenamefont {{LIGO Scientific Collaboration}}\ \emph
  {et~al.}(2023{\natexlab{b}})\citenamefont {{LIGO Scientific Collaboration}},
  \citenamefont {{Virgo Collaboration}},\ and\ \citenamefont {{KAGRA
  Collaboration}}}]{ligo_scientific_collaboration_and_virgo_2023_7890398}%
  \BibitemOpen
  \bibfield  {author} {\bibinfo {author} {\bibnamefont {{LIGO Scientific
  Collaboration}}}, \bibinfo {author} {\bibnamefont {{Virgo Collaboration}}},\
  and\ \bibinfo {author} {\bibnamefont {{KAGRA Collaboration}}},\ }\bibfield
  {title} {\bibinfo {title} {{GWTC-3: Compact Binary Coalescences Observed by
  LIGO and Virgo During the Second Part of the Third Observing Run — O1+O2+O3
  Search Sensitivity Estimates}},\ }\href
  {https://doi.org/10.5281/zenodo.7890398} {10.5281/zenodo.7890398} (\bibinfo
  {year} {2023}{\natexlab{b}})\BibitemShut {NoStop}%
\bibitem [{\citenamefont {Talbot}\ \emph {et~al.}(2019)\citenamefont {Talbot},
  \citenamefont {Smith}, \citenamefont {Thrane},\ and\ \citenamefont
  {Poole}}]{Talbot:2019}%
  \BibitemOpen
  \bibfield  {author} {\bibinfo {author} {\bibfnamefont {C.}~\bibnamefont
  {Talbot}}, \bibinfo {author} {\bibfnamefont {R.}~\bibnamefont {Smith}},
  \bibinfo {author} {\bibfnamefont {E.}~\bibnamefont {Thrane}},\ and\ \bibinfo
  {author} {\bibfnamefont {G.~B.}\ \bibnamefont {Poole}},\ }\bibfield  {title}
  {\bibinfo {title} {Parallelized inference for gravitational-wave astronomy},\
  }\href {https://doi.org/10.1103/PhysRevD.100.043030} {\bibfield  {journal}
  {\bibinfo  {journal} {Phys. Rev. D}\ }\textbf {\bibinfo {volume} {100}},\
  \bibinfo {pages} {043030} (\bibinfo {year} {2019})}\BibitemShut {NoStop}%
\bibitem [{\citenamefont {Speagle}(2020)}]{Speagle_2020}%
  \BibitemOpen
  \bibfield  {author} {\bibinfo {author} {\bibfnamefont {J.~S.}\ \bibnamefont
  {Speagle}},\ }\bibfield  {title} {\bibinfo {title} {{dynesty: a dynamic
  nested sampling package for estimating Bayesian posteriors and evidences}},\
  }\href {https://doi.org/10.1093/mnras/staa278} {\bibfield  {journal}
  {\bibinfo  {journal} {MNRAS}\ }\textbf {\bibinfo {volume} {493}},\ \bibinfo
  {pages} {3132} (\bibinfo {year} {2020})},\ \Eprint
  {https://arxiv.org/abs/https://academic.oup.com/mnras/article-pdf/493/3/3132/32890730/staa278.pdf}
  {https://academic.oup.com/mnras/article-pdf/493/3/3132/32890730/staa278.pdf}
  \BibitemShut {NoStop}%
\bibitem [{\citenamefont {Ashton}\ \emph {et~al.}(2019)\citenamefont {Ashton}
  \emph {et~al.}}]{Ashton:2018jfp}%
  \BibitemOpen
  \bibfield  {author} {\bibinfo {author} {\bibfnamefont {G.}~\bibnamefont
  {Ashton}} \emph {et~al.},\ }\bibfield  {title} {\bibinfo {title} {{BILBY: A
  user-friendly Bayesian inference library for gravitational-wave astronomy}},\
  }\href {https://doi.org/10.3847/1538-4365/ab06fc} {\bibfield  {journal}
  {\bibinfo  {journal} {Astrophys. J. Suppl.}\ }\textbf {\bibinfo {volume}
  {241}},\ \bibinfo {pages} {27} (\bibinfo {year} {2019})},\ \Eprint
  {https://arxiv.org/abs/1811.02042} {arXiv:1811.02042 [astro-ph.IM]}
  \BibitemShut {NoStop}%
\bibitem [{Note1()}]{Note1}%
  \BibitemOpen
  \bibinfo {note} {Though $q\to 0$ is an unphysical region of parameter space
  with no observations, the node at $q=0$ should be thought of as simply a
  parameter necessary to ensure the model is defined over the entire space (any
  distribution must be defined over the full parameter space), and not an
  \protect \textit {a priori} statement that BBHs exist here; indeed the
  $p(q\protect \tmspace +\thickmuskip {.2777em}|\protect \tmspace +\thickmuskip
  {.2777em}m_1, \Lambda )$ smoothed powerlaw is always zero at
  $q=0$.}\BibitemShut {Stop}%
\bibitem [{Note2()}]{Note2}%
  \BibitemOpen
  \bibinfo {note} {This is somewhat at odds with the maximum redshift
  considered in the sensitivity injections of~\cite
  {ligo_scientific_collaboration_and_virgo_2023_7890398}, with $z_{\protect \rm
  max} = 1.9$. We verified our results are unchanged after considering this
  adjustment.}\BibitemShut {Stop}%
\bibitem [{\citenamefont {Kullback}\ and\ \citenamefont
  {Leibler}(1951)}]{kullback1951}%
  \BibitemOpen
  \bibfield  {author} {\bibinfo {author} {\bibfnamefont {S.}~\bibnamefont
  {Kullback}}\ and\ \bibinfo {author} {\bibfnamefont {R.~A.}\ \bibnamefont
  {Leibler}},\ }\bibfield  {title} {\bibinfo {title} {On information and
  sufficiency},\ }\href {https://doi.org/10.1214/aoms/1177729694} {\bibfield
  {journal} {\bibinfo  {journal} {Ann. Math. Statist.}\ }\textbf {\bibinfo
  {volume} {22}},\ \bibinfo {pages} {79} (\bibinfo {year} {1951})}\BibitemShut
  {NoStop}%
\bibitem [{\citenamefont {Lin}(1991)}]{Lin:1991zzm}%
  \BibitemOpen
  \bibfield  {author} {\bibinfo {author} {\bibfnamefont {J.}~\bibnamefont
  {Lin}},\ }\bibfield  {title} {\bibinfo {title} {{Divergence measures based on
  the Shannon entropy}},\ }\href {https://doi.org/10.1109/18.61115} {\bibfield
  {journal} {\bibinfo  {journal} {IEEE Trans. Info. Theor.}\ }\textbf {\bibinfo
  {volume} {37}},\ \bibinfo {pages} {145} (\bibinfo {year} {1991})}\BibitemShut
  {NoStop}%
\bibitem [{\citenamefont {Abbott}\ \emph {et~al.}(2020)\citenamefont {Abbott}
  \emph {et~al.}}]{O3_psds}%
  \BibitemOpen
  \bibfield  {author} {\bibinfo {author} {\bibfnamefont {B.}~\bibnamefont
  {Abbott}} \emph {et~al.} (\bibinfo {collaboration} {KAGRA, LIGO Scientific,
  Virgo, VIRGO}),\ }\bibfield  {title} {\bibinfo {title} {Noise curves used for
  simulations in the update of the observing scenarios paper}} (\bibinfo {year}
  {2020})\BibitemShut {NoStop}%
\bibitem [{\citenamefont {Pratten}\ \emph {et~al.}(2021)\citenamefont {Pratten}
  \emph {et~al.}}]{Pratten:2020ceb}%
  \BibitemOpen
  \bibfield  {author} {\bibinfo {author} {\bibfnamefont {G.}~\bibnamefont
  {Pratten}} \emph {et~al.},\ }\bibfield  {title} {\bibinfo {title}
  {{Computationally efficient models for the dominant and subdominant harmonic
  modes of precessing binary black holes}},\ }\href
  {https://doi.org/10.1103/PhysRevD.103.104056} {\bibfield  {journal} {\bibinfo
   {journal} {Phys. Rev. D}\ }\textbf {\bibinfo {volume} {103}},\ \bibinfo
  {pages} {104056} (\bibinfo {year} {2021})},\ \Eprint
  {https://arxiv.org/abs/2004.06503} {arXiv:2004.06503 [gr-qc]} \BibitemShut
  {NoStop}%
\bibitem [{\citenamefont {Cornish}(2010)}]{Cornish:2010kf}%
  \BibitemOpen
  \bibfield  {author} {\bibinfo {author} {\bibfnamefont {N.~J.}\ \bibnamefont
  {Cornish}},\ }\href@noop {} {\bibinfo {title} {{Fast Fisher Matrices and Lazy
  Likelihoods}}} (\bibinfo {year} {2010}),\ \Eprint
  {https://arxiv.org/abs/1007.4820} {arXiv:1007.4820 [gr-qc]} \BibitemShut
  {NoStop}%
\bibitem [{\citenamefont {Cornish}(2021)}]{Cornish:2021lje}%
  \BibitemOpen
  \bibfield  {author} {\bibinfo {author} {\bibfnamefont {N.~J.}\ \bibnamefont
  {Cornish}},\ }\bibfield  {title} {\bibinfo {title} {{Heterodyned likelihood
  for rapid gravitational wave parameter inference}},\ }\href
  {https://doi.org/10.1103/PhysRevD.104.104054} {\bibfield  {journal} {\bibinfo
   {journal} {Phys. Rev. D}\ }\textbf {\bibinfo {volume} {104}},\ \bibinfo
  {pages} {104054} (\bibinfo {year} {2021})},\ \Eprint
  {https://arxiv.org/abs/2109.02728} {arXiv:2109.02728 [gr-qc]} \BibitemShut
  {NoStop}%
\bibitem [{\citenamefont {Zackay}\ \emph {et~al.}(2018)\citenamefont {Zackay},
  \citenamefont {Dai},\ and\ \citenamefont {Venumadhav}}]{Zackay:2018qdy}%
  \BibitemOpen
  \bibfield  {author} {\bibinfo {author} {\bibfnamefont {B.}~\bibnamefont
  {Zackay}}, \bibinfo {author} {\bibfnamefont {L.}~\bibnamefont {Dai}},\ and\
  \bibinfo {author} {\bibfnamefont {T.}~\bibnamefont {Venumadhav}},\
  }\href@noop {} {\bibinfo {title} {{Relative Binning and Fast Likelihood
  Evaluation for Gravitational Wave Parameter Estimation}}} (\bibinfo {year}
  {2018}),\ \Eprint {https://arxiv.org/abs/1806.08792} {arXiv:1806.08792
  [astro-ph.IM]} \BibitemShut {NoStop}%
\bibitem [{\citenamefont {Essick}(2023)}]{Essick:2023toz}%
  \BibitemOpen
  \bibfield  {author} {\bibinfo {author} {\bibfnamefont {R.}~\bibnamefont
  {Essick}},\ }\bibfield  {title} {\bibinfo {title} {{Semianalytic sensitivity
  estimates for catalogs of gravitational-wave transients}},\ }\href
  {https://doi.org/10.1103/PhysRevD.108.043011} {\bibfield  {journal} {\bibinfo
   {journal} {Phys. Rev. D}\ }\textbf {\bibinfo {volume} {108}},\ \bibinfo
  {pages} {043011} (\bibinfo {year} {2023})},\ \Eprint
  {https://arxiv.org/abs/2307.02765} {arXiv:2307.02765 [gr-qc]} \BibitemShut
  {NoStop}%
\bibitem [{\citenamefont {Broekgaarden}\ \emph {et~al.}(2022)\citenamefont
  {Broekgaarden}, \citenamefont {Stevenson},\ and\ \citenamefont
  {Thrane}}]{Broekgaarden:2022nst}%
  \BibitemOpen
  \bibfield  {author} {\bibinfo {author} {\bibfnamefont {F.~S.}\ \bibnamefont
  {Broekgaarden}}, \bibinfo {author} {\bibfnamefont {S.}~\bibnamefont
  {Stevenson}},\ and\ \bibinfo {author} {\bibfnamefont {E.}~\bibnamefont
  {Thrane}},\ }\bibfield  {title} {\bibinfo {title} {{Signatures of Mass Ratio
  Reversal in Gravitational Waves from Merging Binary Black Holes}},\ }\href
  {https://doi.org/10.3847/1538-4357/ac8879} {\bibfield  {journal} {\bibinfo
  {journal} {Astrophys. J.}\ }\textbf {\bibinfo {volume} {938}},\ \bibinfo
  {pages} {45} (\bibinfo {year} {2022})},\ \Eprint
  {https://arxiv.org/abs/2205.01693} {arXiv:2205.01693 [astro-ph.HE]}
  \BibitemShut {NoStop}%
\bibitem [{\citenamefont {Bavera}\ \emph
  {et~al.}(2021{\natexlab{a}})\citenamefont {Bavera} \emph
  {et~al.}}]{Bavera:2020uch}%
  \BibitemOpen
  \bibfield  {author} {\bibinfo {author} {\bibfnamefont {S.~S.}\ \bibnamefont
  {Bavera}} \emph {et~al.},\ }\bibfield  {title} {\bibinfo {title} {{The impact
  of mass-transfer physics on the observable properties of field binary black
  hole populations}},\ }\href {https://doi.org/10.1051/0004-6361/202039804}
  {\bibfield  {journal} {\bibinfo  {journal} {Astron. Astrophys.}\ }\textbf
  {\bibinfo {volume} {647}},\ \bibinfo {pages} {A153} (\bibinfo {year}
  {2021}{\natexlab{a}})},\ \Eprint {https://arxiv.org/abs/2010.16333}
  {arXiv:2010.16333 [astro-ph.HE]} \BibitemShut {NoStop}%
\bibitem [{\citenamefont {Santini}\ \emph {et~al.}(2023)\citenamefont
  {Santini}, \citenamefont {Gerosa}, \citenamefont {Cotesta},\ and\
  \citenamefont {Berti}}]{Santini:2023ukl}%
  \BibitemOpen
  \bibfield  {author} {\bibinfo {author} {\bibfnamefont {A.}~\bibnamefont
  {Santini}}, \bibinfo {author} {\bibfnamefont {D.}~\bibnamefont {Gerosa}},
  \bibinfo {author} {\bibfnamefont {R.}~\bibnamefont {Cotesta}},\ and\ \bibinfo
  {author} {\bibfnamefont {E.}~\bibnamefont {Berti}},\ }\bibfield  {title}
  {\bibinfo {title} {{Black-hole mergers in disklike environments could explain
  the observed q-\ensuremath{\chi}eff correlation}},\ }\href
  {https://doi.org/10.1103/PhysRevD.108.083033} {\bibfield  {journal} {\bibinfo
   {journal} {Phys. Rev. D}\ }\textbf {\bibinfo {volume} {108}},\ \bibinfo
  {pages} {083033} (\bibinfo {year} {2023})},\ \Eprint
  {https://arxiv.org/abs/2308.12998} {arXiv:2308.12998 [astro-ph.HE]}
  \BibitemShut {NoStop}%
\bibitem [{\citenamefont {Zaldarriaga}\ \emph {et~al.}(2018)\citenamefont
  {Zaldarriaga}, \citenamefont {Kushnir},\ and\ \citenamefont
  {Kollmeier}}]{Zaldarriaga:2017qkw}%
  \BibitemOpen
  \bibfield  {author} {\bibinfo {author} {\bibfnamefont {M.}~\bibnamefont
  {Zaldarriaga}}, \bibinfo {author} {\bibfnamefont {D.}~\bibnamefont
  {Kushnir}},\ and\ \bibinfo {author} {\bibfnamefont {J.~A.}\ \bibnamefont
  {Kollmeier}},\ }\bibfield  {title} {\bibinfo {title} {{The expected spins of
  gravitational wave sources with isolated field binary progenitors}},\ }\href
  {https://doi.org/10.1093/mnras/stx2577} {\bibfield  {journal} {\bibinfo
  {journal} {Mon. Not. Roy. Astron. Soc.}\ }\textbf {\bibinfo {volume} {473}},\
  \bibinfo {pages} {4174} (\bibinfo {year} {2018})},\ \Eprint
  {https://arxiv.org/abs/1702.00885} {arXiv:1702.00885 [astro-ph.HE]}
  \BibitemShut {NoStop}%
\bibitem [{\citenamefont {Bavera}\ \emph
  {et~al.}(2021{\natexlab{b}})\citenamefont {Bavera}, \citenamefont {Zevin},\
  and\ \citenamefont {Fragos}}]{Bavera:2021evk}%
  \BibitemOpen
  \bibfield  {author} {\bibinfo {author} {\bibfnamefont {S.~S.}\ \bibnamefont
  {Bavera}}, \bibinfo {author} {\bibfnamefont {M.}~\bibnamefont {Zevin}},\ and\
  \bibinfo {author} {\bibfnamefont {T.}~\bibnamefont {Fragos}},\ }\href@noop {}
  {\bibinfo {title} {{Approximations to the spin of close Black-hole-Wolf-Rayet
  binaries}}} (\bibinfo {year} {2021}{\natexlab{b}}),\ \Eprint
  {https://arxiv.org/abs/2105.09077} {arXiv:2105.09077 [astro-ph.HE]}
  \BibitemShut {NoStop}%
\bibitem [{\citenamefont {Fuller}\ and\ \citenamefont
  {Lu}(2022)}]{Fuller:2022ysb}%
  \BibitemOpen
  \bibfield  {author} {\bibinfo {author} {\bibfnamefont {J.}~\bibnamefont
  {Fuller}}\ and\ \bibinfo {author} {\bibfnamefont {W.}~\bibnamefont {Lu}},\
  }\bibfield  {title} {\bibinfo {title} {{The spins of compact objects born
  from helium stars in binary systems}},\ }\href
  {https://doi.org/10.1093/mnras/stac317} {\bibfield  {journal} {\bibinfo
  {journal} {Mon. Not. Roy. Astron. Soc.}\ }\textbf {\bibinfo {volume} {511}},\
  \bibinfo {pages} {3951} (\bibinfo {year} {2022})},\ \Eprint
  {https://arxiv.org/abs/2201.08407} {arXiv:2201.08407 [astro-ph.HE]}
  \BibitemShut {NoStop}%
\bibitem [{\citenamefont {Qin}\ \emph {et~al.}(2018)\citenamefont {Qin},
  \citenamefont {Fragos}, \citenamefont {Meynet}, \citenamefont {Andrews},
  \citenamefont {S\o{}rensen},\ and\ \citenamefont {Song}}]{Qin:2018vaa}%
  \BibitemOpen
  \bibfield  {author} {\bibinfo {author} {\bibfnamefont {Y.}~\bibnamefont
  {Qin}}, \bibinfo {author} {\bibfnamefont {T.}~\bibnamefont {Fragos}},
  \bibinfo {author} {\bibfnamefont {G.}~\bibnamefont {Meynet}}, \bibinfo
  {author} {\bibfnamefont {J.}~\bibnamefont {Andrews}}, \bibinfo {author}
  {\bibfnamefont {M.}~\bibnamefont {S\o{}rensen}},\ and\ \bibinfo {author}
  {\bibfnamefont {H.~F.}\ \bibnamefont {Song}},\ }\bibfield  {title} {\bibinfo
  {title} {{The spin of the second-born black hole in coalescing binary black
  holes}},\ }\href {https://doi.org/10.1051/0004-6361/201832839} {\bibfield
  {journal} {\bibinfo  {journal} {Astron. Astrophys.}\ }\textbf {\bibinfo
  {volume} {616}},\ \bibinfo {pages} {A28} (\bibinfo {year} {2018})},\ \Eprint
  {https://arxiv.org/abs/1802.05738} {arXiv:1802.05738 [astro-ph.SR]}
  \BibitemShut {NoStop}%
\bibitem [{\citenamefont {Fuller}\ and\ \citenamefont
  {Ma}(2019)}]{Fuller:2019sxi}%
  \BibitemOpen
  \bibfield  {author} {\bibinfo {author} {\bibfnamefont {J.}~\bibnamefont
  {Fuller}}\ and\ \bibinfo {author} {\bibfnamefont {L.}~\bibnamefont {Ma}},\
  }\bibfield  {title} {\bibinfo {title} {{Most Black Holes are Born Very Slowly
  Rotating}},\ }\href {https://doi.org/10.3847/2041-8213/ab339b} {\bibfield
  {journal} {\bibinfo  {journal} {Astrophys. J. Lett.}\ }\textbf {\bibinfo
  {volume} {881}},\ \bibinfo {pages} {L1} (\bibinfo {year} {2019})},\ \Eprint
  {https://arxiv.org/abs/1907.03714} {arXiv:1907.03714 [astro-ph.SR]}
  \BibitemShut {NoStop}%
\bibitem [{\citenamefont {Rodriguez}\ \emph {et~al.}(2016)\citenamefont
  {Rodriguez}, \citenamefont {Zevin}, \citenamefont {Pankow}, \citenamefont
  {Kalogera},\ and\ \citenamefont {Rasio}}]{Rodriguez:2016vmx}%
  \BibitemOpen
  \bibfield  {author} {\bibinfo {author} {\bibfnamefont {C.~L.}\ \bibnamefont
  {Rodriguez}}, \bibinfo {author} {\bibfnamefont {M.}~\bibnamefont {Zevin}},
  \bibinfo {author} {\bibfnamefont {C.}~\bibnamefont {Pankow}}, \bibinfo
  {author} {\bibfnamefont {V.}~\bibnamefont {Kalogera}},\ and\ \bibinfo
  {author} {\bibfnamefont {F.~A.}\ \bibnamefont {Rasio}},\ }\bibfield  {title}
  {\bibinfo {title} {{Illuminating Black Hole Binary Formation Channels with
  Spins in Advanced LIGO}},\ }\href
  {https://doi.org/10.3847/2041-8205/832/1/L2} {\bibfield  {journal} {\bibinfo
  {journal} {Astrophys. J. Lett.}\ }\textbf {\bibinfo {volume} {832}},\
  \bibinfo {pages} {L2} (\bibinfo {year} {2016})},\ \Eprint
  {https://arxiv.org/abs/1609.05916} {arXiv:1609.05916 [astro-ph.HE]}
  \BibitemShut {NoStop}%
\bibitem [{\citenamefont {Gerosa}\ \emph {et~al.}(2018)\citenamefont {Gerosa},
  \citenamefont {Berti}, \citenamefont {O'Shaughnessy}, \citenamefont
  {Belczynski}, \citenamefont {Kesden}, \citenamefont {Wysocki},\ and\
  \citenamefont {Gladysz}}]{Gerosa:2018wbw}%
  \BibitemOpen
  \bibfield  {author} {\bibinfo {author} {\bibfnamefont {D.}~\bibnamefont
  {Gerosa}}, \bibinfo {author} {\bibfnamefont {E.}~\bibnamefont {Berti}},
  \bibinfo {author} {\bibfnamefont {R.}~\bibnamefont {O'Shaughnessy}}, \bibinfo
  {author} {\bibfnamefont {K.}~\bibnamefont {Belczynski}}, \bibinfo {author}
  {\bibfnamefont {M.}~\bibnamefont {Kesden}}, \bibinfo {author} {\bibfnamefont
  {D.}~\bibnamefont {Wysocki}},\ and\ \bibinfo {author} {\bibfnamefont
  {W.}~\bibnamefont {Gladysz}},\ }\bibfield  {title} {\bibinfo {title} {{Spin
  orientations of merging black holes formed from the evolution of stellar
  binaries}},\ }\href {https://doi.org/10.1103/PhysRevD.98.084036} {\bibfield
  {journal} {\bibinfo  {journal} {Phys. Rev. D}\ }\textbf {\bibinfo {volume}
  {98}},\ \bibinfo {pages} {084036} (\bibinfo {year} {2018})},\ \Eprint
  {https://arxiv.org/abs/1808.02491} {arXiv:1808.02491 [astro-ph.HE]}
  \BibitemShut {NoStop}%
\bibitem [{\citenamefont {Callister}\ \emph
  {et~al.}(2021{\natexlab{b}})\citenamefont {Callister}, \citenamefont {Farr},\
  and\ \citenamefont {Renzo}}]{Callister:2020vyz}%
  \BibitemOpen
  \bibfield  {author} {\bibinfo {author} {\bibfnamefont {T.~A.}\ \bibnamefont
  {Callister}}, \bibinfo {author} {\bibfnamefont {W.~M.}\ \bibnamefont
  {Farr}},\ and\ \bibinfo {author} {\bibfnamefont {M.}~\bibnamefont {Renzo}},\
  }\bibfield  {title} {\bibinfo {title} {{State of the Field: Binary Black Hole
  Natal Kicks and Prospects for Isolated Field Formation after GWTC-2}},\
  }\href {https://doi.org/10.3847/1538-4357/ac1347} {\bibfield  {journal}
  {\bibinfo  {journal} {Astrophys. J.}\ }\textbf {\bibinfo {volume} {920}},\
  \bibinfo {pages} {157} (\bibinfo {year} {2021}{\natexlab{b}})},\ \Eprint
  {https://arxiv.org/abs/2011.09570} {arXiv:2011.09570 [astro-ph.HE]}
  \BibitemShut {NoStop}%
\bibitem [{\citenamefont {Stevenson}(2022)}]{Stevenson:2022hmi}%
  \BibitemOpen
  \bibfield  {author} {\bibinfo {author} {\bibfnamefont {S.}~\bibnamefont
  {Stevenson}},\ }\bibfield  {title} {\bibinfo {title} {{Biases in Estimates of
  Black Hole Kicks from the Spin Distribution of Binary Black Holes}},\ }\href
  {https://doi.org/10.3847/2041-8213/ac5252} {\bibfield  {journal} {\bibinfo
  {journal} {Astrophys. J. Lett.}\ }\textbf {\bibinfo {volume} {926}},\
  \bibinfo {pages} {L32} (\bibinfo {year} {2022})},\ \Eprint
  {https://arxiv.org/abs/2202.03584} {arXiv:2202.03584 [astro-ph.HE]}
  \BibitemShut {NoStop}%
\bibitem [{\citenamefont {Gerosa}\ and\ \citenamefont
  {Berti}(2017)}]{Gerosa:2017kvu}%
  \BibitemOpen
  \bibfield  {author} {\bibinfo {author} {\bibfnamefont {D.}~\bibnamefont
  {Gerosa}}\ and\ \bibinfo {author} {\bibfnamefont {E.}~\bibnamefont {Berti}},\
  }\bibfield  {title} {\bibinfo {title} {{Are merging black holes born from
  stellar collapse or previous mergers?}},\ }\href
  {https://doi.org/10.1103/PhysRevD.95.124046} {\bibfield  {journal} {\bibinfo
  {journal} {Phys. Rev. D}\ }\textbf {\bibinfo {volume} {95}},\ \bibinfo
  {pages} {124046} (\bibinfo {year} {2017})},\ \Eprint
  {https://arxiv.org/abs/1703.06223} {arXiv:1703.06223 [gr-qc]} \BibitemShut
  {NoStop}%
\bibitem [{\citenamefont {Fishbach}\ \emph {et~al.}(2017)\citenamefont
  {Fishbach}, \citenamefont {Holz},\ and\ \citenamefont
  {Farr}}]{Fishbach:2017dwv}%
  \BibitemOpen
  \bibfield  {author} {\bibinfo {author} {\bibfnamefont {M.}~\bibnamefont
  {Fishbach}}, \bibinfo {author} {\bibfnamefont {D.~E.}\ \bibnamefont {Holz}},\
  and\ \bibinfo {author} {\bibfnamefont {B.}~\bibnamefont {Farr}},\ }\bibfield
  {title} {\bibinfo {title} {{Are LIGO's Black Holes Made From Smaller Black
  Holes?}},\ }\href {https://doi.org/10.3847/2041-8213/aa7045} {\bibfield
  {journal} {\bibinfo  {journal} {Astrophys. J. Lett.}\ }\textbf {\bibinfo
  {volume} {840}},\ \bibinfo {pages} {L24} (\bibinfo {year} {2017})},\ \Eprint
  {https://arxiv.org/abs/1703.06869} {arXiv:1703.06869 [astro-ph.HE]}
  \BibitemShut {NoStop}%
\bibitem [{\citenamefont {Tagawa}\ \emph {et~al.}(2021)\citenamefont {Tagawa},
  \citenamefont {Haiman}, \citenamefont {Bartos}, \citenamefont {Kocsis},\ and\
  \citenamefont {Omukai}}]{Tagawa:2021ofj}%
  \BibitemOpen
  \bibfield  {author} {\bibinfo {author} {\bibfnamefont {H.}~\bibnamefont
  {Tagawa}}, \bibinfo {author} {\bibfnamefont {Z.}~\bibnamefont {Haiman}},
  \bibinfo {author} {\bibfnamefont {I.}~\bibnamefont {Bartos}}, \bibinfo
  {author} {\bibfnamefont {B.}~\bibnamefont {Kocsis}},\ and\ \bibinfo {author}
  {\bibfnamefont {K.}~\bibnamefont {Omukai}},\ }\bibfield  {title} {\bibinfo
  {title} {{Signatures of hierarchical mergers in black hole spin and mass
  distribution}},\ }\href {https://doi.org/10.1093/mnras/stab2315} {\bibfield
  {journal} {\bibinfo  {journal} {Mon. Not. Roy. Astron. Soc.}\ }\textbf
  {\bibinfo {volume} {507}},\ \bibinfo {pages} {3362} (\bibinfo {year}
  {2021})},\ \Eprint {https://arxiv.org/abs/2104.09510} {arXiv:2104.09510
  [astro-ph.HE]} \BibitemShut {NoStop}%
\bibitem [{\citenamefont {Fishbach}\ \emph {et~al.}(2022)\citenamefont
  {Fishbach}, \citenamefont {Kimball},\ and\ \citenamefont
  {Kalogera}}]{Fishbach:2022lzq}%
  \BibitemOpen
  \bibfield  {author} {\bibinfo {author} {\bibfnamefont {M.}~\bibnamefont
  {Fishbach}}, \bibinfo {author} {\bibfnamefont {C.}~\bibnamefont {Kimball}},\
  and\ \bibinfo {author} {\bibfnamefont {V.}~\bibnamefont {Kalogera}},\
  }\bibfield  {title} {\bibinfo {title} {{Limits on Hierarchical Black Hole
  Mergers from the Most Negative \ensuremath{\chi} $_{eff}$ Systems}},\ }\href
  {https://doi.org/10.3847/2041-8213/ac86c4} {\bibfield  {journal} {\bibinfo
  {journal} {Astrophys. J. Lett.}\ }\textbf {\bibinfo {volume} {935}},\
  \bibinfo {pages} {L26} (\bibinfo {year} {2022})},\ \Eprint
  {https://arxiv.org/abs/2207.02924} {arXiv:2207.02924 [astro-ph.HE]}
  \BibitemShut {NoStop}%
\bibitem [{\citenamefont {Doctor}\ \emph {et~al.}(2019)\citenamefont {Doctor},
  \citenamefont {Wysocki}, \citenamefont {O'Shaughnessy}, \citenamefont
  {Holz},\ and\ \citenamefont {Farr}}]{Doctor:2019ruh}%
  \BibitemOpen
  \bibfield  {author} {\bibinfo {author} {\bibfnamefont {Z.}~\bibnamefont
  {Doctor}}, \bibinfo {author} {\bibfnamefont {D.}~\bibnamefont {Wysocki}},
  \bibinfo {author} {\bibfnamefont {R.}~\bibnamefont {O'Shaughnessy}}, \bibinfo
  {author} {\bibfnamefont {D.~E.}\ \bibnamefont {Holz}},\ and\ \bibinfo
  {author} {\bibfnamefont {B.}~\bibnamefont {Farr}},\ }\href
  {https://doi.org/10.3847/1538-4357/ab7fac} {\bibinfo {title} {{Black Hole
  Coagulation: Modeling Hierarchical Mergers in Black Hole Populations}}}
  (\bibinfo {year} {2019}),\ \Eprint {https://arxiv.org/abs/1911.04424}
  {arXiv:1911.04424 [astro-ph.HE]} \BibitemShut {NoStop}%
\bibitem [{\citenamefont {Kimball}\ \emph {et~al.}(2020)\citenamefont
  {Kimball}, \citenamefont {Talbot}, \citenamefont {L.~Berry}, \citenamefont
  {Carney}, \citenamefont {Zevin}, \citenamefont {Thrane},\ and\ \citenamefont
  {Kalogera}}]{Kimball:2020opk}%
  \BibitemOpen
  \bibfield  {author} {\bibinfo {author} {\bibfnamefont {C.}~\bibnamefont
  {Kimball}}, \bibinfo {author} {\bibfnamefont {C.}~\bibnamefont {Talbot}},
  \bibinfo {author} {\bibfnamefont {C.~P.}\ \bibnamefont {L.~Berry}}, \bibinfo
  {author} {\bibfnamefont {M.}~\bibnamefont {Carney}}, \bibinfo {author}
  {\bibfnamefont {M.}~\bibnamefont {Zevin}}, \bibinfo {author} {\bibfnamefont
  {E.}~\bibnamefont {Thrane}},\ and\ \bibinfo {author} {\bibfnamefont
  {V.}~\bibnamefont {Kalogera}},\ }\bibfield  {title} {\bibinfo {title} {{Black
  Hole Genealogy: Identifying Hierarchical Mergers with Gravitational Waves}},\
  }\href {https://doi.org/10.3847/1538-4357/aba518} {\bibfield  {journal}
  {\bibinfo  {journal} {Astrophys. J.}\ }\textbf {\bibinfo {volume} {900}},\
  \bibinfo {pages} {177} (\bibinfo {year} {2020})},\ \Eprint
  {https://arxiv.org/abs/2005.00023} {arXiv:2005.00023 [astro-ph.HE]}
  \BibitemShut {NoStop}%
\bibitem [{\citenamefont {Farah}\ \emph
  {et~al.}(2023{\natexlab{b}})\citenamefont {Farah}, \citenamefont {Edelman},
  \citenamefont {Zevin}, \citenamefont {Fishbach}, \citenamefont {Ezquiaga},
  \citenamefont {Farr},\ and\ \citenamefont {Holz}}]{Farah:2023vsc}%
  \BibitemOpen
  \bibfield  {author} {\bibinfo {author} {\bibfnamefont {A.~M.}\ \bibnamefont
  {Farah}}, \bibinfo {author} {\bibfnamefont {B.}~\bibnamefont {Edelman}},
  \bibinfo {author} {\bibfnamefont {M.}~\bibnamefont {Zevin}}, \bibinfo
  {author} {\bibfnamefont {M.}~\bibnamefont {Fishbach}}, \bibinfo {author}
  {\bibfnamefont {J.~M.}\ \bibnamefont {Ezquiaga}}, \bibinfo {author}
  {\bibfnamefont {B.}~\bibnamefont {Farr}},\ and\ \bibinfo {author}
  {\bibfnamefont {D.~E.}\ \bibnamefont {Holz}},\ }\bibfield  {title} {\bibinfo
  {title} {{Things That Might Go Bump in the Night: Assessing Structure in the
  Binary Black Hole Mass Spectrum}},\ }\href
  {https://doi.org/10.3847/1538-4357/aced02} {\bibfield  {journal} {\bibinfo
  {journal} {Astrophys. J.}\ }\textbf {\bibinfo {volume} {955}},\ \bibinfo
  {pages} {107} (\bibinfo {year} {2023}{\natexlab{b}})},\ \Eprint
  {https://arxiv.org/abs/2301.00834} {arXiv:2301.00834 [astro-ph.HE]}
  \BibitemShut {NoStop}%
\bibitem [{\citenamefont {Hogg}(1999)}]{Hogg:1999ad}%
  \BibitemOpen
  \bibfield  {author} {\bibinfo {author} {\bibfnamefont {D.~W.}\ \bibnamefont
  {Hogg}},\ }\href@noop {} {\bibinfo {title} {{Distance measures in
  cosmology}}} (\bibinfo {year} {1999}),\ \Eprint
  {https://arxiv.org/abs/astro-ph/9905116} {arXiv:astro-ph/9905116}
  \BibitemShut {NoStop}%
\end{thebibliography}%

\section{Appendix}

\subsection{Effective Spin Distribution and Primary Mass}
\label{app:chieff_m}

We also search for potential correlation in the population between the $\chieff$ distribution and the primary mass $m_1$. Previous studies have searched for the same correlation (see, e.g., Refs. \cite{Safarzadeh:2020mlb,Biscoveanu:2022qac}) using a linear model for the mean and width in Eq. \ref{eq:general_chieff_correlation}, and discovered weak evidence for a trend in the width of the $\chieff$ distribution. 

\begin{figure}[h!]
    \centering
    \includegraphics[width=0.9\linewidth]{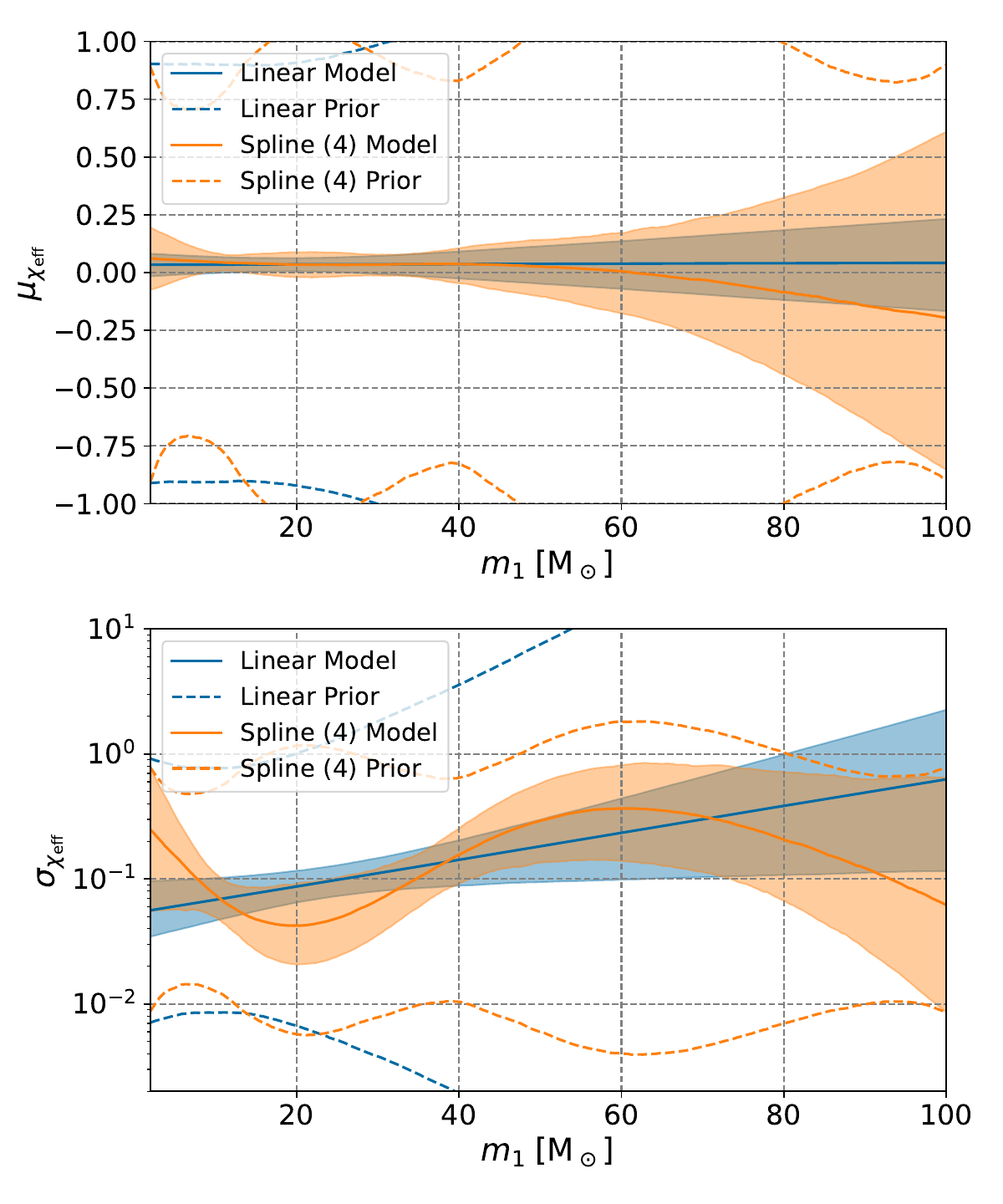}
    \caption{A comparison between the linear model and 4 node spline model for correlation between $\chieff$ and primary mass, inferred using \ac{GWTC-3}. Solid lines represent the median, while the shaded region represents the central 90\% credible interval, and the dashed lines show the boundary of the prior 90\% interval. The upper panel shows the mean of the $\chieff$ Gaussian as a function of primary mass $m_1$, while the lower panel represents the standard deviation of the $\chieff$ distribution.}
    \label{fig:chieff_m_s4_v_linear}
\end{figure}

Using the same spline approach described above, we place nodes between $m_{\rm 1, min} = 2$ and $m_{\rm 1, max} = 100$, linearly spaced in $m_1^{1/3}$. A uniform node spacing doesn't easily allow for structure at $\sim 10-30 \Msun$, which we would like to probe.
Instead, linearly spaced in $m_1^{1/3}$ clusters nodes towards $m_1 \sim 10-30 \Msun$, which we found to be satisfactory. 
\begin{align}
\mu(\theta) &= S\left(m_1\;|\;(2,\mu_{\chieff:0}), ..., (100,\mu_{\chieff:N})\right) \nn \\
\ln \sigma(\theta) &= S\left(m_1\;|\; (2,\ln \sigma_{\chieff:0}), ..., (100,\ln \sigma_{\chieff:N})\right).
\label{eq:chieff_m_spline}
\end{align}
We show a comparison between the spline model with 4 nodes and the linear model, in Fig. \ref{fig:chieff_m_s4_v_linear}. The linear model and spline models appear broadly consistent; indeed the model evidences do not show any preference for one model over another (see the appendix, Fig. \ref{fig:chieff_evidences}. We also show the results for all spline runs in Fig. \ref{fig:chieff_correlations} in the appendix). 
\begin{figure}[h!]
    \centering
    \includegraphics[width=0.9\linewidth]{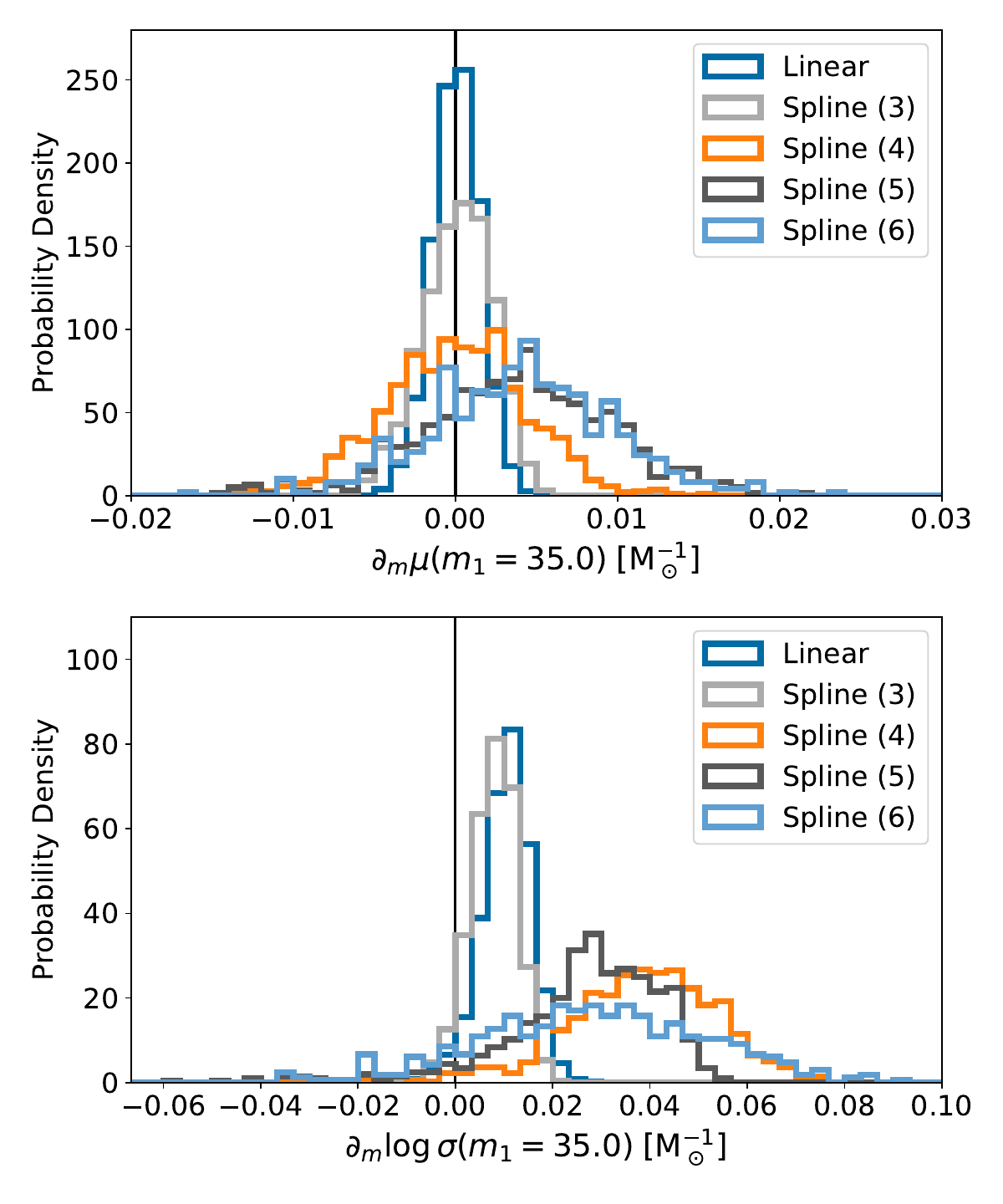}
    \caption{Posteriors on the derivative of the mean and width of the Gaussian as a function of primary mass. Each posterior is broadly consistent, with varying levels of confidence that the slope is greater than zero. The derivative of the mean at $m_0=35\Msun$ has nearly equal support for being positive of negative, while the derivative of the width is positive at $\sim 85-99.7\%$ confidence for each model.}
    \label{fig:chieff_m_slope}
\end{figure}

We also compute the slope of the mean and width at the fiducial value $m_1^* = 35 \Msun$, and here we find stronger evidence for a positive slope in some spline models than in the linear model. We chose $m_1^* = 35 \Msun$ to coincide roughly with the location of the Gaussian ``peak'' in the primary mass distribution~\cite{Talbot:2018cva, KAGRA:2021duu}, and so represents a physically interesting region of parameter space. The linear model finds a broadening at higher primary masses with significance $\sim 93\%$, while the spline models vary in significance, the model with 4 nodes notably exhibits a broadening at $98.7\%$ credence. We show posteriors on the slope at $m_1^* = 35\Msun$ in Fig. \ref{fig:chieff_m_slope}. 

\subsection{Inferences on \ac{GWTC-3}, extra analyses and evidences}
\label{app:extra_analysis}
\begin{figure}
    \centering
    \includegraphics[width=0.9\linewidth]{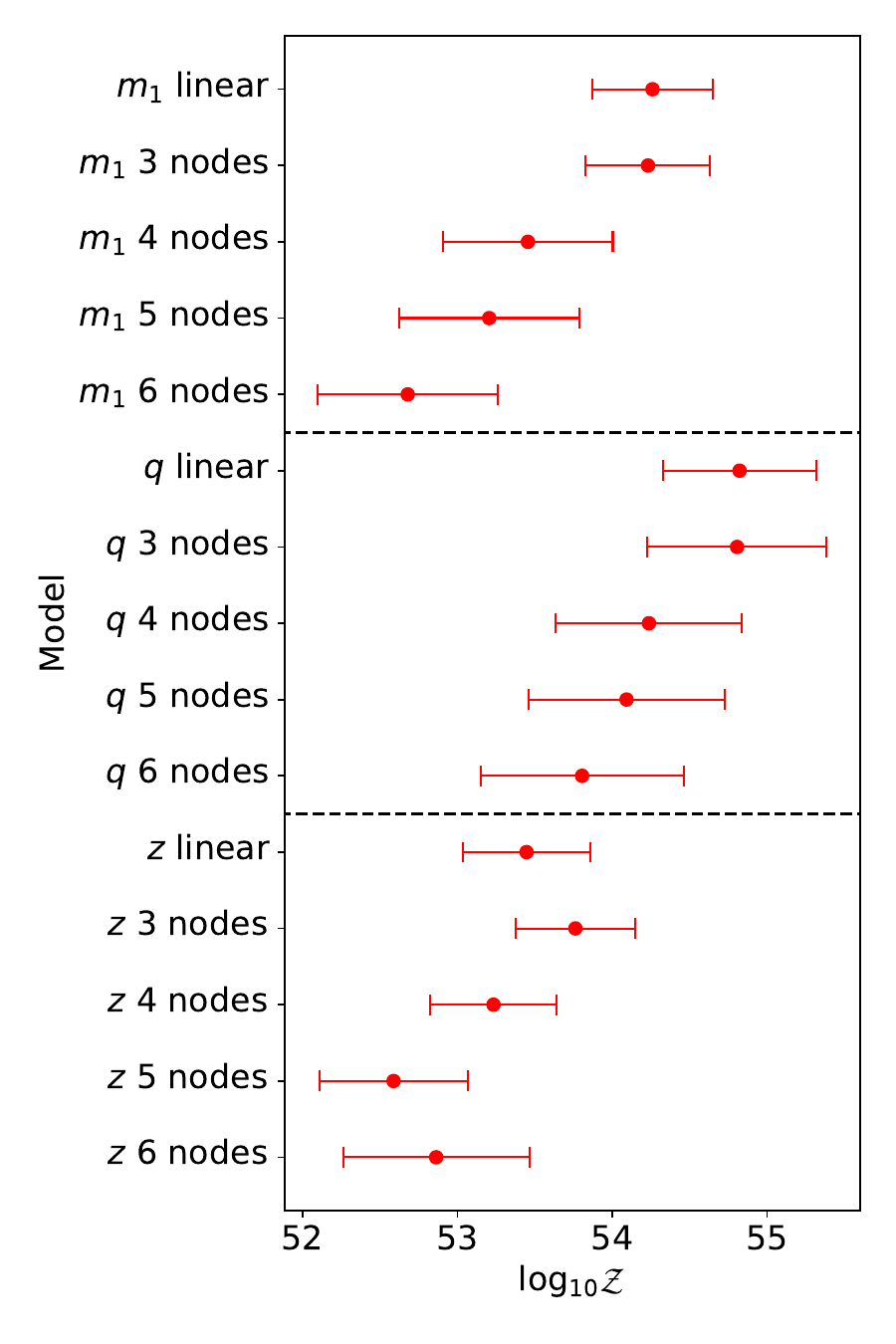}
    \caption{Evidences of the \ac{GWTC-3} catalog given each of the $\chieff$ correlation models. Evidence uncertainties include the average Monte Carlo uncertainty intrinsic to the likelihood estimator, added in quadrature with the nested sampling uncertainty of {\tt Dynesty}.}
    \label{fig:chieff_evidences}
\end{figure}

We also run each correlation inference with 3-6 nodes for the mean and standard deviation spline functions. For the $\chieff-q$ correlation, we show the constraints on the mean and standard deviation functions in Fig. \ref{fig:chieff_evidences}.
We do not observe any strong preference for a including more nodes in the spline correlation functions; there is a generic ``Occam's penalty'' for including more nodes beyond what is necessary to appropriately fit the data. We show the evidence of the \ac{GWTC-3} data given all our correlation models in Fig. \ref{fig:chieff_evidences}; note the evidences are still inconclusive at this stage, though a correlation with mass ratio is somewhat preferred. 

\begin{figure*}
    \centering
    \includegraphics[width=0.95\linewidth]{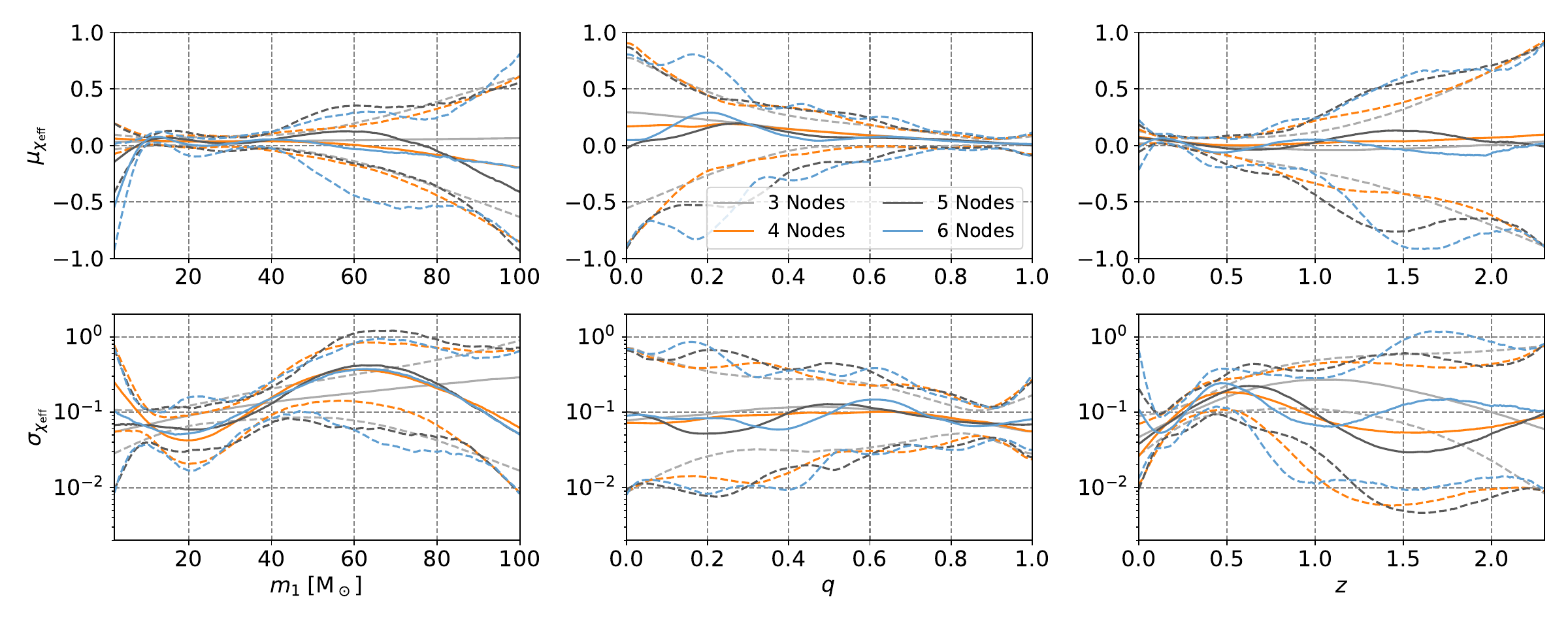}
    \caption{Spline correlations between $\chieff$ and each parameter we studied, primary mass $m_1$ (left column), mass ratio $q$ (center), and redshift $z$ (right). Solid lines represent the median, while the dashed lines represent the boundaries of the central 90\% credible interval. The upper panels represents the mean of the $\chieff$ Gaussian as a function of each parameter, while the lower panels represents the standard deviation of the $\chieff$ distribution.}
    \label{fig:chieff_correlations}
\end{figure*}

\renewcommand{\arraystretch}{1.5}
\begin{table*}
    \centering
    \begin{tabular}{|c|c|c|c|}
       \hline & $m_1$ (at $m_1^* = 35\Msun$) & $q$ (at $q^* = 0.9$) & $z$ (at $z^* = 0.2$)\\
       \hline Linear & \begin{tabular}{@{}c@{}} $p(\partial_m \mu > 0) = 53.9\%$ \\ $p(\partial_m \sigma > 0) = 93.2\%$ \end{tabular} & \begin{tabular}{@{}c@{}} $p(\partial_q \mu > 0) = 2.5\%$ \\ $p(\partial_q \sigma > 0) = 41.7\%$ \end{tabular} & \begin{tabular}{@{}c@{}} $p(\partial_z \mu > 0) = 16.2\%$ \\ $p(\partial_z \sigma > 0) = 92.7\%$ \end{tabular} \\
       \hline Spline (3) & \begin{tabular}{@{}c@{}} $p(\partial_m \mu > 0) = 55.0\%$ \\ $p(\partial_m \sigma > 0) = 89.5\%$ \end{tabular} & \begin{tabular}{@{}c@{}} $p(\partial_q \mu > 0) = 13.4\%$ \\ $p(\partial_q \sigma > 0) = 33.0\%$ \end{tabular} & \begin{tabular}{@{}c@{}} $p(\partial_z \mu > 0) = 14.2\%$ \\ $p(\partial_z \sigma > 0) = 95.6\%$ \end{tabular} \\
       \hline Spline (4) & \begin{tabular}{@{}c@{}} $p(\partial_m \mu > 0) = 60.5\%$ \\ $p(\partial_m \sigma > 0) = 98.7\%$ \end{tabular} & \begin{tabular}{@{}c@{}} $p(\partial_q \mu > 0) = 18.7\%$ \\ $p(\partial_q \sigma > 0) = 44.5\%$ \end{tabular} & \begin{tabular}{@{}c@{}} $p(\partial_z \mu > 0) = 19.6\%$ \\ $p(\partial_z \sigma > 0) = 98.0\%$ \end{tabular} \\
       \hline Spline (5) & \begin{tabular}{@{}c@{}} $p(\partial_m \mu > 0) = 78.3\%$ \\ $p(\partial_m \sigma > 0) = 95.7\%$ \end{tabular} & \begin{tabular}{@{}c@{}} $p(\partial_q \mu > 0) = 21.7\%$ \\ $p(\partial_q \sigma > 0) = 58.1\%$ \end{tabular} & \begin{tabular}{@{}c@{}} $p(\partial_z \mu > 0) = 17.0\%$ \\ $p(\partial_z \sigma > 0) = 88.8\%$ \end{tabular} \\
       \hline Spline (6) & \begin{tabular}{@{}c@{}} $p(\partial_m \mu > 0) = 77.8\%$ \\ $p(\partial_m \sigma > 0) = 93.6\%$ \end{tabular} & \begin{tabular}{@{}c@{}} $p(\partial_q \mu > 0) = 30.0\%$ \\ $p(\partial_q \sigma > 0) = 65.5\%$ \end{tabular} & \begin{tabular}{@{}c@{}} $p(\partial_z \mu > 0) = 5.9\%$ \\ $p(\partial_z \sigma > 0) = 98.6\%$ \end{tabular} \\
       \hline 
    \end{tabular}
    \caption{Credibility of a positive slope in the evolution of the mean (denoted $\delta \mu > 0$) at the fiducial value, and of a positive slope in the evolution of the width (denoted $\delta \sigma > 0$). The column on the left is the model assumed, where ``Spline ($N$)'' refers to the spline model using $N$ nodes. In each cell, the top line represents the credibility of a positive slope in the mean, and the bottom line represents the same quantity for the width.}
    \label{tab:slope_significance}
\end{table*}

\subsection{Nonlinear Correlation in $\chieff-z$}
\label{app:nonlinear_z}
\begin{figure}
    \centering
    \includegraphics[width=0.9\linewidth]{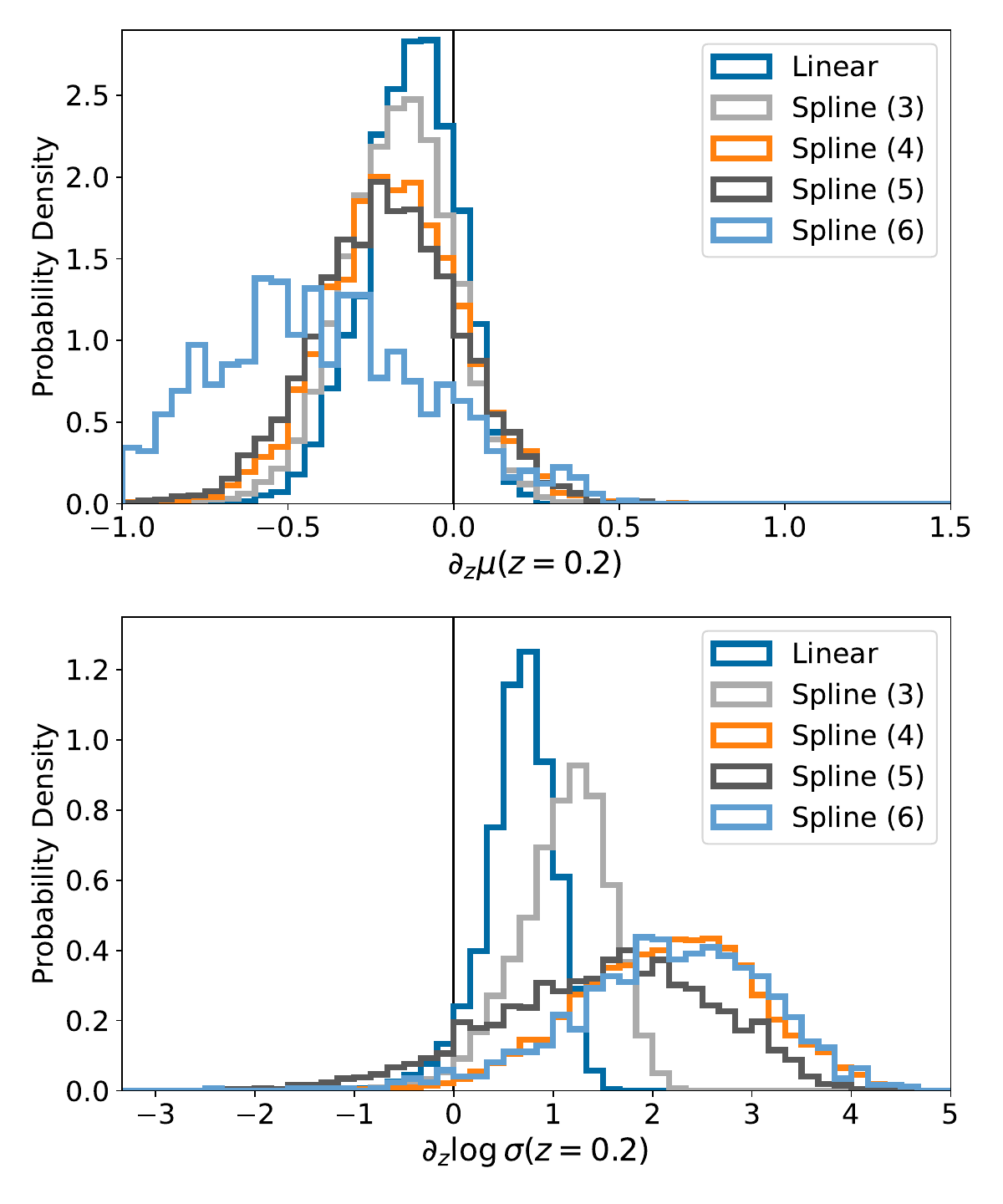}
    \caption{Posteriors on the derivative of the mean and width of the Gaussian with respect to redshift at $z^* = 0.2$, inferred on \ac{GWTC-3} data. Each posterior is consistent, with varying levels of confidence that the slope is greater than zero.}
    \label{fig:chieff_z_slope}
\end{figure}

To study the potential risk of inferring a nonlinear correlation a \ac{GWTC-3}-like catalog from a linearly correlated Universe, we used 12 sets of 69 detections drawn from a uncorrelated Universe observed with Hanford and Livingston in O4-like PSDs \cite{O3_psds}, in a similar process to the method presented in \S \ref{sec:o4_projections}. These synthetic events were injected from a linearly correlated Universe (namely, uncorrelated with slope zero), with a mean $\mu_{\chieff} = 0.06$ and width $\sigma_{\chieff} = 0.11$, and the same thresholds, waveforms, and PE techniques described in \S \ref{sec:o4_projections}. Then, we infer the $\chieff$ correlation with respect to the redshift $z$ for each synthetic catalog, using the method we introduced in \S \ref{sec:GWTC3_results}. We compute the instantaneous slope at the first three nodes for each hyperposterior sample, and can then estimate the \ac{JS} divergence between the posteriors of the slopes at neighboring nodes. 

We also show the slope posteriors at a fiducial value $z^*=0.2$ in Fig. \ref{fig:chieff_z_slope}. We avoid choosing $z^*=0$, as the data is uninformative in the limit of $z\to 0$ for the same reason it is uninformative for $q\to 0$. While some of the 69 BBH events are closer than others, none of them are consistent with being at $z\to 0$, and indeed it is baked into the population model where the probability density at a given redshift is proportional to the differential comoving volume $p(z) \propto \frac{dV_c}{dz}$ to have zero support at $z\to 0$ \cite{Hogg:1999ad, Fishbach:2018edt}. In other words, a flexible $z-\chieff$ correlation model, in particular spline models with appropriate node density and placement, will increase in uncertainty as $z\to 0$. Indeed, we observe this to some degree in each spline model in Fig. \ref{fig:chieff_correlations}, but it is especially clear in the spline model with 5 nodes. 

We also should note the significance of broadening with redshift using the linear model is only 92.7\%, while Ref.~\cite{Biscoveanu:2022qac} inferred a broadening at 98.6\% credibility. We checked various differences between our analyses, and discovered that this difference arises from how selection effects are estimated. In particular, Ref.~\cite{Biscoveanu:2022qac} assumes a semi-analytic threshold on the network SNR $\rho_{\rm net} > 9$ for O1 and O2 events, while we use a threshold of $\rho_{\rm net} > 10$, following Ref.~\cite{KAGRA:2021duu}.

\subsection{Node Placement and Bias in Mock Catalogs}
\label{app:node_placement_bias}
In \S \ref{sec:o4_projections} we study a Universe with a nonlinear correlation in the $\chieff-q$ plane. We find that, when we fix the x-axis positions of the nodes to be different than the true node positions, we may recover some bias in the inference.

Specifically, the results in Fig. \ref{fig:chieff_q_mock_catalog_s4_v_linear} highlight one of the drawbacks for using spline models. Fitting a spline model to a smooth function often results in extraneous structure, bumps that the spline function cannot remove entirely due to its functional form \cite{Farah:2023vsc}. In the case of the mock catalog we simulated, the correlation is created using a spline node placement (Table \ref{tab:true_hyper}) different from the spline node placement scheme we recover with. Because of this, the spline cannot simultaneously fit the correlation in the region of near equal mass ratio while also effectively fitting the region of more extreme mass ratios, and so it must compromise with a suboptimal fit across the parameter space. Hence the spline model ``wiggles'' when it ideally should be smooth.  

Because we are not yet in the limit of infinitely many nodes (we use 4 nodes in this inference), the spline functions are not perfectly flexible. More nodes are inherently more flexible, and so would presumably do a better job at fitting the correlation. Similarly, using a correct node placement scheme should also improve the fit. Indeed, we find that when we analyze the mock catalog using the same node positions as the true correlation (Table \ref{tab:true_hyper}), the inferred correlation is closer to the truth. This highlights the importance of either running with multiple node counts and placement schemes, or marginalizing over the node x-axis positions as well.


\end{document}